\documentclass[a4paper,superscriptaddress,nofootinbib]{revtex4-1}
\usepackage{epsfig}
\usepackage{amsfonts}
\usepackage{amsmath}
\usepackage{amssymb}
\usepackage{amsthm}
\usepackage{array}
\usepackage{booktabs}
\usepackage{color}
\usepackage{graphicx}
\usepackage{hyperref}
\usepackage{multirow}
\usepackage{slashed}
\usepackage{subfigure}
\usepackage{theorem}
\usepackage{textcomp}

\allowdisplaybreaks[1]

\numberwithin{equation}{section}

\newcommand{\be}{\begin{equation}}
\newcommand{\ee}{\end{equation}}
\newcommand{\beq}{\begin{equation}}
\newcommand{\eeq}{\end{equation}}
\newcommand{\bea}{\begin{eqnarray}}
\newcommand{\eea}{\end{qnarray}}

\def\<{\left\langle}
\def\>{\right\rangle}

\begin{document}

\begin{flushright}
 \hspace{3cm} LAPTH-008/16
\end{flushright}

\title{Drell-Yan production of multi $Z'$-bosons at the LHC within \break Non-Universal ED and 4D  Composite Higgs Models}

\author{Elena Accomando}
\email[E-mail: ]{e.accomando@soton.ac.uk}
\affiliation{School of Physics \& Astronomy, University of Southampton,
        Highfield, Southampton SO17 1BJ, UK}
\affiliation{Particle Physics Department, Rutherford Appleton Laboratory, 
       Chilton, Didcot, Oxon OX11 0QX, UK}

\author{Daniele Barducci}
\email[E-mail: ]{daniele.barducci@lapth.cnrs.fr}
\affiliation{LAPTh, Universit{\'e}\ Savoie Mont Blanc, CNRS, B.P.110, F-74941 Annecy-le-Vieux, France}

\author{Stefania De Curtis}
\email[E-mail: ]{stefania.decurtis@fi.infn.it}
\affiliation{INFN, Sezione di Firenze, and Department of Physics and Astronomy, University of Florence, Via G. Sansone 1, 50019 Sesto Fiorentino, Italy}

\author{Juri Fiaschi}
\email[E-mail: ]{juri.fiaschi@soton.ac.uk}
\affiliation{School of Physics \& Astronomy, University of Southampton,
        Highfield, Southampton SO17 1BJ, UK}
\affiliation{Particle Physics Department, Rutherford Appleton Laboratory, 
       Chilton, Didcot, Oxon OX11 0QX, UK}

\author{Stefano Moretti}
\email[E-mail: ]{S.Moretti@soton.ac.uk}
\affiliation{School of Physics \& Astronomy, University of Southampton,
        Highfield, Southampton SO17 1BJ, UK}
\affiliation{Particle Physics Department, Rutherford Appleton Laboratory, 
       Chilton, Didcot, Oxon OX11 0QX, UK}

\author{C.H. Shepherd-Themistocleous}
\email[E-mail: ]{S.Moretti@soton.ac.uk}
\affiliation{School of Physics \& Astronomy, University of Southampton,
        Highfield, Southampton SO17 1BJ, UK}
\affiliation{Particle Physics Department, Rutherford Appleton Laboratory, 
       Chilton, Didcot, Oxon OX11 0QX, UK}

\begin{abstract}
\noindent
The Drell-Yan di-lepton production at hadron colliders is by far the preferred channel to search for new heavy spin-1 particles. Traditionally, such searches have exploited the Narrow Width Approximation (NWA) for the signal, thereby neglecting the effect of the interference between the additional $Z'$-bosons and the Standard Model $Z$ and $\gamma$. Recently, it has been established that both finite width and interference effects can be dealt with in experimental searches while still retaining the model independent approach ensured by the NWA.
This assessment has been made for the case of popular single $Z'$-boson models currently probed
at the CERN Large Hadron Collider (LHC). In this paper, we test the scope of the CERN machine in relation to the above issues for some benchmark multi $Z'$-boson models. In particular, we consider  Non-Universal Extra Dimensional  (NUED) scenarios and the 4-Dimensional Composite Higgs Model (4DCHM), both predicting a multi-$Z'$ peaking structure.
We conclude that in a variety of cases, specifically those in which the leptonic decays modes of one or more of the heavy neutral gauge bosons are suppressed and/or significant interference effects exist between these or with the background, especially present when their decay widths are significant, traditional search approaches based on the assumption of rather narrow and isolated objects might require suitable modifications to extract the underlying dynamics.  
\end{abstract}

\pacs{NN.NN.NN.Cc}

\maketitle

\tableofcontents

\setcounter{footnote}{0}

%%%%%%%%%%%%%%%%%%%%%%%%%%%%%%%%%%%%%%%%%%%
\section{Introduction}\label{sec:intro}

Extra neutral uncoloured spin-1 particles, usually called $Z^\prime$
and $\gamma^\prime$, are a common feature of Beyond the Standard Model
(BSM) scenarios of Electro-Weak Symmetry Breaking (EWSB), which can
arise from general extensions of its gauge group motivated by Grand
Unified Theories (GUTs) \cite{Accomando:2010fz, Langacker:2008yv}, Kaluza Klein (KK) excitations of SM gauge fields in extra dimensions \cite{Antoniadis:1990ew,Antoniadis:1993jp,Antoniadis:1992fh,Antoniadis:1994yi,Benakli:1995ut}, models of compositeness \cite{Contino:2010rs}, Technicolor \cite{Belyaev:2008yj}  and some variants
of Supersymmetric theories to name but a few. Typically, at hadron
colliders, such objects are searched for via Drell-Yan (DY) production
into two leptons: $pp(\bar p) \to \gamma, Z, Z^\prime ,\gamma^\prime
\to \ell^+\ell^-$, where $\ell=e,\mu$. This channel has the advantages
of low backgrounds and good mass resolutions.  The latest limits of
relevance here are those obtained by the CMS
collaboration~\cite{Khachatryan:2014fba} using data collected during
Run I at the CERN Large Hadron Collider (LHC) with $\sqrt{\hat{s}}= 8$
TeV and full integrated luminosity ${\cal L}\simeq 20$ $fb^{-1}$. The
data analysis has enabled the extraction of mass bounds at around $2.5$ TeV
on several different $Z'$-bosons predicted by a variety of $Z^\prime$
models. Such limits are derived by searching for a resonance (a
so-called `bump search') in the cross section as a function of the
invariant mass of dileptons, $M_{ll}$. The analysis is performed
under the assumption that the resonance is relatively narrow, so that
Finite Width (FW) and interference effects of the new heavy $Z'$-boson
with the SM $\gamma$ and $Z$ vector bosons can be neglected in the
first instance. The signal rate is typically estimated using the
so-called Narrow Width Approximation (NWA) and the signal line-shape
is modelled by a Breit-Wigner function convoluted with a Gaussian
function, which is used to describe the dilepton mass resolution. The results
are presented as 95\% Confidence Level (C.L.) upper bounds on the
$Z'$-boson production cross section times the $Z'$ branching ratio,
normalized to the SM cross section at the $Z$-boson peak. In modelling
the signal in this way, the CMS collaboration adopts an approach through which
model independent limits on the cross section can be derived (see
Ref. ~\cite{Khachatryan:2014fba} for details). These can in turn be
interpreted as constraints on the mass of the $Z^\prime$-boson
pertaining to a specific model (i.e., the model dependence is only
contained in the dilepton Branching Ratio (BR) of the assumed $Z'$).
The interpretation of the experimental results within any theoretical framework is complicated by effects such as FW and/or interference. This is explicitly shown in Ref.~\cite{Accomando:2013sfa}, where these depend on the model being considered and can significantly affect the theoretical predictions. FW and interference effects have been studied by a number of authors also in different processes (see for example Refs. ~\cite{Dittmar:1996my, Rizzo:2007xs,Petriello:2008zr,Papaefstathiou:2009sr,Rizzo:2009pu,Chiang:2011kq, Accomando:2011eu, Choudhury:2011cg, Boos:2006xe, deBlas:2012qp}). For heavy neutral vector boson production in Drell-Yan, only within narrow width $Z^\prime$-boson scenarios can the deviations
from the NWA due to the above two phenomena be safely neglected at least to
some degree.  Following the recommendation of
\cite{Accomando:2013sfa}, when interpreting the derived 95\%
C.L.\ upper bound on the BSM cross section to extract the mass limits
within a specific model, the CMS collaboration restricts the
integration range of the differential cross section to the invariant
mass of the dilepton pair in the window $|M_{ll}-M_{Z^\prime}|\le
0.05\ \times E_{\rm LHC}$ where $E_{\rm LHC}$ is the collider energy
and $M_{Z^\prime}$ the hypothetical pole mass of the new
$Z'$-boson. This mass range (also called in the literature {\it
  optimal} or {\it magic}) is designed so that the systematic errors
in neglecting the model-dependent FW and interference effects (between
$\gamma , Z, Z'$) are kept below $O(10\% )$ for a large class of
narrow $Z'$-boson models and for the full range of $Z^\prime$-boson
masses that can be reached at the current LHC Run II. This approach
allows for a straightforward interpretation of the data analysis
results in the context of any theory predicting a narrow width
$Z'$-boson.  In the case of a wide $Z'$-boson, the prescription of
Ref. \cite{Accomando:2013sfa} is no longer appropriate and use should
be made of a model-dependent analysis where FW and/or
interference effects exist. A search for high mass resonances decaying to
dilepton final states has also been performed by the ATLAS collaboration 
and bounds on the $Z'$-boson mass have been extracted under the assumption of 
a number of specific $Z'$-boson models. When interpreting their
experimental results within the same theory, CMS and ATLAS exhibit good
agreement on the obtained exclusion limits. Typically, ATLAS considers
theoretical models predicting resonances that are narrow relative to
the detector resolution. In such cases, interference effects are not
taken into account. The exception to this is the class of Minimal $Z'$
Models for which large coupling strengths, and hence larger widths,
are considered.  In this case, interference effects are included
explicitly in the ATLAS analysis (see Ref.~\cite{Aad:2014cka} and
references therein for details).

The experimental analyses have been designed to address the $Z'$-boson
search in the single-resonance scenario and prescriptions to bridge
the data analysis results and the theoretical interpretation within
explicit $Z'$-boson theories have been given. It is the purpose of
this paper to analyse the above phenomenology within scenarios
predicting multiple $Z'$-bosons. Herein, further challenges appear as,
in several well-motivated theoretical models, such $Z'$ states can be
quite close in mass and mix with each other so that various scenarios
might emerge. Two such resonances may be wide and close enough in mass 
so as to appear as a single broad resonance in the dilepton invariant mass
spectrum. These resonances may interfere strongly with each other
and/or with the SM background, thereby further blurring the usual
procedures adopted in profiling a possible excess.
We illustrate these features within two classes of
models: the Non-Universal Extra Dimensional (NUED) scenario and the
4-Dimensional Composite Higgs Model (4DCHM). The first belongs to
the multi-$Z'$ weakly coupled class of theories while the second is an
example of a strongly interacting theory. We show that NUED extra
gauge bosons can be searched for and theoretically interpreted using
the traditional techniques currently employed by the CMS
experiment. In contrast, the 4DCHM requires a modified approach
for setting limits on masses and/or couplings of the extra heavy gauge
bosons.

The content of our work is as follows\footnote{A brief account of this
can be found in Ref. \cite{Accomando:2015cva}.}. In
Sec. \ref{sec:ED}, we introduce the first multi-$Z'$-boson model used
as benchmark, that is the NUED model, and some of its most recent
variations. This model predicts a tower of KK-excitations of the SM
$\gamma$ and $Z$-bosons: $\gamma_{KK}^{n}$ and $Z_{KK}^{n}$ where $n$
indicates the tower's level. In Sec. \ref{subsec:searches_ED}, we
describe its phenomenology, focussing on the first level KK modes as
they might be produced in the DY channel at the present LHC Run II. We
show that the standard experimental setup adopted by CMS for the
"bump" hunt works well in searching for these objects. Moreover, the
95\% C.L.\ upper limit on the BSM cross section given by the
experimentalists as a result of their dilepton spectrum analysis can
be directly and unambiguously interpreted within such a model in order
to extract mass bounds on $\gamma_{KK}^{1}$ and $Z_{KK}^{1}$. These
extra gauge bosons are indeed expected to be rather narrow.
Finally, in Sect. \ref{subsec:profile_ED}, we show that if the BSM
giving rise to any any new observed resonances were to be NUED, then a
novel experimental strategy will be required to correctly interpret
the data.

We do the same in Sect.~\ref{sec:4DCHM} within the 4DCHM. This
scenario predicts five extra heavy $Z'$-bosons, two of which might be
active in the DY process. We initially describe its first principles
and particle content, then in Sect. \ref{sec:4DCHM_pheno} we address
its phenomenological consequences. We thus illustrate the type of
signatures one might expect to observe in data acquired at the LHC
during Run II.  We start by describing the most popular representation
which is adopted for general CHMs, that is the single $Z'$-boson
reduction. In Sect. \ref{sec:4DCHM-reduction}, we consider a
simplified version of the 4DCHM where only the $Z'_3$ boson is
active. We then compare our single-resonant scenario to the
literature. In addition, we discuss the validity of the NWA in this
reduced context. In Sect. \ref{sec:4DCHM-limits}, we illustrate the
complete picture of the 4DCHM, opening up the full multi-resonant
structure of the model. We derive the exclusion limits on mass and
coupling strength of the new gauge bosons from the data collected
during the past LHC Run I with 8 TeV energy and luminosity ${\cal L}\simeq 20 fb^{-1}$. In doing so, we raise the issue of
implementing a modified experimental fitting procedure accounting for
the multi-$Z'$-boson signal. A compressed peaking structure and the
presence of a sizeable dip before the resonances, owing to the
interference between the new vector bosons and the SM background, are
in fact notable features of this novel signal shape that might require
a dedicated approach. This is presented in
Sect. \ref{sec:4DCHM-signal}, where we show in detail the type of
signatures which could be produced during Run II at the LHC. We point
out where the default fitting method could be modified. Finally, in
Sect. \ref{sec:4DCHM-profile}, we address the question of how to
identify the signal as coming from a CHM. We then summarise and
conclude in Sect.~\ref{sec:summa}.

\section{The ED model}\label{sec:ED}

One of the flawed parts of the SM  concerns the understanding of the gravitational interactions. Such interactions in fact destroy the renormalisability of the theory and give rise to the hierarchy problem. As these quantum gravity effects seem to imply the existence of extended objects living in more than four dimensions, a possible solution to these problems is provided by a scenario of large EDs and a low scale quantum gravity in the TeV region \cite{ArkaniHamed:1998rs,Antoniadis:1998ig}. Within this scenario, a natural question is how to detect the EDs. The answer can only be given for specific classes of models, as it depends on the details of the realisation of the EDs and the way known particles emerge inside them. The theoretical scenario analysed here is based on the  model of Refs. \cite{Antoniadis:1990ew,Antoniadis:1993jp,Antoniadis:1992fh,Antoniadis:1994yi,Benakli:1995ut}, when embedded in the  framework described in Refs. \cite{ArkaniHamed:1998rs,Antoniadis:1998ig}. This setup  is called the NUED model. Here, all SM fermions are totally localised on the brane whereas  all SM gauge bosons are fully propagating into the bulk.  One of its simplified versions, called NUED(EW), predicts that only the EW SM gauge bosons are allowed to propagate in the EDs as proposed in Ref. \cite{Bella:2010sc}. Our study is representative of both models, the original and the simplified one.

\noindent
In these two scenarios, two fundamental energy scales play a major role. The first one, $M_s=l_s^{-1}$, is related to the inner structure of the basic objects of the theory, that we assume to be elementary strings. Their point-like behaviour is viewed as a low-energy phenomenon: above $M_s$ the string oscillation modes get excited making their true extended nature manifest. The second important scale, $R^{-1}$, is associated with the existence of a higher dimensional space: above $R^{-1}$ new dimensions open up and particles, called KK excitations, can propagate in them. The number of EDs, $D$, which are compactified on a $D$-dimensional torus, can be as big as six \cite{ArkaniHamed:1998rs} or seven \cite{Antoniadis:1998ig}. Here, we consider a NUED model in 5 dimensions. The particle content of this model can be described as follows. The gravitons, represented as closed strings, can propagate in the whole higher-dimensional space, 3+$d_\parallel$+$d_\perp$. Here, 3+$d_\parallel$ defines the longitudinal dimension of the big brane, which contains the small 3D brane where the observed SM particles live. The symbol $d_\perp$ indicates the EDs transverse to the big brane, which are felt only by gravity. The SM gauge bosons, represented as open strings, can propagate only on the (3+$d_\parallel$)-brane. The SM fermions are localised on the 3D brane, which intersects the (3+$d_\parallel$)-dimensional one. They do not propagate in EDs (neither $d_\parallel$ nor $d_\perp$), hence they do not have KK-excitations.

\noindent
From this picture it is clear that in our scenario $D=d_\parallel=1$. Assuming periodic conditions on the wave functions along each compact direction, the states propagating in the $(4+D)$-dimensional space are seen from the 4D point of view as a tower of states having a squared mass:
\begin{equation}
M^2_{KK}\equiv M^2_{\vec n} = m_0^2 +\frac {n^2}{R^2} \, ,
\label{KKdef}
\end{equation}
with $m_0$ the 4D mass and $n$ a non-negative integer. The states with $n \neq 0$ are called KK states. Since, in the class of NUED models in 5D, KK modes exist only for the gauge bosons, while fermions have no KK states, obviously, the particle content is very different from the ordinary SM inventory. The fermionic sector remains practically unchanged, but for each gauge boson we encounter a zero mode, together with a tower of complementary particles of higher mass, $M_{KK}$. The usual interpretation in terms of 4D particles is that the zero modes are the known SM gauge bosons, while the KK states are their heavier copies. Hence, more explicitly, in the NUED model all $SU(3)\times SU(2)\times U(1)$ SM gauge bosons propagate into the bulk 5D space and therefore have KK-excitations. In the more recent NUED(EW) construct, only the $SU(2)\times U(1)$ EW gauge bosons can propagate in the compactified ED and acquire KK excitations \cite{Bella:2010sc}. In both models the fermionic content is totally confined on the 3D brane. These two scenarios share most part of their features and just differ for the gluon contribution to fully hadronic or semileptonic processes at the LHC which are not addressed in our analysis. So, our results are  valid in both scenarios.

\noindent
Assuming that leptons and quarks are localised on the brane is quite a distinctive feature of the class of NUED models, giving rise to well defined predictions. An immediate consequence of the localisation is that fermion interactions preserve the momenta in the four-dimensional world but violate the energy-momentum conservation along the additional fifth dimension. One can thus produce single KK excitations, for example, via $f\bar{f^\prime}\rightarrow V_{KK}^{(n)}$ where $f,f^\prime$ are fermions and $V_{KK}^{(n)}$ represents a massive KK excitation of $W,Z,\gamma ,g$ gauge bosons. Conversely, gauge boson interactions conserve the momenta along all 4+1 dimensions, making the self-interactions of the kind $VV\rightarrow V_{KK}^{(n)}$ forbidden. Owing to these interactions, KK states or their indirect effects could have been detected at LEP and/or LHC in principle. An updated review on both indirect and direct exclusion limits on KK-particles, predicted within the class of NUED models, can be found in Ref. \cite{Accomando:2015rsa}. The indirect limits come from the EW Precision Tests (EWPTs) at LEP, as the presence of KK excitations can in principle affect the computation of the low-energy precision observables through the (re-)definition of the Fermi constant, $G_F$, weak mixing angle and masses of the SM vector bosons. 

\noindent
The constraints on $M_{KK}$ extracted from the EWPTs have a strong dependence on the realisation of the scalar sector in the 5D NUED model(s). There are no physical considerations dictating that the Higgs boson should be a brane field or the zero mode of a bulk field. It is very common in the literature to consider a scenario where both these options are realised, the discovered Higgs being a mixture of these. The relative contribution of the two fields is parametrised by $\tan\beta=\frac{<\phi_2>}{<\phi_1>}$ or, equivalently, $\sin\beta$. Here, $\phi_1$ is the bulk field, so for $\sin\beta=0$ we are in a model with only a bulk Higgs state. The important point here is that, for $\sin\beta\ne 0$, the Vacuum Expectation Value (VEV) of the brane field can cause mixing between the different modes of the gauge bosons and the weak eigenstates are no longer mass eigenstates. The ensuing diagonalisation to determine the mass eigenvalues leads to a model dependent redefinition of gauge boson masses and couplings, which receive additional corrections from the KK states due to the rotation in state space. The strength of these corrections depends on the contribution of the brane Higgs field, being proportional to powers of $\sin\beta$. These effects induce additional corrections to the EWPT observables measured at LEP. Depending on the Higgs sector realisation, the indirect limit from LEP is therefore $R^{-1}\ge$ 3.8--5.4 TeV. This has left very little room for KK states discovery at the 7, 8 TeV LHC. During the past run, direct searches performed with total integrated luminosity ${\cal L} = 20~fb^{-1}$ have been able to set exclusion bounds comparable to those coming from EWPTs. The analysed processes are the Drell-Yan $Z_{KK}^{(1)}, \gamma_{KK}^{(1)}$ production in both di-lepton and di-jet channels. Tab. \ref{table:bounds} summarises the present indirect and direct bounds on the mass of the KK states within the two considered frameworks. In the table, the blank entries in the first two rows indicate that the corresponding bounds have not been extracted yet. From Tab. \ref{table:bounds}, one can deduce that the search window in the future RunII at the upgraded LHC is $R^{-1}\ge 3.8$ TeV for NUED model(s) in 5D. 

\begin{table}[tbp]
\centering
\begin{tabular}{c c c c c}
\hline
\hline 
Model & $\sin\beta$ & EWPT & LHC ($pp\rightarrow l^+l^-$) & LHC ($pp\rightarrow jj)$ \\ 
[0.5ex]
\hline
NUED & 0.45 & 3.8 TeV & 3.8 TeV &  - \\
NUED & 1.0 & 5.4 TeV & 3.8 TeV &  - \\
NUED(EW) & 0.45 & 3.8 TeV & 3.8 TeV & 3.25 TeV \\
NUED(EW) & 1.0 & 5.4 TeV & 3.8 TeV & 3.25 TeV \\
[1ex]
\hline
\end{tabular}
\caption {Summary of EWPTs and LHC (8 TeV and ${\cal L} = 20 fb^{-1}$) 95\% CL exclusion bounds on the mass of KK excitations of SM gauge bosons within the NUED  model and its simplified version NUED(EW) as described in the text.}
\label{table:bounds}
\end{table}

\noindent
In the forthcoming two subsections, we study the DY channel which can be mediated by the KK excitations of the SM neutral gauge bosons, $Z_{KK}^{(n)}$ and $\gamma_{KK}^{(n)}$, where $n$ defines the excitation number of the resonance in the tower. For each level of the ED tower of states, the two resonances are very close in mass so their spectrum would appear degenerate in any experimental search. In order to validate our numerical procedures in view of our LHC RunII studies, we first re-obtain independently (some of) the experimental limits quoted in Tab. \ref{table:bounds}. In Sect. \ref{subsec:searches_ED}, we then assess the scope of the LHC upgrade in excluding or discovering the NUED models considered here via the DY signature.. We shall, in particular, perform this analysis taking into account FW and interference effects as mentioned in the introduction.

\subsection{DY Process: present bounds and "bump" searches}
\label{subsec:searches_ED}

In this section, we derive discovery prospects and exclusion limits at the present LHC RunII with 13 TeV energy and a luminosity ranging from ${\cal L} = 30~fb^{-1}$, that is the integrated luminosity which one expects to collect in the next two years, and ${\cal L} = 300~fb^{-1}$ that is the total design luminosity. We consider the DY process giving rise to electron and muon pairs in the final states. This process can be mediated by the KK excitations of the SM neutral gauge bosons:
\begin{equation}
pp\rightarrow \gamma , Z, \gamma_{KK}^n, Z_{KK}^n\rightarrow l^+l^-
\end{equation}
with $l=e,\mu$. As the mass bounds on the KK modes coming from the LHC RunI are pretty high,  only the first level of the ED tower of KK states has some chance to be detected (or excluded) at the ongoing LHC RunII. We thus limit our analysis to the production and decay of the extra $\gamma_{KK}^1, Z_{KK}^1$ from now on called simply $\gamma_{KK}, Z_{KK}$.

\noindent
Before entering the details of the analysis, we first carry out some preliminary exercises illustrating the phenomenology induced by the possible existance of EDs. In particular, we would like to underline the effects coming from FW and interference of the extra gauge bosons with the SM ones on the signal shape. The FW effects are what typically one expects for (rather) narrow resonances, as the width of the NUED and NUED(EW) extra gauge bosons is 
below $\sim 6\%$ of their mass: $\Gamma_{\gamma_{KK}}/M_{\gamma_{KK}}=4.2\%$ and $\Gamma_{Z_{KK}}/M_{Z_{KK}}=6\%$. In contrast, in Fig.~1a, we clearly see that the effect of the interference between the KK modes (first level of the ED tower) and the SM gauge bosons on the signal line-shape is quite distinctive of NUED models. We can in fact notice the presence of a pronounced dip (a sort of inverted peak)  appearing before the resonant structure around the pole mass of the new gauge bosons. The contributions of the different components to the total differential cross section are visible in Fig.~1b. One feature of this model is here explicit: there is no individual contribution shaping the inverse peak (positioned at around 2.2 TeV for this benchmark point), rather the latter emerges as a global dynamics due to a cumulative effect driven  by the various negative contributions coming from the interferences between the SM neutral bosons with their associated KK excitations. This happens because we are in the case of maximal interference since the chiral couplings of the heavy excitations are the same as those of their SM counterpart, up to a rescaling factor of $\sqrt{2}$. We further anticipate that, even if highly model-dependent, this is a common feature of multi-$Z^\prime$ models as we discuss in the next sections.

\begin{figure}[t]
\centering
\includegraphics[width=0.45\linewidth]{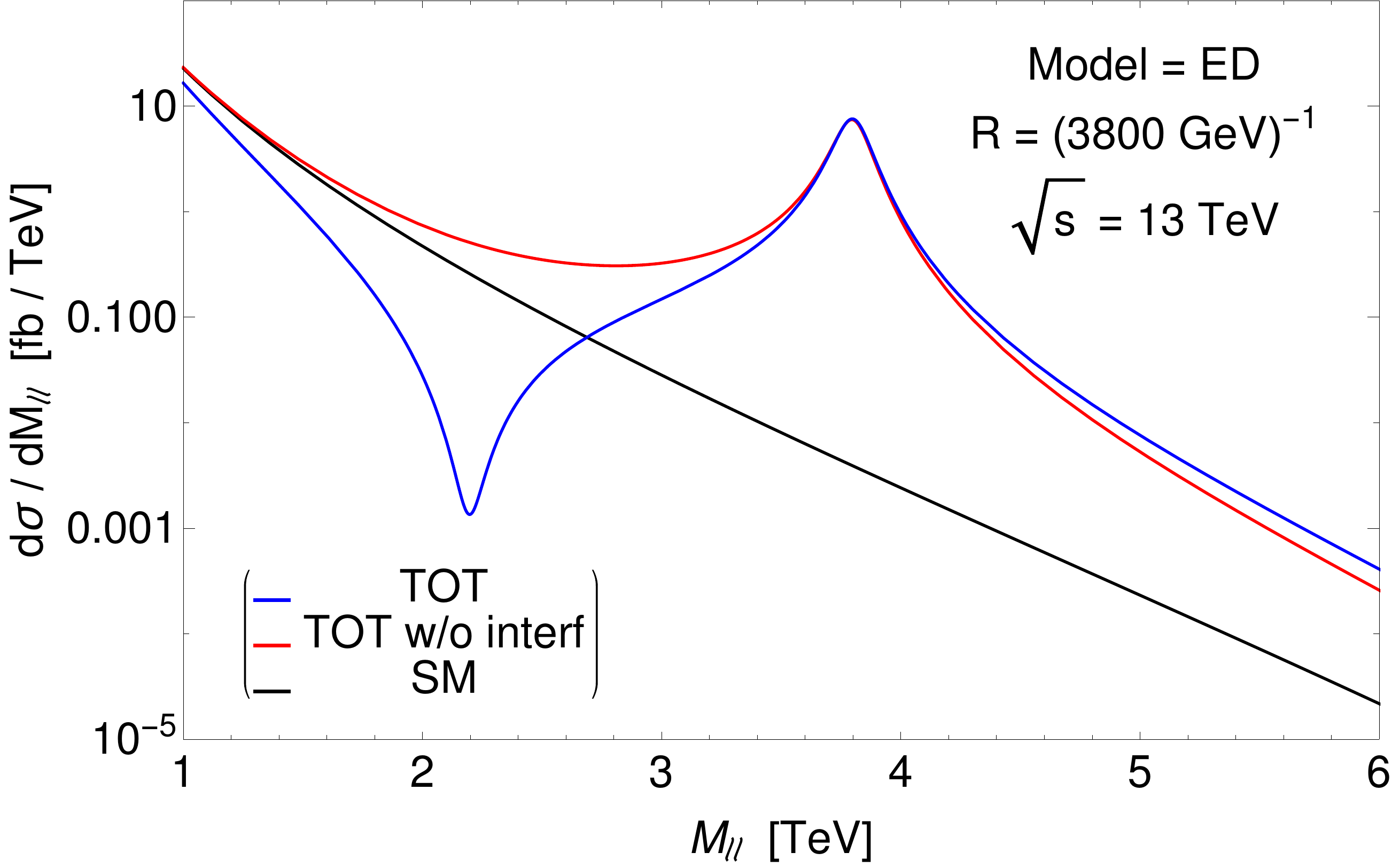}{(a)}
\protect{\label{fig:ED_XS}}
\includegraphics[width=0.45\linewidth]{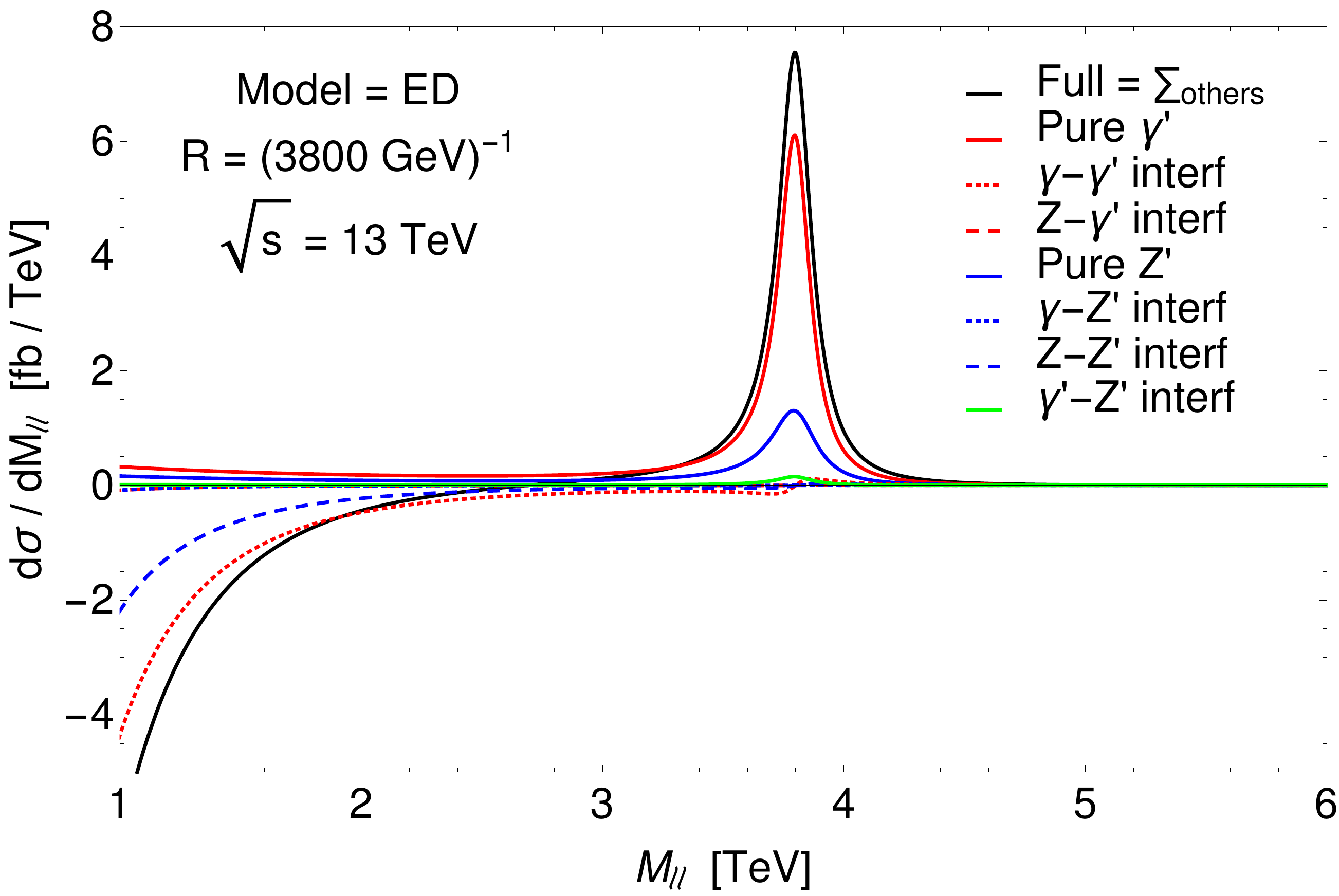}{(b)}
\protect{\label{fig:ED_XS_components}}
\caption{(a) Differential cross section as a function of the di-lepton invariant mass for the NUED model with $1/R~=~3.8~$ TeV. The blue line represents the full result while the red line does not include interference effects. (b) Same as in plot (a) for each individual contribution to the total differential cross section. The color code is described in the legend.}
\end{figure}

\begin{figure}[t]
\centering
\includegraphics[width=0.45\linewidth]{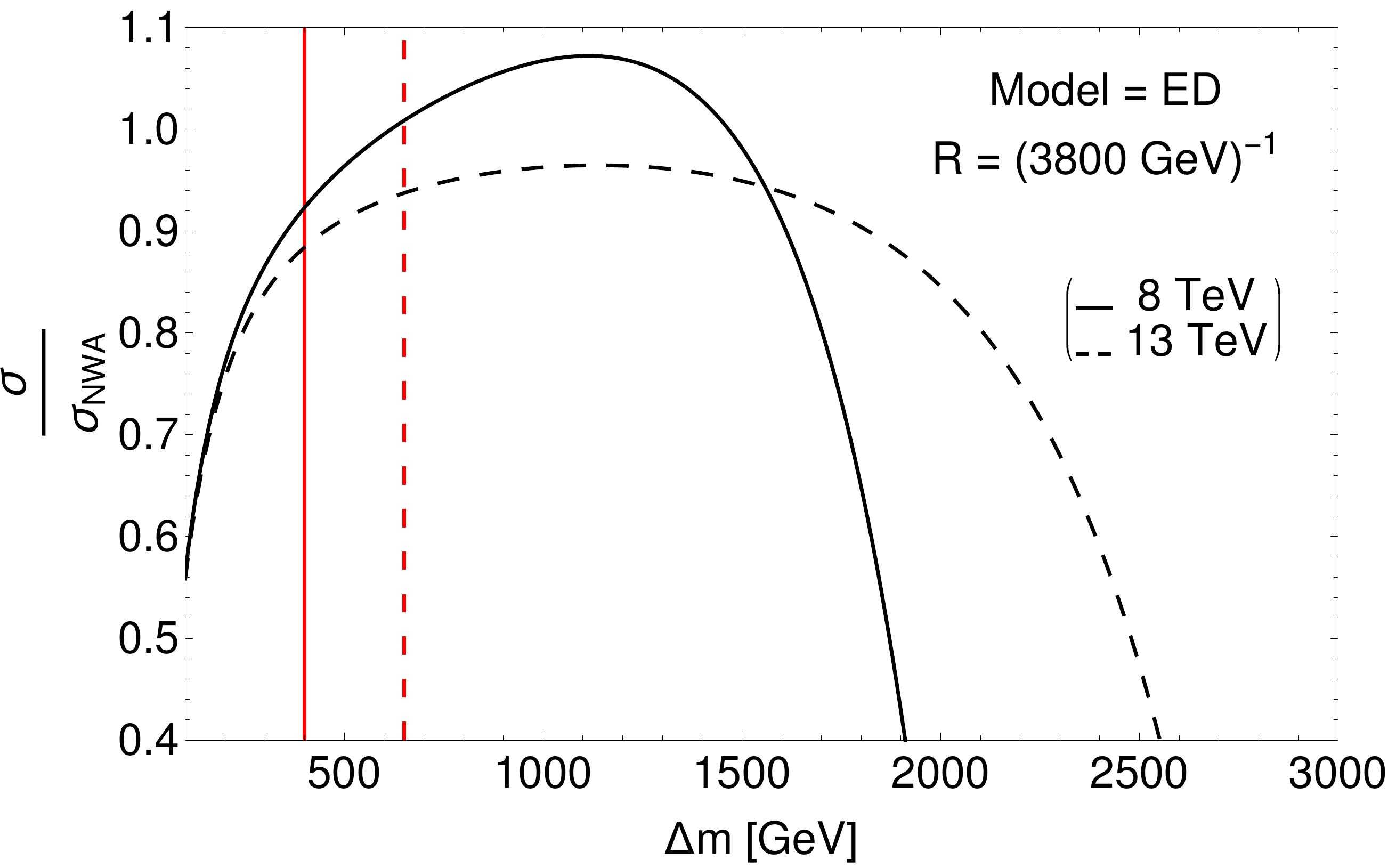}
\includegraphics[width=0.45\linewidth]{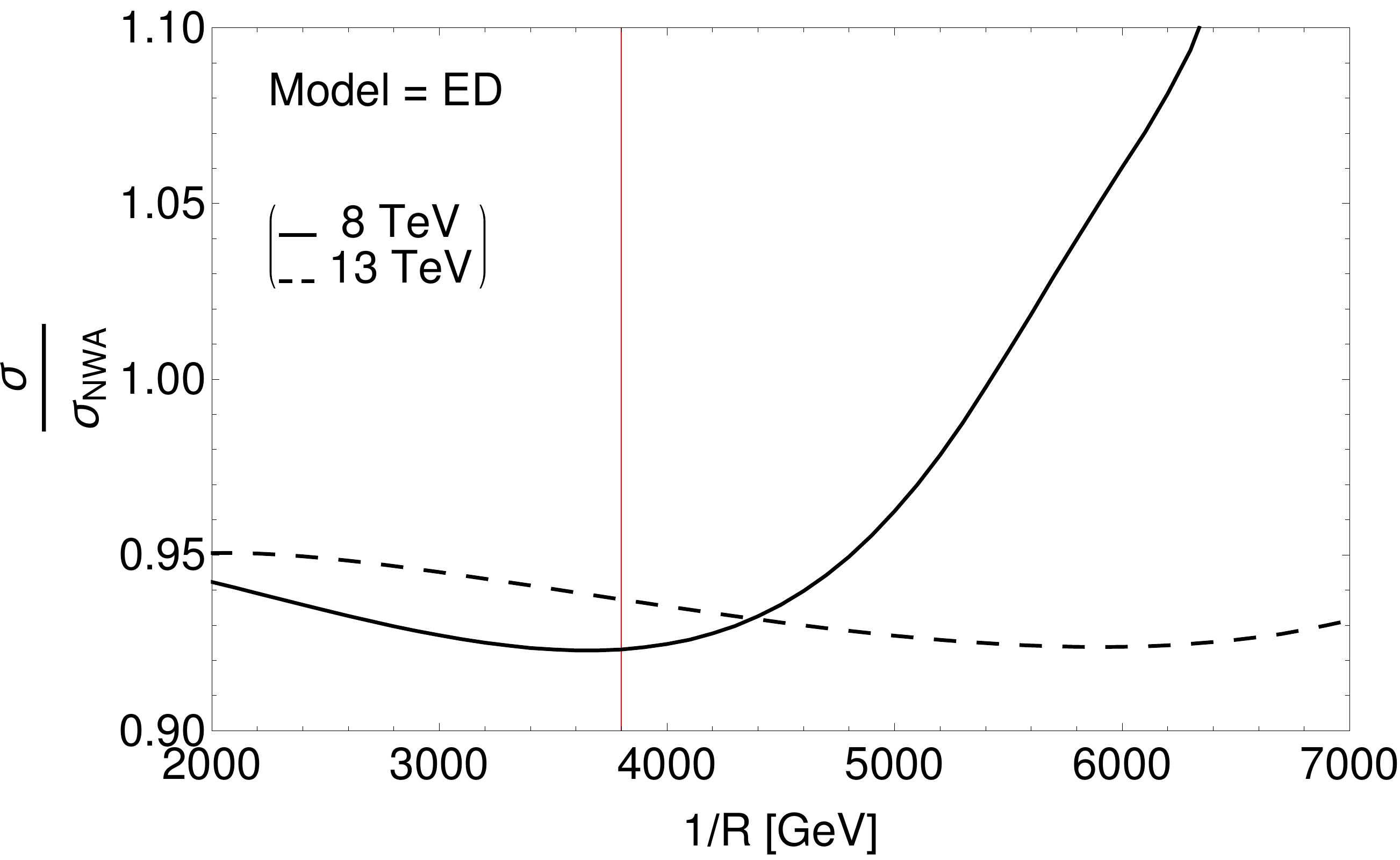}
\label{fig:ED_R}
\caption{(a) Ratio of the complete cross section over the NWA result as a function of the symmetric integration interval taken around the $Z'$ pole mass, $|M_{ll}-M_{Z'}|\le \Delta m$, for the NUED model with $R^{-1}$=3.8 TeV. We consider the LHC at 8 TeV (solid line) and 14 TeV (dashed line). The red vertical lines represent the integration interval $|M_{ll}-M_{Z'}|\le 5\% E_{\rm LHC}$ adopted by CMS at the 8 TeV LHC (solid) and 14 TeV LHC (dashed), when extracting the $Z'$ mass limits within a given model by crossing the computed theoretical cross section with the 95\% C.L. upper bound on the BSM cross section derived from the data analysis.
 (b) Ratio of the complete cross section integrated over the mass interval $|M_{ll}-M_{Z'}|\le 5\% E_{\rm LHC}$ over the NWA result as a function of the inverse of the compactified extra dimension length $R$. The red vertical line represents the actual limit on $R^{-1}$ according to Ref.~\cite{Accomando:2015rsa}.}
\label{fig:ED_XS_resolution}
\end{figure}

\noindent
Despite large interferences could happen before the appearance of the resonant peak, whose position and magnitude strongly depend on this specific model, the extraction of mass bounds on the KK resonances can still be performed in a model independent way up to a large extent. Altogether, in fact, the model dependent FW and interference effects can be kept below $O(10\%)$ of the total cross section when we integrate the dilepton spectrum in the invariant mass interval $|M_{ll}-M_{Z'}|\le 5\% \ \ E_{\rm LHC}$ around the hypothetical pole mass, $M_{Z'}$, of the two (almost) degenerate KK excitations belonging to the first level of the ED tower of states. Here, $E_{\rm LHC}$ is the collider energy. This integration interval has been proposed in Ref.~\cite{Accomando:2013sfa} for computing the total theoretical cross section within a large class of single $Z'$-boson models. Later, it has been adopted by the CMS collaboration in interpreting the results of the data analysis of dimuon and dielectron mass spectra at the past LHC RunI ~\cite{Khachatryan:2014fba}. In order to extract limits on the mass of extra $Z'$-bosons, the total theoretical cross section is computed in that dilepton invariant mass interval around the peak and then crossed with the 95\% C.L. upper bound on the BSM cross section derived from the experimental data analysis. For sake of clarity, let us recall the strategy adopted by CMS when searching for narrow $Z'$-bosons which are expected to appear with a well defined line-shape over a smooth SM background. A notable feature of this analysis is in fact that limits can be extracted in a (quasi) model-independent way to enable straigthforward interpretation in any model predicting a narrow resonant structure. 

\noindent
The first characterising element is that the analysis assumes as generic shape for the signal a Breit-Wigner convoluted with a Gaussian resolution function. Low mass tails, due to PDF's and model dependent effects like FW and interference of the extra $Z'$ boson with the SM $\gamma$ and $Z$, are not considered. The analysis is, by design, not sensitive to potential tails of the signal and the magnitude of such tails is much less than the SM background. Attempting to make  the experiment sensitive to the tails would moreover render the analysis model dependent, thereby automatically restricting the coverage of theoretical models where one could extract mass bounds in a consistent way. The CMS approach consists thus in modelling the signal via a function which is common to a large class of models predicting a single, rather narrow, $Z'$-boson. This generic signal-shape (a Breit-Wigner normalized to the total cross section computed in NWA) more closely resembles the exact result shown in Fig. 1a, where both FW and interference effects are accounted for, than the FW approximation. This latter displays a tail at low invariant masses which is in fact almost completely washed out when adding in the interference effects. This result is common to a large class of narrow single $Z'$-boson scenarios, as extensively discussed in Ref.~\cite{Accomando:2013sfa}. As to the SM background, the CMS collaboration represents its shape by a functional form whose parameters are fixed via a fit to the Monte Carlo (MC) SM background estimate. Its rate is normalised to the data. The normalization of the SM background is performed in a window of the dilepton spectrum taken around the hypothetical $Z'$-boson pole mass. The extremes of this mass range are set in such a way that a minimum of about 400 events are collected there. It should be noted that if a sizeable interference dip appears at rather low dilepton invariant masses, as in NUED models, it could affect the estimate of the SM background shape, a priori, and its normalization. This would suggest to modify the present selection of the mass region where the SM background is normalized to data and shift it away from the peak.

\begin{figure}[t]
\centering
\includegraphics[width=0.45\linewidth]{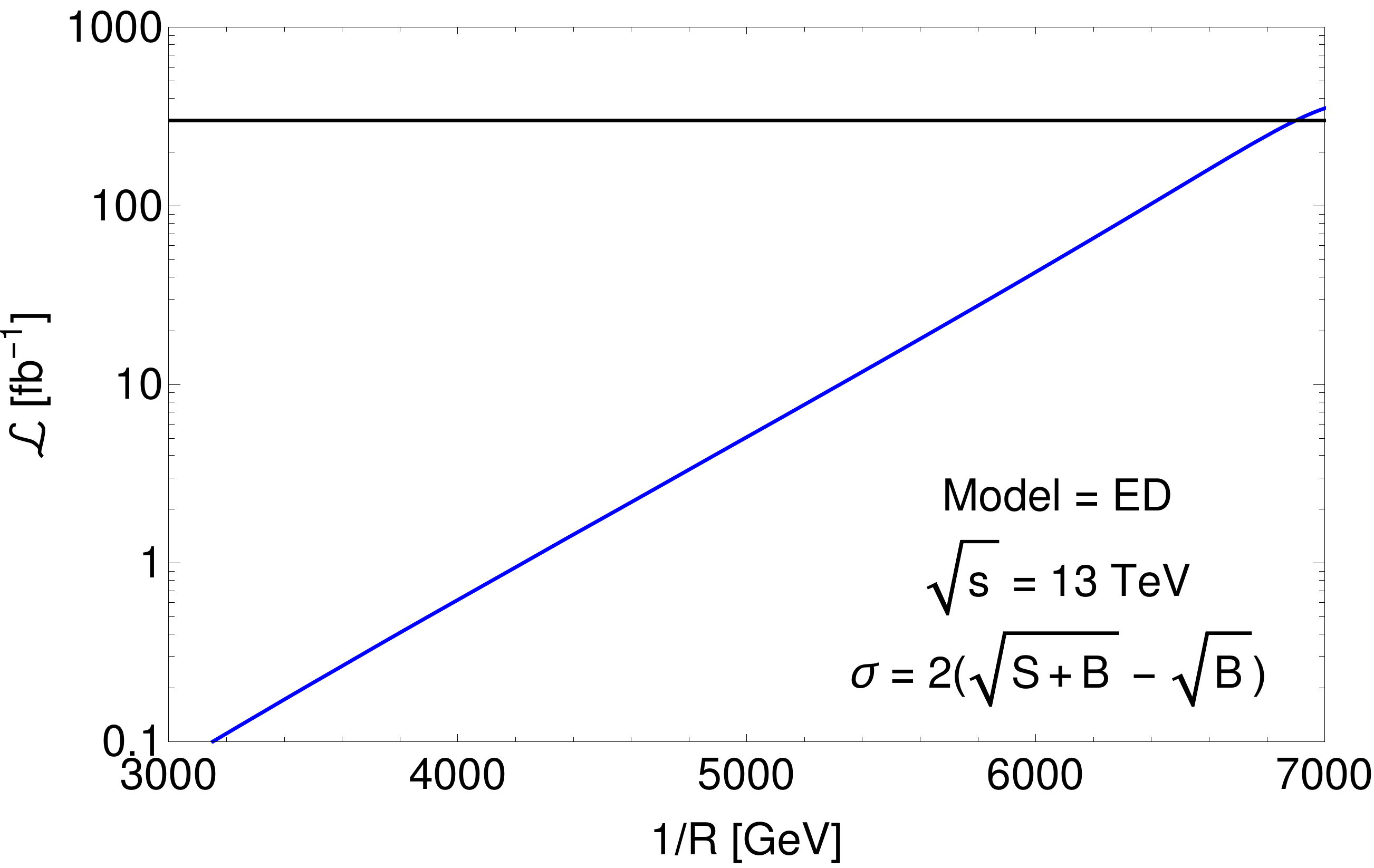}{(a)}
\includegraphics[width=0.45\linewidth]{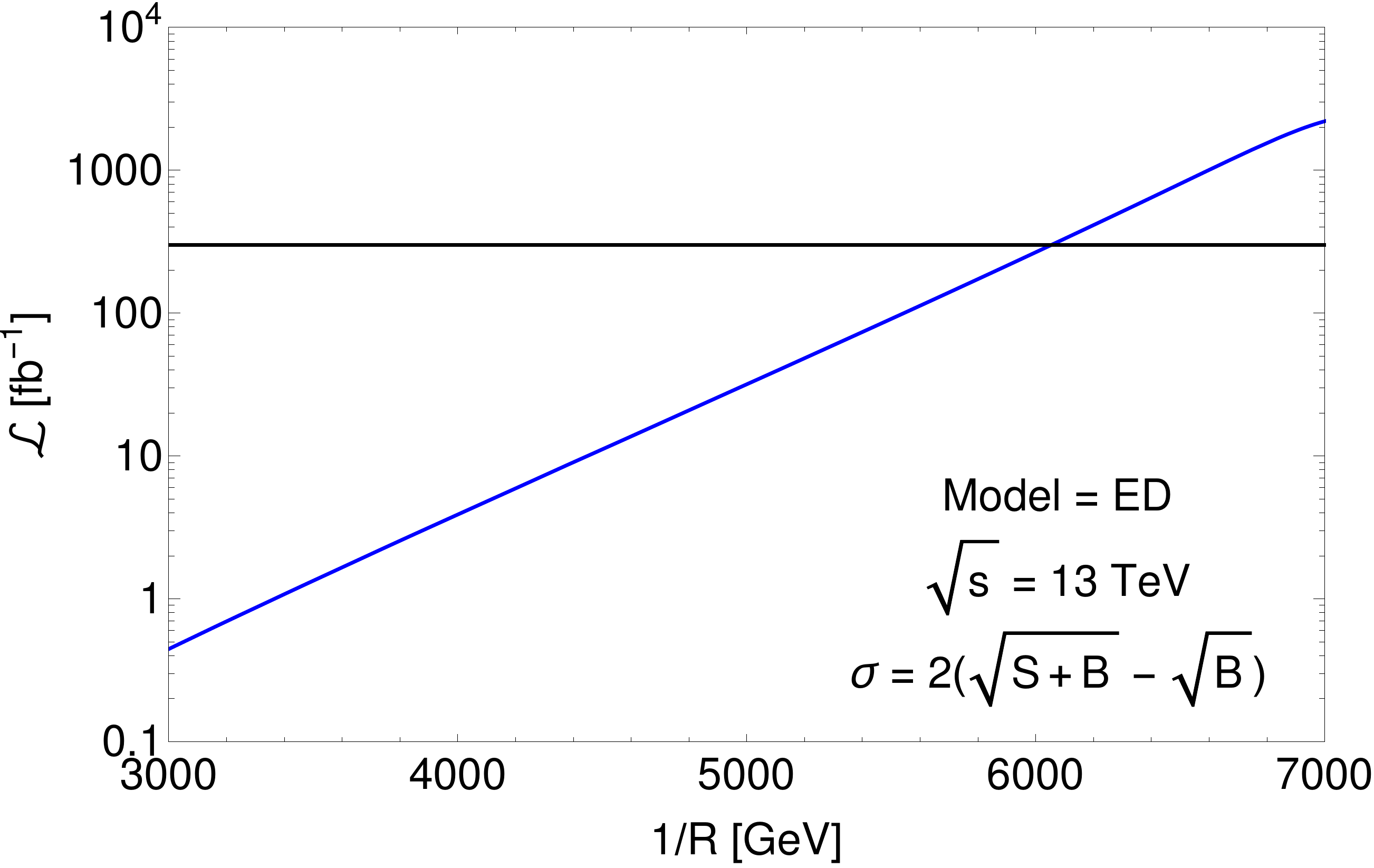}{(b)}
\caption{Required luminosity for the exclusion (plot (a) $\alpha=2$) and discovery (plot (b) $\alpha=5$) limits in the NUED model as a 
function of the inverse length of the compactified extra dimension. The horizontal lines are fixed to be ${\cal L} = 300~fb^{-1}$, which is the design luminosity that will be achieved at the end of RunII. They intercepts the curve of the model giving an exclusion (discovery) limit $1/R> 6.9 (6)$~TeV.
}
\label{fig:ED_Limits_13TeV}
\end{figure}

\noindent
Having the functional forms for the $Z'$-boson signal and the SM background, an extended unbinned likelihood function for the spectrum of di-lepton invariant masses is then constructed. If no evidence for BSM physics is observed, the 95\% C. L. upper bound on the cross section is derived. This result can then be used to extract limits on the $Z'$ boson parameters, i.e., mass and possibly couplings, from a number of new physics models. However, a key point is that, in order to perform a consistent interpretation of the data, the theoretical cross section within any given model must be computed by minimising the model-dependent effects as well. This is just the role of the integration range advocated in Ref. \cite{Accomando:2013sfa} for computing the total cross section in any given model. This range consists in a symmetric dilepton invariant mass interval around the hypotetical pole mass of the new vector boson(s). It is in particular designed to be independent of the individual characteristics of the $Z'$-signal, e.g. mass and width, and on the specific features of the experiment so to constitute a generic setup for data interpretation within a large class of $Z'$ models. It is in fact expressed in terms of the sole collider energy: $|M_{ll}-M_{Z'}|\le 0.05~E_{\rm LHC}$ with $E_{\rm LHC}$ the collider energy and $Z'$ a generic extra heavy gauge boson. As shown in Ref.~\cite{Accomando:2013sfa}, within this mass interval and for a big enough  BR$(Z'\rightarrow l^+l^-)$, the model dependent features of the $Z'$-boson signal can be negleted up to an $O(10\%)$ theoretical accuracy in all theories predicting a narrow $Z'$-boson. 

\noindent
The notable outcome is that the theoretical cross sections of the $Z'$ bosons predicted within a variety of models belonging to the $E_6$, LR (Left-Right) and Sequential Standard Model (SSM) class of theories can all directly be compared with the 95\% C. L. upper bound on the $Z'$-boson  cross section resulting from the experimental analysis performed by the CMS collaboration. This procedure thus allows one extracting exclusion bounds on the mass of the various $Z'$ bosons at once, without requiring dedicated analyses.  In this respect, the multi-resonant NUED models behave exactly in the same way as the singly resonant $E_6$, LR and SSM models. By virtue of this feature, limits on the degenerate multi-resonant KK modes can thus be extracted from the  CMS data analysis of the di-lepton spectra, directly and unambiguously. To support this statement, in Fig.~\ref{fig:ED_XS_resolution}a, we plot the ratio of the full cross section over the NWA result as a function of the symmetric integration region around the pole mass of the degenerate KK excitations belonging to the first level of the ED tower of states. The vertical red line represents the dilepton invariant mass interval proposed in Ref.~\cite{Accomando:2013sfa} for computing the total theoretical cross section. This shows that, if one restricts the integration region, the difference between complete cross section and NWA result, normalized to the NWA cross section, is indeed  below $O(10\%)$. The NWA cross section shown in Fig.~\ref{fig:ED_XS_resolution} is the sum of the two NWA cross sections corresponding to the two KK excitations of the SM photon and $Z$ boson. We have also verified that the latter result does not change with respect to the length scale $R$ of the extra dimension. In Fig.~\ref{fig:ED_XS_resolution}b, we plot the deviation of the full result from that one in NWA as a function of the parameter $1/R$. Here the red vertical line represents the actual limit on $1/R$ according to Ref. \cite{Accomando:2015rsa}. We find deviations below $O(10\%)$ in the full range of $R$ values which can be explored at the LHC RunII. This is due to the fact that the interference pattern is such that the interference effects become sizeable at low invariant masses, away enough from the resonant peak. This feature is extremely model dependent. In the next sections, we will see that in composite Higgs models the position of the interference with respect to the resonant peak will be indeed completely different, motivating different approaches for the multi-$Z'$ hunt at the LHC and leading to different conclusions.

\noindent
Remaining within ED models, the prescription of \cite{Accomando:2013sfa} works perfectly. We have thus (re-)calculated the present bound on the mass of the KK excitations of SM photon and $Z$ boson. As previously mentioned, the most recent bound from the LHC data at 7, 8 TeV gives $1/R > 3.8~$ TeV~\cite{Accomando:2015rsa}. We have been able to reproduce this limit using the CMS setup (for details see Ref.~\cite{Khachatryan:2014fba}), hence validating our code and procedure. In the calculation we have included Next-to-Next-to-Leading-Order (NNLO) corrections and we have combined the two channels ($e^+e^-, \mu^+\mu^-$) with their individual acceptances and efficiencies as quoted in~\cite{Khachatryan:2014fba}. 
\begin{figure}[t]
\centering
\includegraphics[width=0.45\linewidth]{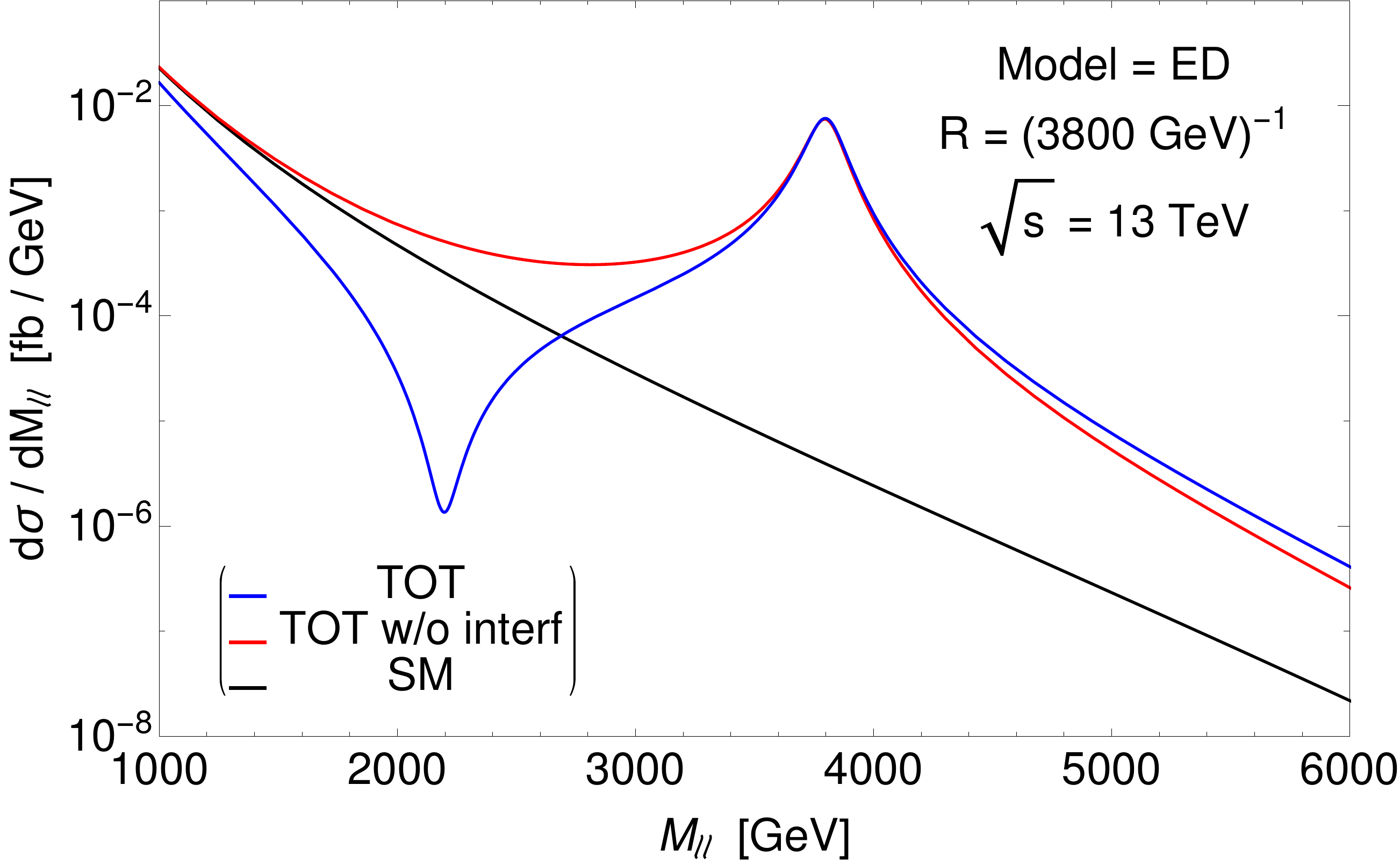}{(a)}
\includegraphics[width=0.45\linewidth]{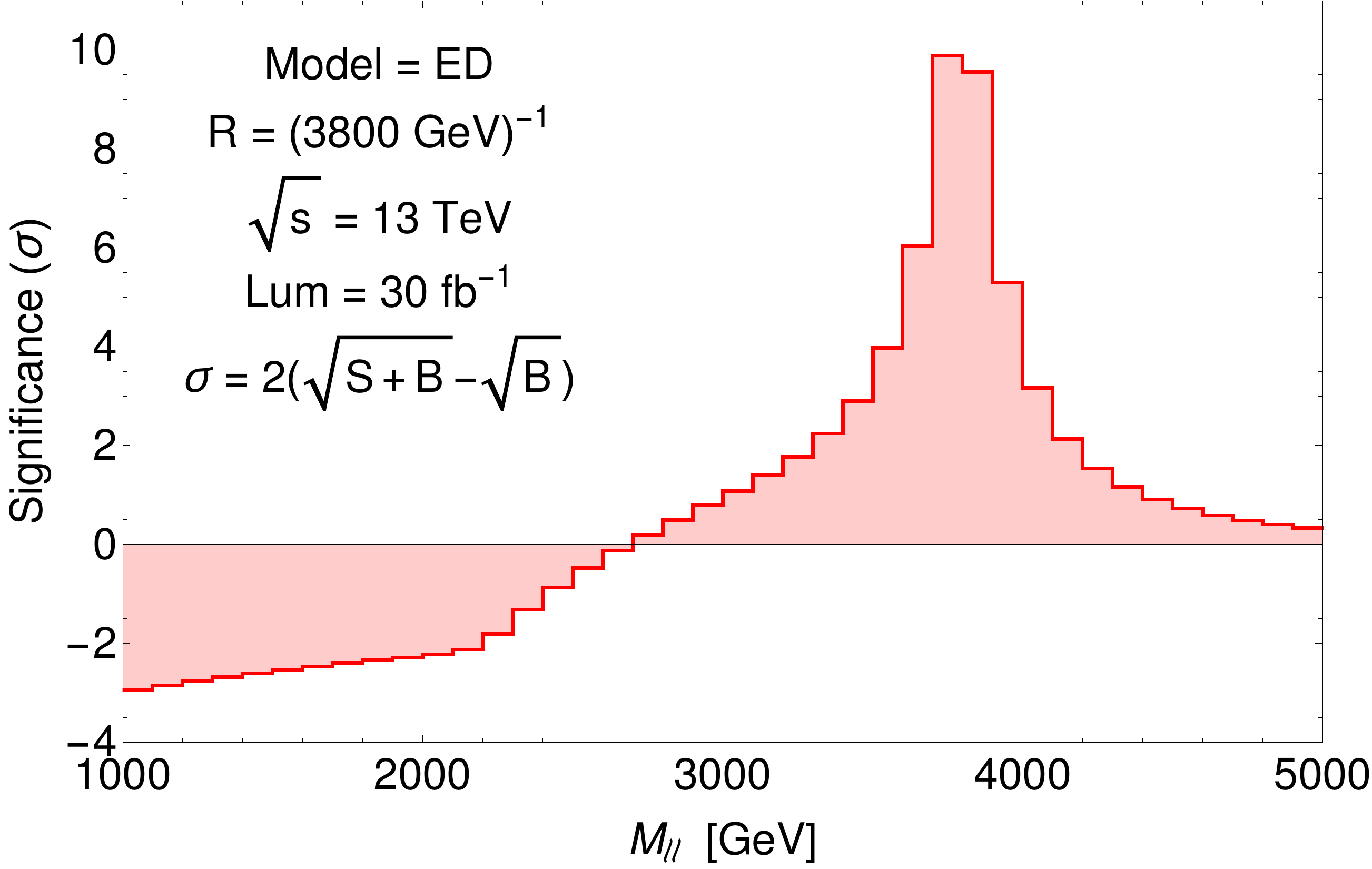}{(b)}
\includegraphics[width=0.45\linewidth]{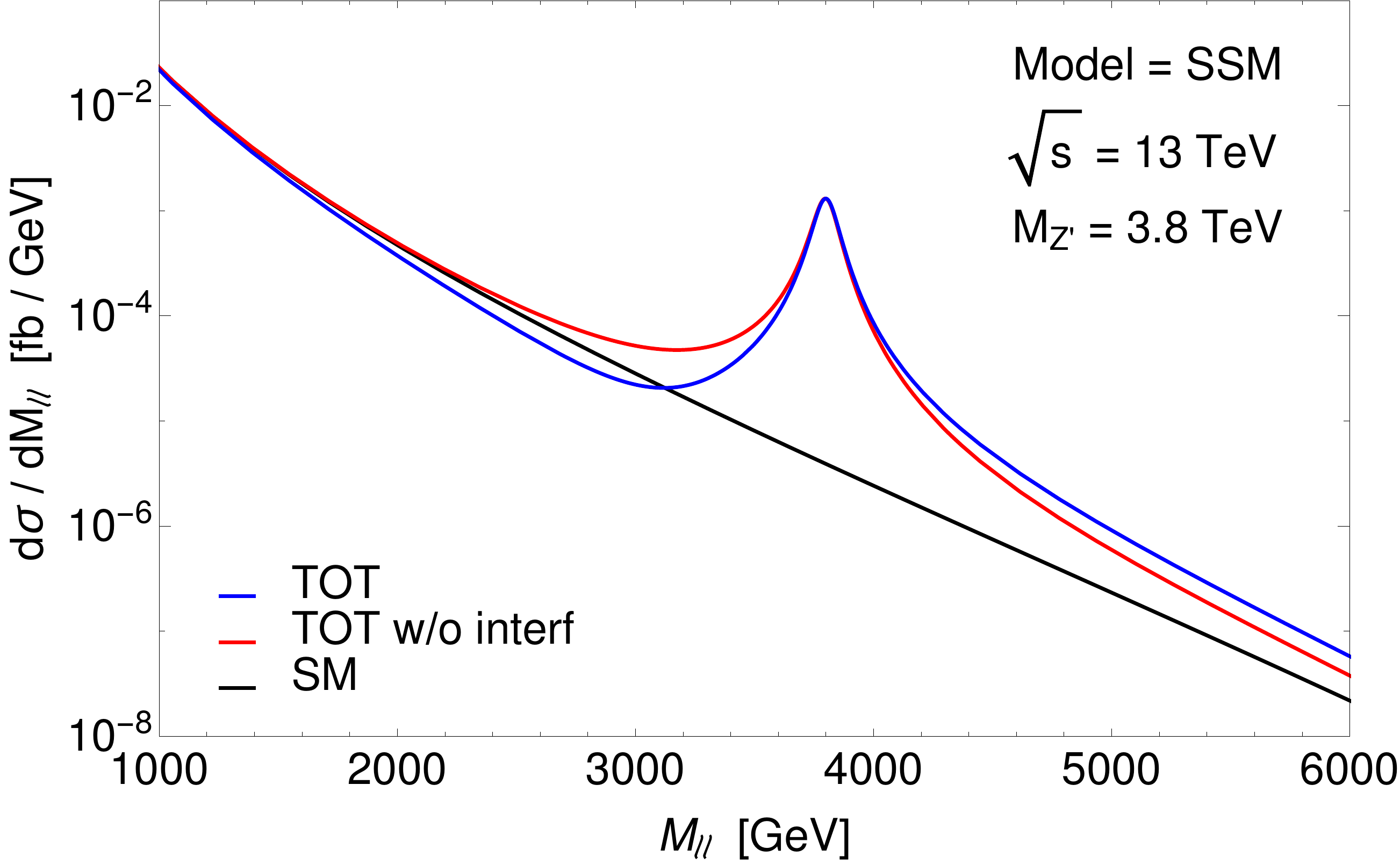}{(c)}
\includegraphics[width=0.45\linewidth]{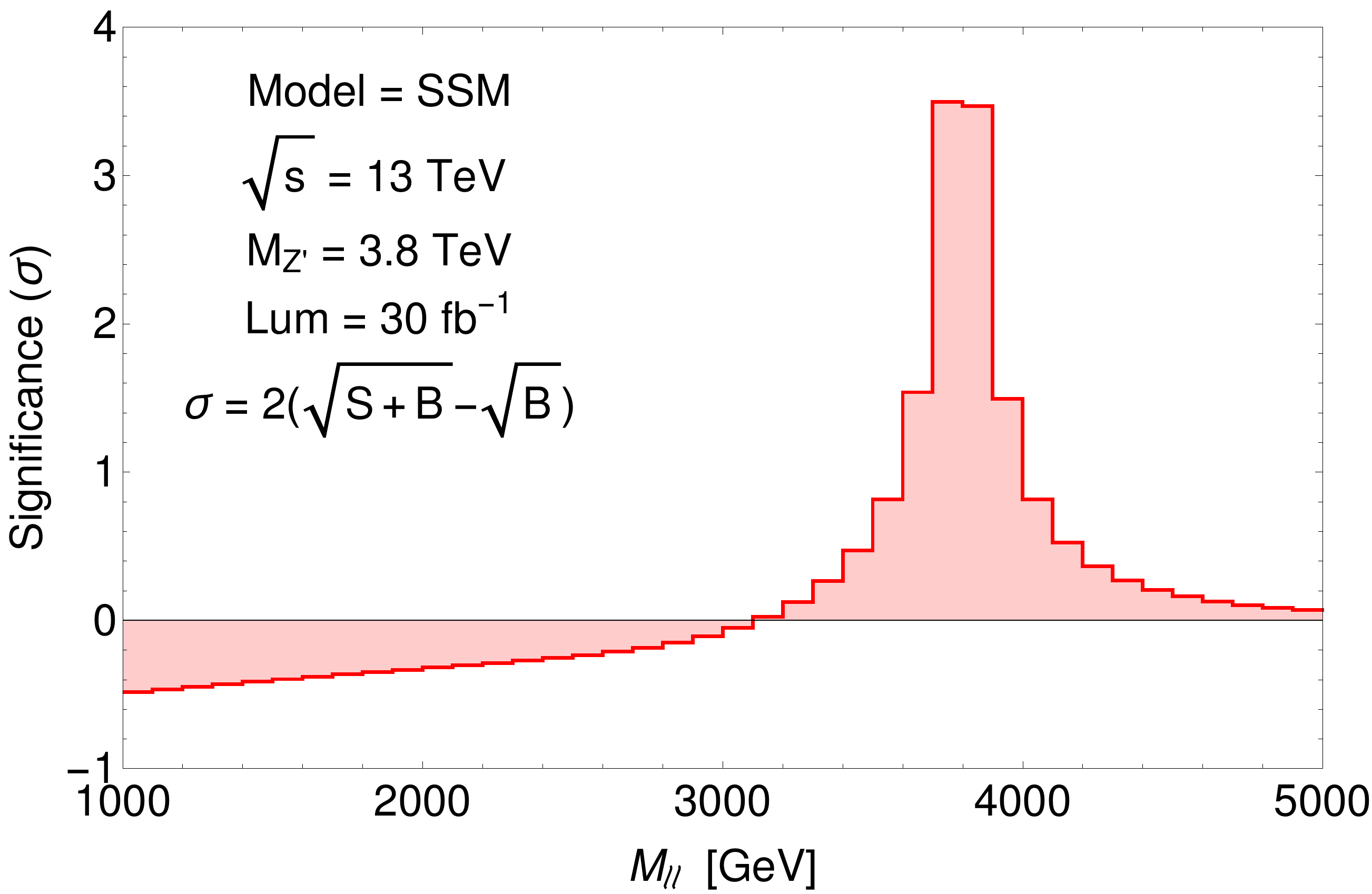}{(d)}
\caption{(a) Signal shape within the NUED models for $R^{-1}=3800$ GeV. (b) Significance corresponding to the signal in plot (a), assuming a luminosity $\mathcal{L}\simeq~30~fb^{-1}$.
Same as plot (a) for the SSM with $M_{Z'}=3800$ GeV. (d) Same as plot (b) for the SSM with $M_{Z'}=3800$ GeV.}
\label{fig:signal_NUEDvsSSM}
\end{figure}
We will now apply the same setup when discussing the prospects of discovery and exclusion of KK excitations at the 13 TeV LHC RunII.
The results of this analysis are presented in Fig.~\ref{fig:ED_Limits_13TeV}, where the left panel (a) shows the projected exclusion bounds and the right one (b) the projected discovery potential as a function of the collected luminosity at the ongoing 13 TeV run of the LHC. The horizontal lines are fixed to be ${\cal L} = 300~fb^{-1}$, which is the design luminosity that will be achieved at the end of RunII. We find that, in absence of any signal, one will be able to push the exclusion limit on the mass of the KK excitations of the photon and $Z$ boson up to $1/R > 6.9$~TeV. In the positive case of a signal, we will be able to claim the discovery of the first EW neutral states of the ED tower up to around $6$ TeV. This analysis has been performed in the traditional way valid for narrow resonances which can be represented as a Breit-Wigner line-shape standing over a low SM background. Such an approach would not be appropriate in two cases: if the branching of the $Z'$-boson(s) into electron and muon pairs were not high enough to generate a cross section much bigger than the SM background and/or if the new resonance(s)  were rather wide. 

\noindent
Assuming a new resonance has been discovered in the described bump hunt during the LHC RunII  at low luminosity, the next step would be tracking the underlying theory predicting such a particle. In the next sub-section, we thus concentrate on profiling the new resonance(s) during a successive LHC run at higher luminosities. We focus on a situation where a (degenerate) peak is clearly seen at large di-lepton invariant masses in the standard bump search. Under this circumstance, the dip at low invariant masses could be used to characterise the signal in such a way to confirm EDs as the underlying BSM scenario we considered (NUEDs). We shall do this in the next sub-section.

\subsection{DY Process: profiling KK modes}
\label{subsec:profile_ED}

\begin{figure}[t]
\centering
\includegraphics[width=9cm]{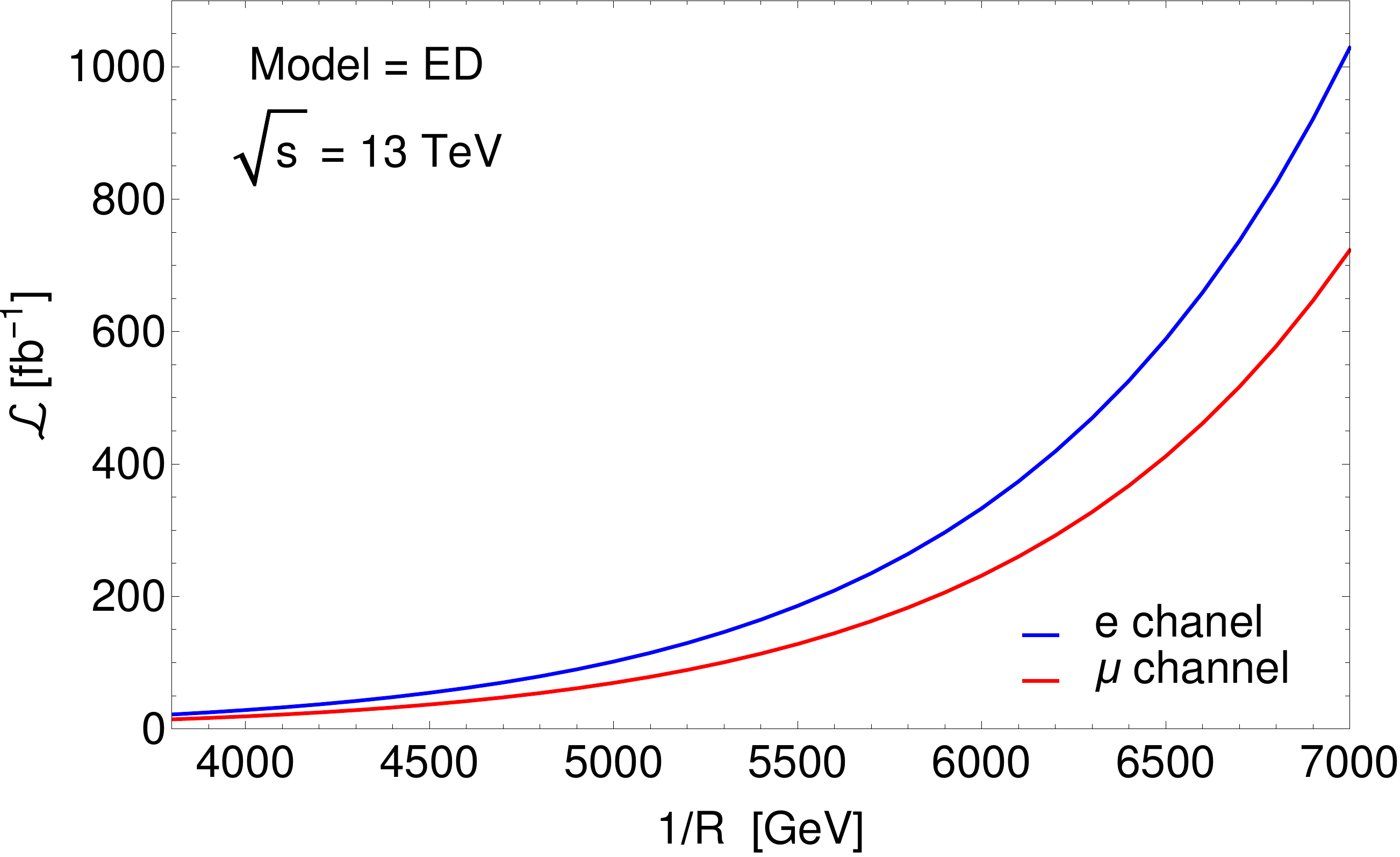}
\caption{Required luminosity for excluding the background hypothesis at 95\% C.L. owing to the depletion of events caused by new physics. The blue (red) line represents the dielectron (dimuon) final state for which the corresponding experimental acceptance and efficiency have been implemeted.}
\label{fig:ED_Dip}
\end{figure}

We address here the question of how to profile the KK excitations in the case of their  discovery in the standard bump search. A distinctive features of NUED models is the apparance of a sizeable dip before the resonant structure, as already pointed out in Refs. \cite{Accomando:1999sj, Bella:2010sc, Boos:2011ib}. This characteristic is common to multi-$Z'$-boson  models, even if the distance between dip and peak is highly model dependent, and helps disantangling them from singly resonant $Z'$-boson scenarios. This behaviour is quantified in Figs.~\ref{fig:signal_NUEDvsSSM}(a) and \ref{fig:signal_NUEDvsSSM}(b)  where we compare the signal shape predicted by the NUED models and the SSM, which is used as the primary benchmark by the LHC experimental collaborations. As one can see, the depletion of events is much more pronounced and concentrated in a smaller region before the peak in the multi-$Z'$ boson case. The statistical significance of the dip is indeed much bigger in the NUED models, as shown in Figs.~\ref{fig:signal_NUEDvsSSM}(c) and \ref{fig:signal_NUEDvsSSM}(d). Of course, the significance scales with the luminosity. In Fig.~\ref{fig:ED_Dip}, we show the integrated luminosity that is required to exclude the SM background hypothesis at 95\% C.L., owing to the depletion of events caused by the destructive interference between the new $Z_{KK}$ and $\gamma_{KK}$ bosons and their SM counterparts, as a function of the KK mode mass.
These contours have been evaluated by integrating the differential cross section in a symmetric invariant mass window around the dip, taken between the point where the new resonance peak(s) crosses the SM background and the symmetric counterpart. As one can see, for the design luminosity ${\cal L} = 300~fb^{-1}$, the dip could be detectable for all KK-mode masses that can be possibly discovered at the LHC RunII thus allowing to interpret the data accordingly and pin down the existence of EDs.
\par\noindent
Another way of profiling a resonance(s), very known in the literature, is to introduce the Forward-Backward Asymmetry (AFB). A detailed analysis of the features and uncertainties on this observable has already been performed in our previous work \cite{Accomando:2015cfa} for a large class of single $Z^\prime$-boson scenarios. There, the results are quite promising. We have thus applied our study of the AFB observable within the NUED model(s). In this case, unfortunately, the conclusions are not that good. The shape of the AFB distribution as a function of the dilepton invariant mass is in fact not statistically significant for the designed luminosities achievable at the LHC in the near future. Different types of asymmetry, measured in a different channel, could however play that role. In Ref. \cite{Accomando:2013dia} it has been shown that, 
combining the charge and spin polarization asymmetries in the $t\bar t$ channel, one could identify the presence of the two quasi-degenerate states $\gamma_{KK}$ and $Z_{KK}$ in a resonant signal at the LHC. The measurement of such asymmetries would then allow one to distinguish the quasi-degenerate double resonant spectrum, predicted by the NUED model(s), from a "standard' single $Z'$-boson  that could present a similar signal in a bump hunt analysis.

\section{The CHM scenario}\label{sec:4DCHM}

One way to alleviate the hierarchy problem present in the SM, which manifests itself through the appearance of quadratically divergent radiative corrections to the Higgs mass, therefore implying a huge degree of fine tuning if the SM is extrapolated up to the Planck scale, is to protect the mass of the scalar with a symmetry. This is in fact the same mechanism through which in the SM the fermion and gauge boson masses are shielded from these virtual corrections, that is, by  means of a chiral and gauge symmetry, respectively, while the scalar mass is left unprotected. Supersymmetry (SUSY) is the most common manifestation of this paradigm. The boson-fermion symmetry present in the theory guarantees the stability of the Higgs mass via cancellations between the top and the stop loop contributions to the two point function of the Higgs (pole) mass. However, this is not the only solution. An alternative proposal is to assume that the Higgs boson is a composite state, arising from some unspecified strong dynamics at a scale higher than the EW one. In order to realise the Higgs boson as a spinless light state (that is, lighter than other resonances that might be present in the strong sector), it can be postulated to be a Nambu-Goldstone Boson (NGB) arising from the spontaneous breaking of a global symmetry in the strong sector. The NGB  will eventually acquire a (small) mass through an explicit, but weak, breaking of the global symmetry, becoming a Pseudo-NGB (PNGB). This automatically solves the hierarchy problem since all the radiative corrections affecting the Higgs mass will be saturated at the composite scale, that is, its mass will not be sensitive to virtual effects above it, and also it agrees with the historical pattern that has so far seen all the (pseudo)scalar particles known in
Nature to be composite states. This idea goes back to the '80s~\cite{Kaplan:1983fs} and strongly resembles the dynamics with which it is possible to explain the lightness of the pions with respect to other mesons like  the $\rho$'s, that is, by postulating the former to be a PNGB of the spontaneous breaking of the QCD chiral symmetry.

\noindent
One of the most economical breaking patterns that can be imagined is the one that develops just four PNGBs, that is, the minimum number to be identified with the SM Higgs doublet. Together with the requirement of a custodial symmetry to protect the EW $\rho$ parameter from large deviations, this automatically leads to the choice of $SO(5)/SO(4)$ as the most simple realisation of the PNGB paradigm. This coset choice was introduced and discussed in~\cite{Agashe:2004rs}.
Beside a theoretical appeal, the importance of such a theory is that it is testable at the LHC. If the hierarchy problem is in fact solved by a new strong dynamics, this will manifest itself also  through new resonances that should be, for a reason of fine tuning, around the TeV scale. This is in fact the case for the copies of the SM quarks (especially of the top quark, given the dominant role it plays in the virtual corrections to the Higgs mass), that are called  \emph{top partners}, which are expected to be at an energy scale which is actually presently being tested at the LHC.
In general, also copies of the SM gauge bosons might be present, although at a mass higher that the spin 1/2 states, due to their contribution to the EW oblique observables, which push these states to a somewhat higher, nevertheless accessible, mass scale.

\noindent
Despite the assumptions made in the experimental searches, that usually allow for the presence of just one extra particle in order to derive limits which are as model-independent as possible, it is important to stress that in realistic CHM realisations these states are present with a higher multiplicity and this is valid both for spin 1/2 and for spin 1 resonances. This feature might cause model-dependent behaviour from the pure sum of the various signal contributions, up to more involved interference effects between these states and the SM, or between themselves.
In order to quantify these effects we choose a specific composite Higgs realisation and, in this work, we will focus on the so called 4DCHM proposed in~\cite{DeCurtis:2011yx}. The 4DCHM can be described as two non-linear $\sigma$-models, one for the $SO(5)/SO(4)$ breaking pattern while the other for the $SO(5)_L\otimes SO(5)_R/SO(5)_{L+R}$ one. This construction develops 10+4 NGBs, 10 of which will be absorbed adding a complete $SO(5)$ multiplet of resonances living in the $Adj[SO(5)]$, giving therefore rise to 10 massive degrees of freedom, identified with 4 neutral and 6  charged (conjugated) spin 1 physical states. The remaining 4 play the role of the Higgs fields.

\noindent
Let us briefly describe the characteristics of the 4 neutral spin-1 states, which are the subject of this work
\footnote{Actually, in order to guarantee a correct hypercharge assignment to the SM fermions, an extra $U(1)_X$ group needs to be added, bringing
to 5 the number of neutral resonances. Under the assumption of equal couplings for the $SO(5)$ and $U(1)_X$ groups (adopted in~\cite{Barducci:2012kk} as a specific parameter choice of the model described in~\cite{DeCurtis:2011yx}), two of the mass eigenstates can be redefined to be the ones aligned with the hypercharge direction, $T^Y=T^{3R}+T^X$, and the orthogonal combination respectively. Under this assumption, the latter will not couple to the constituents of the proton and we will therefore neglect it throughout our analysis (this happens in the minimal realisation where just the third generation of fermions mixes with the extended sector).}. 
The group $SO(4)$ is isomorphic to $SU(2)_L\otimes SU(2)_R$ and two resonances do correspond to the neutral component of the $(\textbf{3,1})$ and $(\textbf{1,3})$ triplets, degenerate in mass before the explicit breaking of the $SO(4)$ global symmetry. The other two neutral resonances  arise from the neutral component of the $SO(5)/SO(4)$ coset, with a mass $\sqrt{2}$ times higher than the ones just described. However, just one of these states will couple to the light fermions, reducing therefore the number of resonances playing a role in this analysis to 3. We refer the reader to Ref.~\cite{DeCurtis:2011yx} (see also \cite{Barducci:2012kk} for additional $Z'$ studies in DY channels) for a complete description of the model. Here, we just extract the mass scale of the states of interest for our phenomenological discussion. Neglecting the $SO(4)$ explicit breaking, the masses of the $SO(4)$ and $SO(5)/SO(4)$ resonances are given by $fg_\rho$ and $\sqrt{2}fg_\rho$, respectively, where $f$ is the
(compositeness)  scale of the spontaneous strong symmetry breaking and $g_\rho$ the gauge coupling of the extra $SO(5)$ group.
The explicit breaking of the $SO(4)$ symmetry will occur by introducing the SM $SU(2)_L\otimes U(1)_Y$ interactions, with coupling strength $g_0$ and $g_{0Y}$, respectively. This will cause a linear mixing between the SM $W^3_L$, $Y$ and the neutral component of the $(\textbf{3,1})$ and $(\textbf{1,3})$ triplet of $SU(2)_L\otimes SU(2)_R$, generating therefore a positive shift of the masses of these extra states. The mixing angles between the SM and extra states will be approximately $\theta\sim g_0/g_\rho$ and $\psi\sim g_{0Y}/g_{\rho}$. This will make these two states to acquire a mass $fg_\rho/\cos\theta$ and $fg_\rho/\cos\psi$, while further corrections of the order of $\xi=v^2/f^2$ will appear after EWSB, being $v$ the SM Higgs VEV. These EWSB effects will be the only source of corrections to the mass of the coset resonances, which will retain therefore a mass of $\sqrt{2}f g_\rho$, modulo corrections of order $\xi$. 
The squared masses of the interested gauge bosons at $\mathcal{O}(\xi)$ are given by~\cite{Barducci:2012kk}:
\begin{equation}
\begin{split}
 \begin{split}
& M^2_{Z^\prime_2}\simeq \frac{ m_\rho^2}{c_\psi^2} (1-\frac{s_\psi^2 c_\psi^4}{4 c_{2\psi}}\xi),\\
& M^2_{Z^\prime_3}\simeq \frac{ m_\rho^2}{c_\theta^2} (1-\frac{s_\theta^2 c_\theta^4}{4 c_{2\theta}}\xi),\\
& M^2_{Z^\prime_5}\simeq 2 m_\rho^2 \left[1+\frac 1 {16} (\frac 1 {c_{2\theta}}+\frac 1{2 c_{2 \psi}})\xi\right],
\end{split}
\end{split}
\label{eq:masses}
\end{equation}
with $\tan\theta=g_0/g_\rho$ and $\tan\psi=\sqrt{2}g_{0Y}/g_{\rho}$. (The numbering is due to the fact that 
$Z^\prime_{1,4}$ are the states which are inert for the purpose of this work.) With similar considerations it is possible to derive the couplings of these resonances to the light quarks and leptons. We report these expressions, derived at $\mathcal{O}(\xi)$, in Appendix~\ref{app}. Note, however, that in all our results both the masses and relevant couplings have been derived in a numerical way, without relying on any expansion approximation.

\noindent
Beside masses and couplings to SM fermions, of great relevance for this analysis are the widths of the extra gauge boson resonances. They can easily vary from a few percent of the masses of the $Z^\prime$s up to values comparable with the masses themselves.
Recall in fact that, generally in CHMs, extra fermions ( {\it top partners}) are present. They are coupled to the extra vector bosons, with a coupling strength $\propto g_\rho$, where the proportionality factor will be given by a combination of mixing angles, which will rotate the gauge states into the physical  ones. It is  therefore easy to understand that, if the new gauge bosons can decay into a pair of heavy fermions, the partial width in these final states can indeed  be larger than the one into SM fermions, since $g_0,g_{0Y}\ll g_\rho$. This has been studied in, {\it e.g.}, \cite{Barducci:2012kk}, where it is shown that the width of the extra resonances can be considered as a free parameter, depending essentially only on the mass scale of the top partners.

\noindent
In order to present our results for this multi $Z^\prime$ model, we need  to assess what are the current constraints on the mass spectrum of the 4DCHM arising from LEP, SLC, Tevatron and LHC data. As it is well known, extra gauge bosons give a positive contribution to the Peskin-Takeuchi $S$ parameter, which will set a limit on the masses of these extra states and hence on the compositeness scale $f$. Following the guidance of~\cite{Grojean:2013qca} we can say that a choice of $f>$~750 GeV and $m_{Z^\prime}>$~2 TeV can be considered as safe in order to prevent large corrections to the $S$ parameter. Corrections to the $T$ parameter are  slightly more involved, since they strongly depend on the extra fermionic content of the model but it can be estimated that a value of the {top partner} masses bigger than 800 GeV can be a choice compatible with the EWPTs. While a complete calculation of the EW oblique parameters in the framework of the 4DCHM is  beyond the scope of this work, we want to stress that the previous two estimates are indeed sufficient in order to study the phenomenology of relevance for our analysis. This $\sim$~2 TeV bound on the $Z^\prime$s mass is somewhat comparable with the one that can be obtained recasting LHC searches for narrow high mass di-lepton and $WZ$ resonances.
These searches set in fact a limit for a $Z^\prime$ with SM couplings to the light quarks and leptons around 2.5 TeV~\cite{Aad:2014cka,Aad:2014pha}. After  rescaling our signal rates, taking into account different couplings and BRs to the di-lepton final states, as well as summing over the possible contributions of the $Z^\prime_{2,3,5}$, we have that these searches set a mass limit of $\sim$~2 TeV for the masses of the (quasi) degenerate narrow $Z^\prime_{2,3}$ (the bound weakens for large width resonances, see later). We will use this value as a limit on the $Z^\prime$ masses.

\noindent
Direct searches for extra quarks are also relevant in constraining the 4DCHM parameter space. For example, the CMS limits on pair produced {top partners}, decaying into third generation quarks plus a SM boson, varies from  800 GeV~\cite{Chatrchyan:2013wfa} in the case of an exotic fermion with electric charge 5/3 to 782 GeV~\cite{Chatrchyan:2013uxa} and 785 GeV~\cite{CMS-PAS-B2G-13-003} in the case of extra fermions with the top and bottom quark electric charge, respectively\footnote{Note that in the latter two cases the extra quarks can decay either via charged or neutral currents. The reported bounds are the most stringent ones considering all  possible BR combinations.}\footnote{ While this work was in its completion phase, CMS released new limits on the mass of the 5/3 charged quark obtained with early 13 TeV data. These limits, depending on the chiral structure of the {\it top partner}, span from 940 to 960 GeV~\cite{CMS:2015alb}}. While, in principle, different extra fermions can feed the same final states giving rise to higher exclusion bounds, we will keep, as lower limits on their masses, the ones just mentioned. This is motivated by the fact that, for a $Z^\prime$ with a mass larger than 2 TeV, it is enough to have a {top partner} not lighter than 1 TeV in order to have the aforementioned effects of the extra fermions onto the $Z^\prime$s widths.
For this reason, in presenting our results, beside fixing the $Z^\prime$ masses above the 2 TeV value, the $\Gamma/M$ ratio of each state will be arbitrarily taken.

\subsection{The 4DCHM phenomenology}\label{sec:4DCHM_pheno}

 Here we present the phenomenology of the 4DCHM in the DY channel at the LHC. This channel could contain, a priori, the production and decay of all five extra heavy vector bosons predicted by the 4DCHM. However, the lightest BSM neutral resonance, $Z^\prime_1$, is inert  and the $Z^\prime_4$ is not coupled to first and second generation fermions. In the following therefore we will neglect these extra states and focus on the three remaing ones: $Z^\prime_2$, $Z^\prime_3$ and $Z^\prime_5$. The mediators of the leptonic DY channel, which give rise to dieletron and dimuon final states, are depicted as follows:
\begin{equation}
pp\rightarrow \gamma , Z, Z'_2, Z'_3, Z'_5\rightarrow l^+l^- 
\end{equation}
with $l=e, \mu$. For the experimental setup, we take as a reference the last analysis of dilepton spectra performed by the CMS collaboration in its search for exotic signatures given in Ref.~\cite{Khachatryan:2014fba}. Our values of the acceptance-times-efficiency factor for electrons and muons are based on that publication. In order to illustrate the phenomenology of the 4DCHM, we select the three benchmark points shown in Tab.~\ref{tab:benchmarks}. We moreover proceed by successive steps of incresing complexity.

\noindent
To begin with, we note that the lightest relevant vector boson, $Z^\prime_2$, is less coupled to all the fermions than the $Z^\prime_3$ (see~\cite{Barducci:2012kk} for the analytical expressions of the relevant couplings). Moreover, the heaviest resonance $Z^\prime_5$ is both too heavy and weakly coupled to the proton constituents to be produced at a significant rate. For these reasons, to a first approximation, one can consider a scenario where just one extra heavy $Z^\prime$ is produced, that is, the $Z^\prime_3$. This framework could not be fully representative of a general CHM as it is missing the possible multi-resonant structure of such theories with the corresponding interference effects. Nonetheless, it is adopted in the literature (see for istance Ref. \cite{Pappadopulo:2014qza}) as a first stage towards the complete picture. The framework where only the $Z^\prime_3$ might be observed at the LHC is a part of the parameter space which already contains some notable features of CHMs. In the next sub-section, we study this reductive but already explicative scenario. 

\begin{table}
  \begin{tabular}{|c||c|c|c|c|c|}
    \hline
      Benchmark & $f$ [GeV] & $g_\rho$ & $M_{Z_2}$ [GeV] & $M_{Z_3}$ [GeV] & $M_{Z_5}$ [GeV]\\
    \hline
      F &  1200 & 1.75 & 2192 & 2258 & 2972\\
    \hline
      G & 2900 & 1.00 & 3356 & 3806 & 4107\\
    \hline
      H & 700 & 3.00 & 2129 & 2148 & 2971\\
    \hline
  \end{tabular}
  \caption{4DCHM parameter space points associated to the benchmarks F, G and H mentioned in the text. }
  \label{tab:benchmarks}
\end{table}

\subsubsection{4DCHM: the singly resonant $Z'_3$ reduction and the NWA}\label{sec:4DCHM-reduction}

In this sub-section, we analyse a simplified version of the 4DCHM, where only one extra gauge boson can be detected at the LHC. Taking the $Z'_3$ boson as the new heavy spin-1 resonance is the most natural choice for this setup, as previously explained, and exploiting its features represents a useful term of comparison with the literature (see Ref. ~\cite{Pappadopulo:2014qza, deBlas:2012qp, delAguila:2010mx}) and a valid warming up in view of the study of the full picture. In studying its specific properties, first of all we check whether the commonly used NWA could be a viable method for computing the theoretical cross section in this singly resonant framework. We use benchmark F of Tab. \ref{tab:benchmarks} (which is essentially the (f) point corresponding to Fig. 12 of Ref.~\cite{Barducci:2012kk}: see Tabs.~19 and 22 therein for its features). We take the ratio $x=\Gamma_{Z^\prime_3}/M_{Z^\prime_3}$ to be 5\%. This quantity is a free parameter in our model. It can range from very low values ($x\simeq 1\%$) to much higher values ($x\simeq 20\%$) depending on the opening of some decay channels for the $Z^\prime_3$ boson, such as a decay into new heavy fermions. We then vary the dilepton invariant mass window around the $Z^\prime_3$ pole mass, $|M_{ll}-M_{Z^\prime_3}|\le \Delta m$, and compare three quantities: the complete cross section, the cross section without the interference term between the new  $Z^\prime_3$ boson and the SM $Z$ and $\gamma$ and the pure $Z^\prime_3$ signal computed in NWA.

\begin{figure}[t]
\centering
\includegraphics[width=0.45\linewidth]{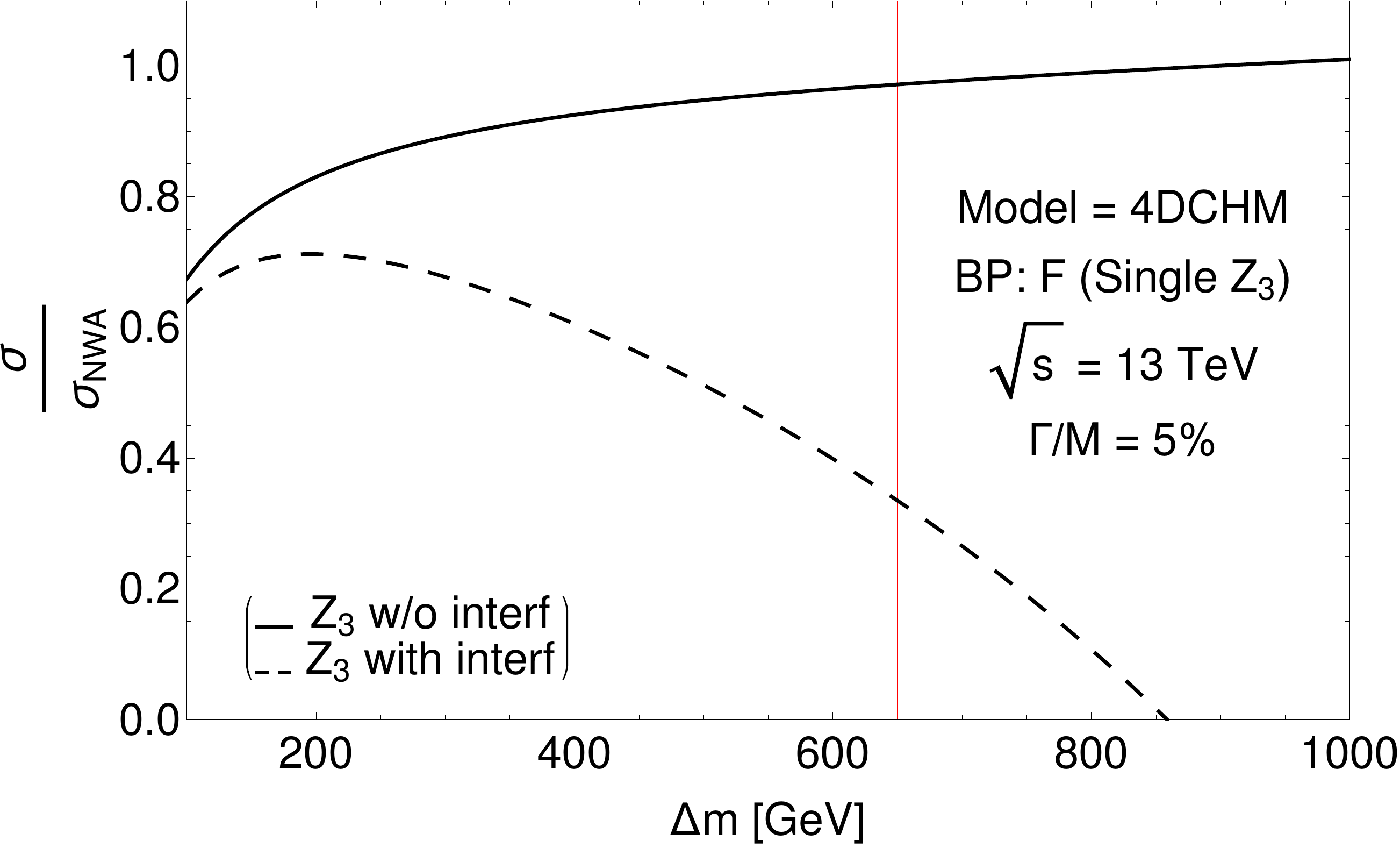}{(a)}
\includegraphics[width=0.45\linewidth]{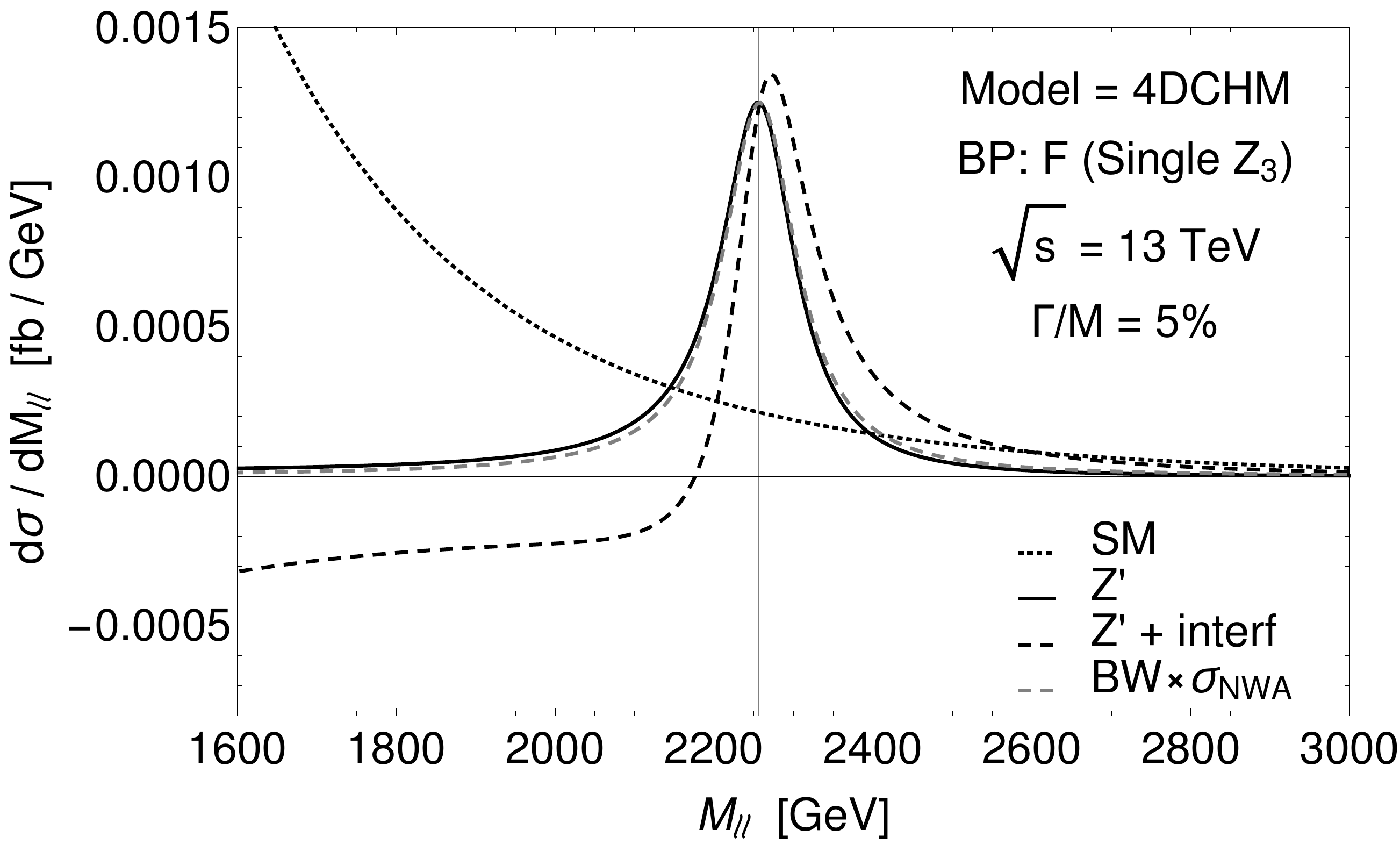}{(b)}
\caption{4DCHM benchmark F of Tab.~\ref{tab:benchmarks} considering only the $Z^\prime_3$ boson as active with $M_{Z^\prime_3}=2258$ GeV and $\Gamma_{Z^\prime_3}/M_{Z^\prime_3} = 5\%$.\break  (a) Differential cross section integrated in a dilepton invariant mass window $\Delta m$ around the $Z^\prime_3$ pole mass and normalized to the result computed in NWA. The solid line represents the ratio between the signal cross section, computed by taking into account only FW effects, and its NWA. The dashed line displays the ratio between the signal cross section, computed by accounting for both FW and interference effects, and its NWA. The vertical red line flags up the optimal mass interval which keeps the interference and FW effects below $O(10\% )$ in the case of narrow single-$Z^\prime$ models belonging to the $E_6$, LR and SSM class of theories~\cite{Accomando:2013sfa}.
(b) Differential cross section in the dilepton invariant mass $M_{ll}$. The solid line represents the signal in FWA. The dashed line is the complete signal, including both FW and interference effects. The grey dashed line shows the Breit-Wigner line shape normalized to the total cross section in NWA. The dotted line is the SM background. The two vertical dashed lines represent the position of the maximum in the first two cases.}
\protect{\label{fig:Wulzer}}
\end{figure}

As shown in Fig.~\ref{fig:Wulzer}a, integrating the differential cross section over a dilepton invariant mass region equivalent to three widths ($\Delta m\simeq 340$ GeV) would be enough to reproduce the NWA in absence of interference effects (solid line). These latter terms  completely change this picture. The dashed line shows that in presence of interference one can never reproduce the NWA result within a few percent accuracy. The minimum difference between the complete result and the NWA is around 30\% in this representative case and happens for a rather  narrow integration window, i.e., $|M_{ll}-M_{Z^\prime_3}|\le 200$ GeV. Unfortunately, not even the more sophisticated approach of Ref.~\cite{Accomando:2013sfa}, indeed designed for narrow single-$Z'$ models and working rather well for the multi-resonant NUED model(s) (see Sec. \ref{subsec:searches_ED}), seems to be applicable in the present context. As exemplified by the red vertical line, the difference here between the NWA and the full result is of ${\cal O}(70\%)$.

The wider the integration window, the bigger the discrepancy. This means that the interference contribution is overall destructive when one computes the total cross section. This feature is displayed in Fig.~\ref{fig:Wulzer}b where we plot the dilepton invariant mass distribution with and without interference. The purpose of this plot is to illustrate  the change in the shape of the signal that one obtains when considering the contribution of the $Z^\prime_3$ alone (with its finite width) and when adding the interference with the SM background. The latter produces a typically negative(positive) correction below(above) the $Z^\prime_3$ pole. Also notice the $\approx15$ GeV shift of the maximum  between the two curves. The main message here is that the interference {\it eats} part of the "bump" at lower masses and shifts the maximum of the curve to higher values of the dilepton spectrum. The resonant structure is thus no longer symmetric but has a sharp edge on the left-hand side of the peak. No matter what the selected mass window around the pole mass is, the signal depletion will persists and the complete result will never match the NWA. 

In conclusion, the crude NWA is not the correct mathematical tool to be used within the 4DCHM.
This has two immediate consequences. First of all, the theoretical cross section for the signal cannot be computed in NWA when crossing it with the 95\% C.L. upper bound on the BSM cross section derived by the experimental collaborations in order to extract bounds on the mass and/or couplings of the extra heavy gauge bosons. If doing so, the mass limits would be in fact  overestimated. For this particular case, already a theoretical error of +30\% on the cross section evaluated via NWA would imply a positive shift in the $Z'_3$ mass bound of around 160 GeV. If one adopted the mass interval $|M_{ll}-M_{Z^\prime_3}|\le 5\% E_{\rm LHC}$, presently used  by the CMS collaboration when interpreting the results of the data analysis within narrow single $Z'$-boson models (see red line in Fig.~\ref{fig:Wulzer}a), the shift would increase to 450 GeV. Hence, we should caution against simplistic approaches using the NWA, thereby implicitly assuming  that FW and interference effects are negligible. This is not appropriate for CHMs in general. Neither the cross section nor the peak position coincide in the two cases. Consequently, neither the exclusion limit nor the discovery estimate in mass would be accurate. 

The additional consequence of the non-applicability of the NWA is that the signal shape assumed by the CMS experimental collaboration, that is a Breit-Wigner convoluted with a Gaussian resolution function, is not always appropriate for this model. The interference indeed may distort this symmetrical function around the hypothetical $Z'$-boson pole mass. This effect might have consequences in the shape analysis of the dilepton invariant mass spectrum which is performed via the likelihood approach as summarised in Sect. \ref{subsec:searches_ED}. For $Z'$-boson that are not very narrow, characterized for example by a ratio $\Gamma_{Z^\prime_3}/M_{Z^\prime_3}=5\%$ as in Fig.~\ref{fig:Wulzer}b, the line-shape distorsion of the signal would be observable as the invariant mass resolution in the dielectron channel is smaller/comparable to that. In this circumstance, such an alteration of the signal compared to the Breit-Wigner hypothesis would affect the limit setting procedure, in presence of data points. This aspect could be improved rather easily. We do not attempt any modification here but just highlight it for a possible future use.

\begin{figure}[t]
\begin{center}
\centering
\subfigure[]{
\includegraphics[width=.45\textwidth]{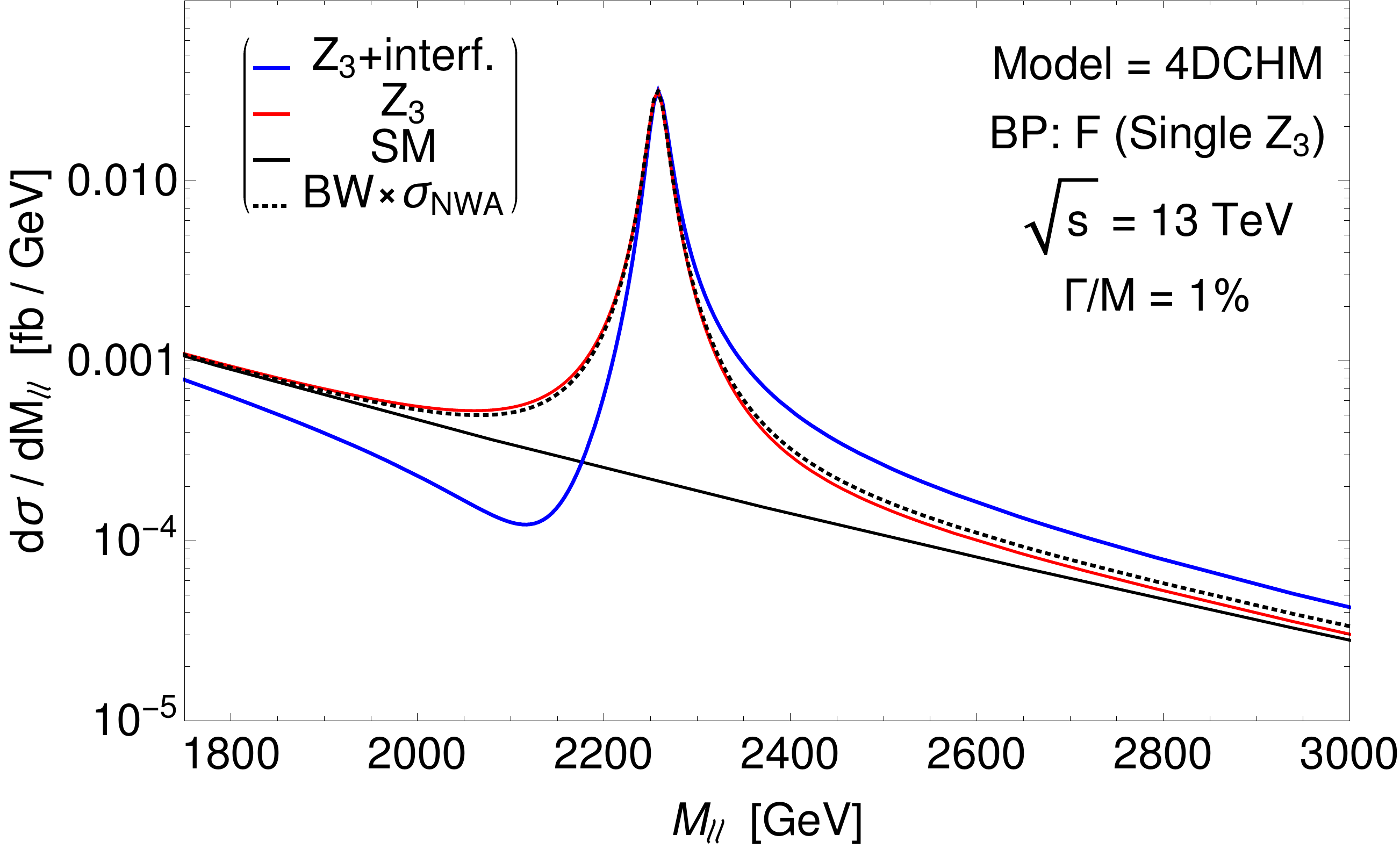}
\label{}
}
\subfigure[]{
\includegraphics[width=.45\textwidth]{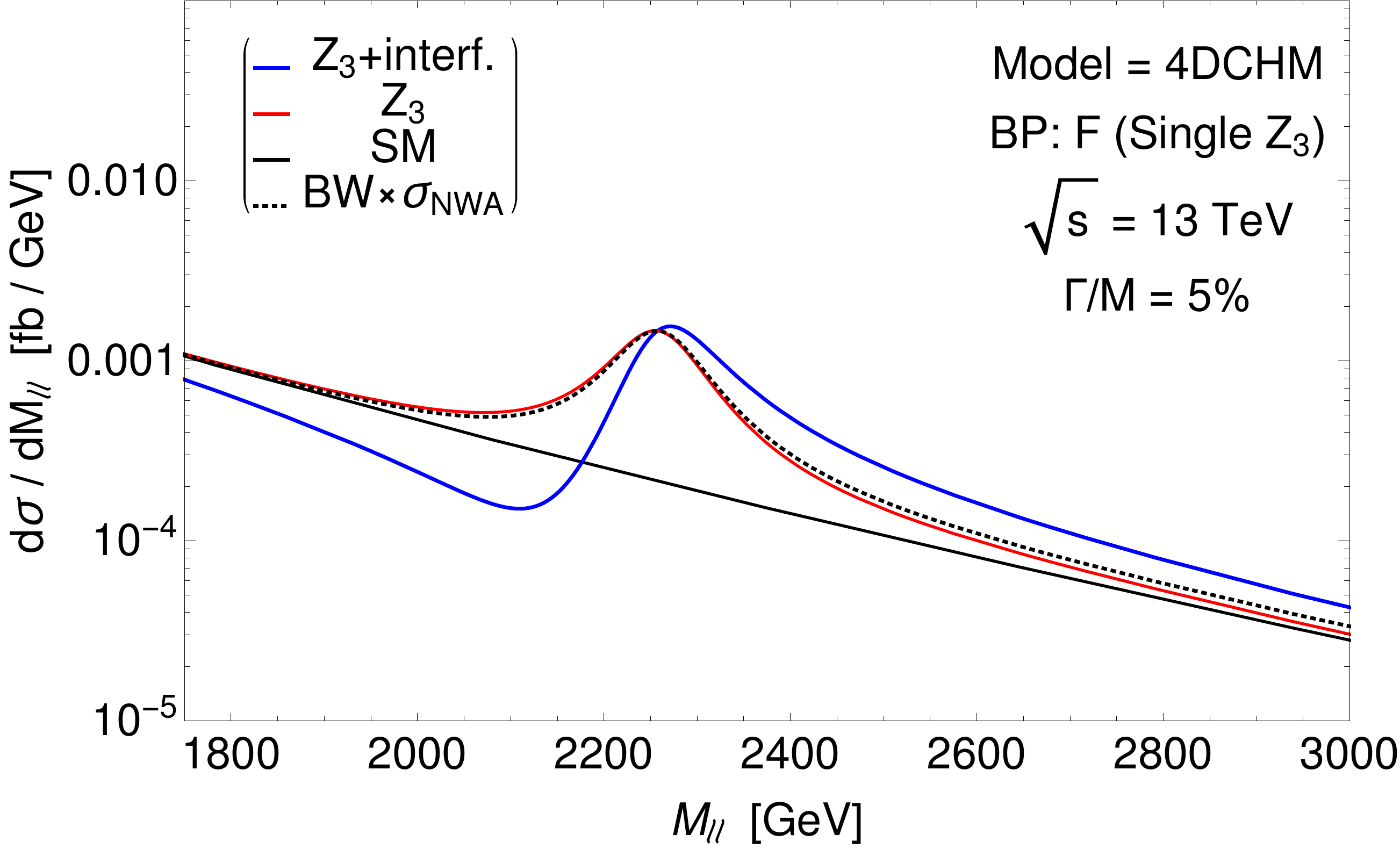}
\label{}
}
\subfigure[]{
\includegraphics[width=0.45\textwidth]{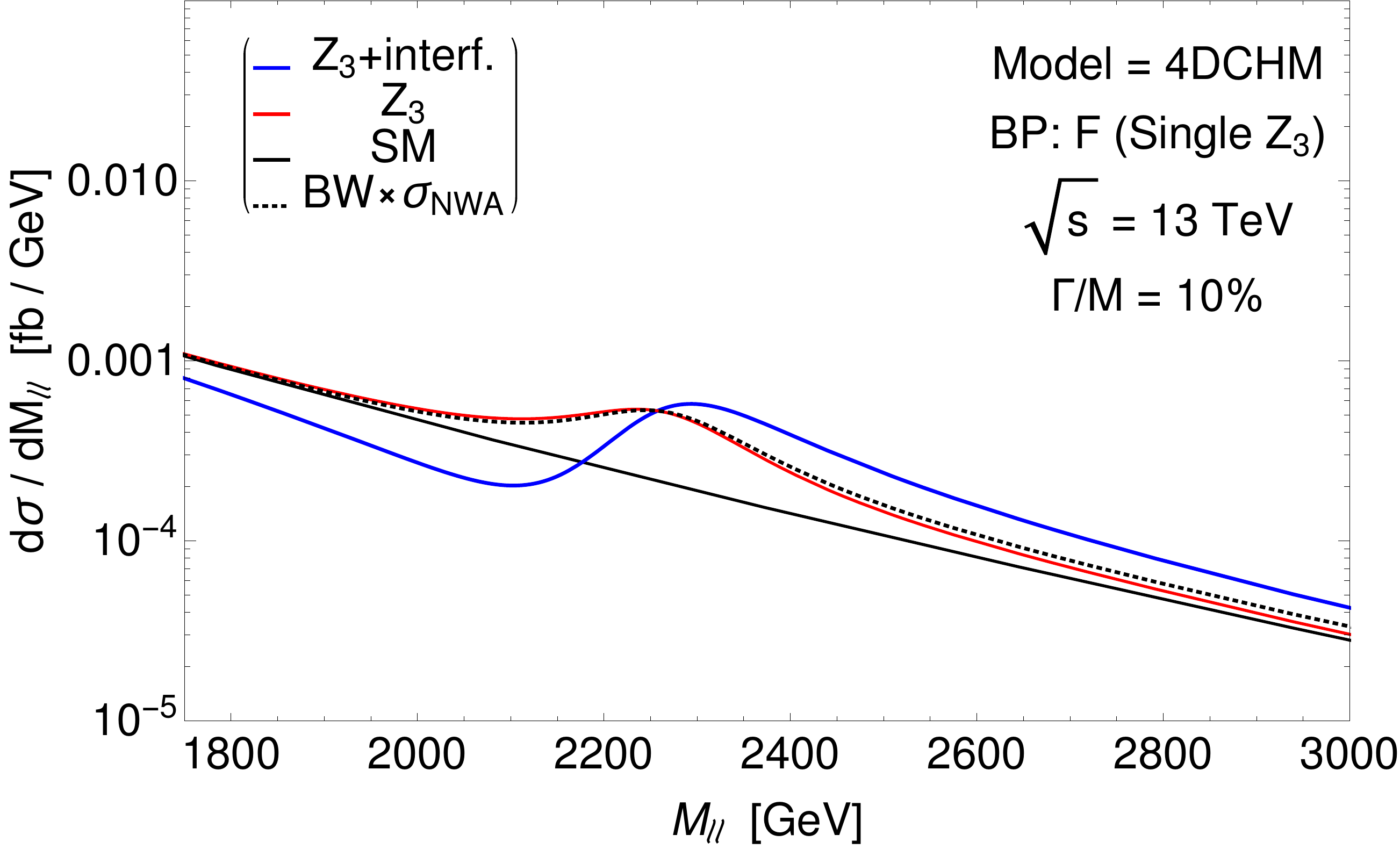}
\label{}
}
\subfigure[]{
\includegraphics[width=0.45\textwidth]{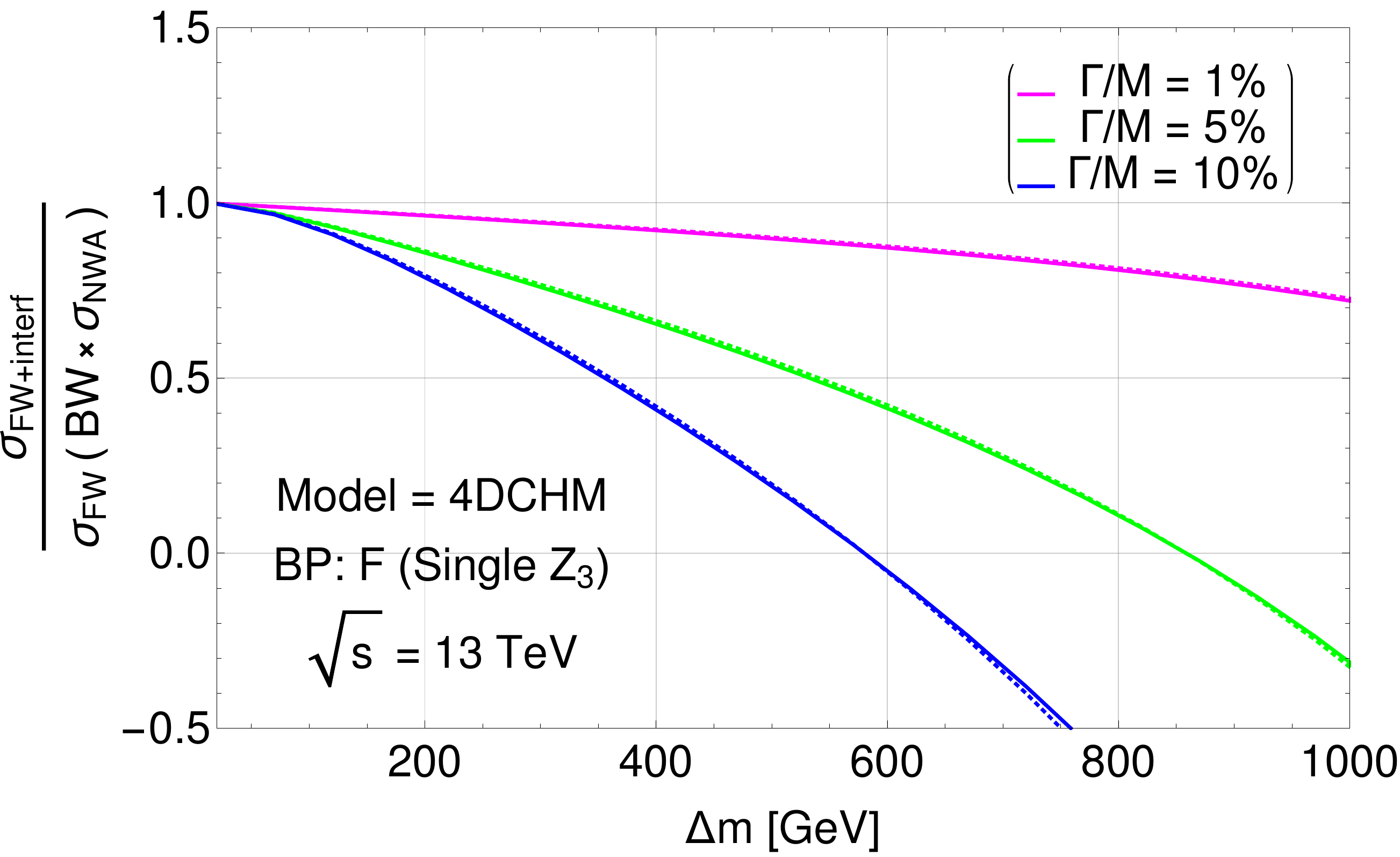}
\label{}
}
\caption{(a) Differential cross section in the dilepton invariant mass for the benchmark point F in Tab.~\ref{tab:benchmarks} in the reduced singly-resonant approximation (only $Z'_3$ is included). We consider a very narrow $Z'_3$ boson with $\Gamma_{Z^\prime_3}/M_{Z^\prime_3} = 1\%$. (b) Same as plot (a) for a medium ratio $\Gamma_{Z^\prime_3}/M_{Z^\prime_3} = 5\%$. (c) Same as plot (c) for a wide $Z'_3$ boson with $\Gamma_{Z^\prime_3}/M_{Z^\prime_3} = 10\%$. (d) Solid lines: ratio between the full signal cross section for the $Z^\prime_3$ boson corresponding to benchmark F, including both Finite-Width and interference effects, and its pure FW approximation as a function of the integration interval for the three different choices of the ratio $\Gamma_{Z^\prime_3}/M_{Z^\prime_3}$. Dashed lines: same but considering as  normalization the BW approximation for the signal line shape normalized to the total cross section in NWA.}
\label{fig:interval}
\end{center}
\end{figure}
 
A more considerate approach than in standard NWA studies (established in Ref.~\cite{Langacker:1984dc} and adopted in innumerable publications since, see Ref.~\cite{Accomando:2013sfa} for a mini-review) is taken in Ref.~\cite{Pappadopulo:2014qza}, where issues regarding FW and interference effects are dwelt upon. The authors consider a generic theoretical framework that is supposed to reproduce a large class of explicit models predicting a single $Z'$-boson. This scenario captures all the common features of a variety of theories, by using a simplified Lagrangian that depends on a small set of free parameters (as already proposed in ~\cite{delAguila:2010mx}). Relations are then used to convert these free parameters into the specific properties of the various models. In this way, all the $Z'$-bosons predicted by different theories can be simulated via a unique simplified setup. When considering the dilepton spectrum in a DY-process, the realm of validity of such a framework corresponds however only to a dilepton invariant mass range equivalent to one width around the $Z'$-boson pole mass. This is the signal region where off-shell effects, that is FW and interference, affecting the limit setting procedure can be considerably reduced. The authors thereby advise to perform the $Z'$-boson analysis within such a restricted search window in order to derive model-independent bounds. To support this statement, for the predicted new resonance, they compare the full cross section with the NWA result rescaled by a factor that accounts for the fraction of the Breit-Wigner shaped signal included in the integration interval equivalent to one $Z'$-boson width. This is done in the spirit of reaching a consistent theoretical interpretation of the experimental results that, as we know, are obtained under the assumption that the signal shape is a Breit-Wigner normalized to the total cross section computed in NWA. The effect of this rescaling is thereby meant to align the 95\% C. L. upper bound on the BSM cross section, which is given in NWA, with the full theoretical cross section (or its FW approximation) computed in an integration range where FWA and Breit-Wigner line-shape (normalized to the NWA result) coincide up to some extent. According to Ref. ~\cite{Pappadopulo:2014qza}, this interval is exactly the mass window $|M_{ll}-M_{Z^\prime_3}|\le \Gamma_{Z^\prime_3}$ where their simplified model is valid.

\noindent
Before attempting any comparison with what is done experimentally, we import here this general procedure and try to apply it to our case on a theoretical/computational basis. In Figs. \ref{fig:interval}a-c, we plot the dilepton spectrum comparing three cases: when the signal is obtained in FWA (red line), when it is computed by taking into account both FW and interference effects (blue line) and when it is described by a simple Breit-Wigner function normalized to the total cross section in NWA (black dotted line). We consider three benchmark scenarios ranging from a very narrow $Z'$-boson (Fig. \ref{fig:interval}a), a medium-large $Z'$-boson (Fig. \ref{fig:interval}b) and a wide $Z'$-boson (Fig. \ref{fig:interval}c). We observe that the Breit-Wigner shaped signal resembles quite closely the FW approximation for all three $Z'$-boson widths. The major alteration of the signal shape comes from interference effects. These latter distort the symmetrical distribution of the signal and shift the maximum to higher invariant mass values. The effect increases with the $Z'$-boson width. To quantify the change produced by the FW and interference contributions on the total cross section, in Fig. \ref{fig:interval}d, we plot the full integrated cross section of the $Z^\prime_3$ signal, including both FW and interference effects, normalized to the $Z^\prime_3$ signal rate in FWA, as function of the integration interval in the dilepton invariant mass for three values of the $Z^\prime_3$ width:  $\Gamma_{Z^\prime_3}/M_{Z^\prime_3} = 1\% , 5\% , 10\%$. The plotted ratio should point out the weight of the interference contribution to the total cross section in the given integration range. The superimposed dotted lines display the ratio of the full signal cross section over the  Breit-Wigner shaped signal normalized to the NWA result. The difference with respect to the solid lines is neglegible. As one can see in Fig. \ref{fig:interval}d, for very narrow resonances ($\Gamma_{Z^\prime_3}/M_{Z^\prime_3} = 1\%$), both FW and Breit-Wigner (BW) shaped signal work well within a pretty extended region (much bigger than one $Z'$-boson width). For resonances with an intermediate value of the width ($\Gamma_{Z^\prime_3}/M_{Z^\prime_3} = 5\%$), the agreement between full and FW (or BW) results is better than $O(10\% )$ only if one restricts the mass integration interval to a region equivalent to one $Z'$-boson width. For $\Gamma_{Z^\prime_3}/M_{Z^\prime_3} = 10\%$, the agreement between full and FW (or BW)   cross sections drops down. Even within an integration interval equivalent to  $\Gamma_{Z^\prime_3}$, the discrepancy is in fact of the order of 25\%. 
Summarizing, our finding is that the FW approximation and the BW shaped signal work pretty well for very narrow resonances or if the distance between the crossing point and the resonant peak is bigger than at least one $Z'$-boson width. In this case, in fact, the interference is far apart enough so not to alter the FW (or BW) line shape of the signal. The problem is that this condition is not always satisfied within the 4DCHM. The width is in fact a free parameter and can assume values between 1\% and 20\%, resonably.

The message of Ref.~\cite{Pappadopulo:2014qza} is that for generic CHMs the experimental collaborations should restrict the search window in a mass range equivalent to one $Z'$-boson  width in order to avoid modeling FW and interference contributions that are proper of the specific theory. Under this stringent condition and independently on the value of  $\Gamma_{Z^\prime_3}/M_{Z^\prime_3}$, limits on the mass of the extra $Z'$-boson predicted within a large class of explicit models could be derived, at once, from the 95\% C. L. upper bounds on the BSM cross section in a consistent and model-independent way. Our findings are different: model dependent effects can be neglected only for very narrow $Z'$ bosons ($\Gamma_{Z^\prime_3}/M_{Z^\prime_3}\le 3\%$) and, in this case, one does not need to restrict the search window within one $Z'$-boson width. The different conclusion we have
is due to the fact that we include the interference contribution to the differential (or total) cross section consistently (similarly to ~\cite{deBlas:2012qp}) while the authors of Ref. ~\cite{Pappadopulo:2014qza} parametrize its rate via an overall factor which can vary between -1 and +1, their focus being the FW approximation.

\noindent
All this should make clear that it is not really feasible to extract limits in the 4DCHM (and realistic CHMs in general) by using the NWA approach when interpreting the results of direct searches. For these models, the more accurate FW (or BW) approach is not advisable either, as it fails to reproduce the complete cross section generally for intermediate to wide $Z'$-bosons. Only for narrow $Z'$s, the pure FW approximation (or the BW line-shape normalized to the total cross section in NWA) can be used safely. In any case, one should not retain the signal only within one $Z'$-boson width. This procedure is not needed for narrow $Z'$-bosons. Moreover, it would be intrinsically model dependent, as the width depends on the specific theory, and not generally consistent with a realistic data analysis which scans over the full dilepton spectrum in order to maximize the statistics and the sensitivity to New Physics.

For sake of clarity, let us briefly recall the experimental procedure applied when setting mass limits. As summarised in Sect.~\ref{subsec:searches_ED}, the experimental analyses perform an unbinned likelihood fit over the dilepton invariant mass spectrum. The likelihood function is constructed under the hypothesis that the resonant signal should stand up over a smooth SM background shape. For mass scales beyond 2 TeV, that is the region of interest in recent and ongoing searches where one might expect to observe a possible new vector boson, the number of SM background events is rather small at the present luminosities. To be more quantitative, in the past LHC RunI at 8 TeV, the SM background was identically zero at the collected luminosity ${\cal L}$= 20 fb$^{-1}$ for $M_{ll}\ge$ 1.8 TeV. Under this circumstance, the possible depletion of events, next to the resonant peak and due to intereference effects, has no impact at all on the $Z'$-search. No matter what these model-dependent effects will be, the narrow $Z'$-boson  signal in the dilepton invariant mass distribution will always appear as a well defined "bump" standing over a zero SM background (this is true also when the SM background is not zero but is sub-dominant compared to the signal). For $Z'$-boson masses expected to be at scales where the SM background is (almost) zero one can thereby perform a model-independent analysis up to a large extent. The results hence allow to extract bounds on the $Z'$ boson mass in a variety of different models, including the CHMs, under the assumption that the predicted heavy neutral boson is narrow. In order to compare the theoretical cross section with the 95\% CL upper bound on the BSM cross section derived from the data analysis, one thus need to compute the full integral under the "bump" which means considering an integration region different from one natural width, generally. 

For wider resonances, the FW (or BW) signal hypothesis is no longer valid owing to the increasing interference effects that alter the signal line-shape. A modified and dedicated approach should be taken in direct searches. Under this condition, when deriving mass limits we strongly advise to compute the complete cross section within general CHMs. Moreover, if a discovery in the usual "bump" hunt should take place during an early run at low luminosity, the problem of interpreting it would come next. In a later run at higher luminosity, one would hope to profile such a new resonance in order to pin down the underlying theory. In doing so, one should analyse the signal shape, which in the 4DCHM is highly characterized by interference effects. For this purpose, independently on the $Z'$-boson width, the FW approach only (with no modelling of the true interference) would simply fail.

\begin{figure}[t]
\centering
\subfigure[]{
\includegraphics[width=.45\textwidth]{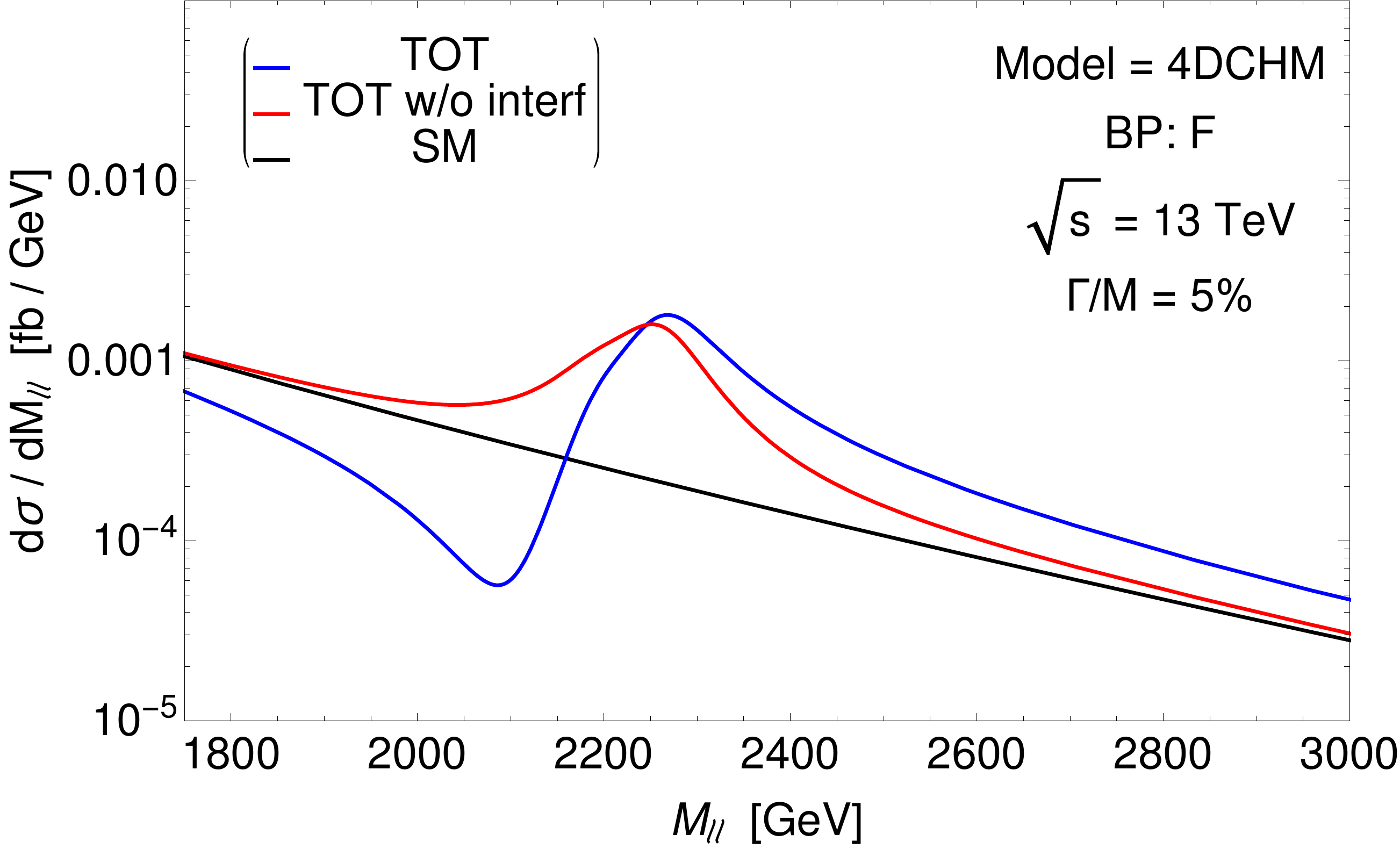}
\label{}
}
\subfigure[]{
\includegraphics[width=.45\textwidth]{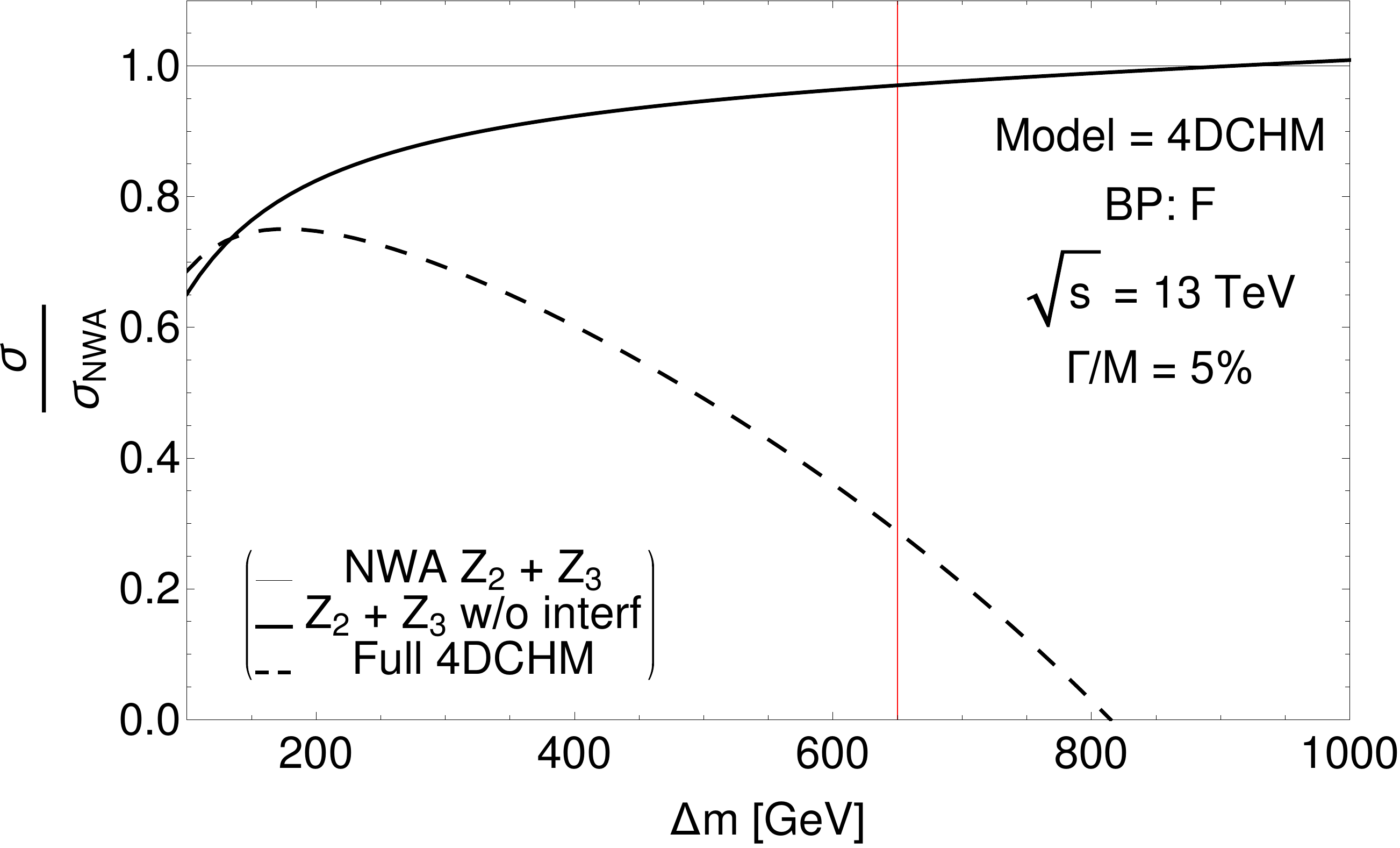}
\label{}
}
\caption{(a) Differential cross section in the dilepton invariant mass for the benchmark point F in Tab.~\ref{tab:benchmarks} within the complete 4DCHM i.e. double-resonant $Z^\prime_{2,3}$ scenario. (b) Ratio of the full signal cross section for the $Z^\prime_{2,3}$ bosons corresponding to benchmark F within the complete 4DCHM scenario (dashed line) 
and the two resonances FWA (solid line) over the NWA result as a function of the symmetric integration interval around the peak. The vertical red line represents the CMS adopted optimal cut which keeps the interference and FW effects below 10\% in the case of narrow single $Z^\prime$ models \cite{Accomando:2013sfa}.}
\label{fig3}
\end{figure}

As we aim to be as general as possible, we therefore evaluate the full (differential) cross section accounting for both FW and interference when discussing the 4DCHM phenomenology in the following sections. As a final remark, we would like to point out that the single $Z'$ boson reduction of CHMs can be partial, as it is for the 4DCHM. Being applicable only to  restricted regions of the parameter space, it  cannot be representive of the full dynamics of a CHM. In the next sub-section, we therefore analyse the complete version of the 4DCHM, which gives rise to a multi-resonant peaking structure. 

\subsubsection{Multi-$Z'$ 4DCHM: direct limits}\label{sec:4DCHM-limits}

We now analyse the complete 4DCHM and study the impact of its multi-resonant structure for $Z'$ searches at the LHC. We first derive the direct limits on mass and couplings of the new $Z^\prime_{2,3}$ bosons at the past LHC RunI with 7, 8 TeV energy and integrated luminosity ${\cal{L}}\simeq 20~fb^{-1}$. 
In a CHM with low mass spectra, like the 4DCHM we are presenting here, the FW and interference effects discussed in the previous section are potentially even more complicated, owing to the presence of multi-resonant spin~1 states. In this section, we consider the complete 4DCHM where both the  $Z^\prime_2$ and $Z^\prime_3$ bosons are produced in Drell-Yan. The third active resonance, $Z^\prime_5$, is much heavier and thus difficult to produce, ultimately giving a very negligible contribution to the dilepton invariant mass spectrum which can be explored at the LHC RunII. For these reasons and ease of computation, we thus continue neglecting the $Z^\prime_5$ resonance.
 
The inclusion of the $Z^\prime_2$ boson does not alter the conclusions drawn for the simplified singly-resonant scenario, not qualitatively at least, only quantitatively. In Fig. \ref{fig3}a, we plot the dilepton invariant mass spectrum as predicted in the complete double-resonant $Z^\prime_{2,3}$ 4DCHM. This distribution should be compared to the corresponding $Z^\prime_3$ spectrum  in Fig. \ref{fig:interval}b, computed in the singly-resonant reduction. As one can see, the major difference concerns the dip before the resonant peak(s), the "bump" being pretty unchanged. The dip, already visible in the single $Z^\prime_3$ boson case, gets in fact accentuated in the complete 4DCHM. Incidentally, one may notice that none of the individual terms representing the interference between the various gauge bosons (SM and beyond) is responsible for the full effect. They all contribute equally and this feature is general to the 4DCHM parameter space. As previously stated, the negative contributions before the resonant peak(s) coming from interference spoil again the result in NWA (or in the FW approach, for that matter). This can be seen in Fig. \ref{fig3}(b)  where we have repeat the previous exercise of plotting the ratio between the full signal cross section and its NWA as a function of the integration interval, $\Delta m$, around the $Z'$ pole mass. In the double-resonant case the NWA is defined as the sum of the two individual NWAs for the two $Z^\prime_{2,3}$ bosons. We find again appreciable differences between full and NWA cross sections, which are comparable to or larger than those  appearing in the single-resonant $Z^\prime_3$ scenario. Hence, the conclusions are same as before. One should avoid using the NWA within the 4DCHM (and similar CHMs) when computing the theoretical cross section to derive limits on the mass of the new gauge bosons. The FWA could be used pretty safely, but only for very narrow $Z'$-bosons. These two approximations would however be not applicable in the analysis of the signal shape for profiling the new resonances in case of discovery. In the following, we will thereby perform a complete calculation of (differential) cross sections.

\begin{figure}
\centering
\includegraphics[width=0.40\textwidth]{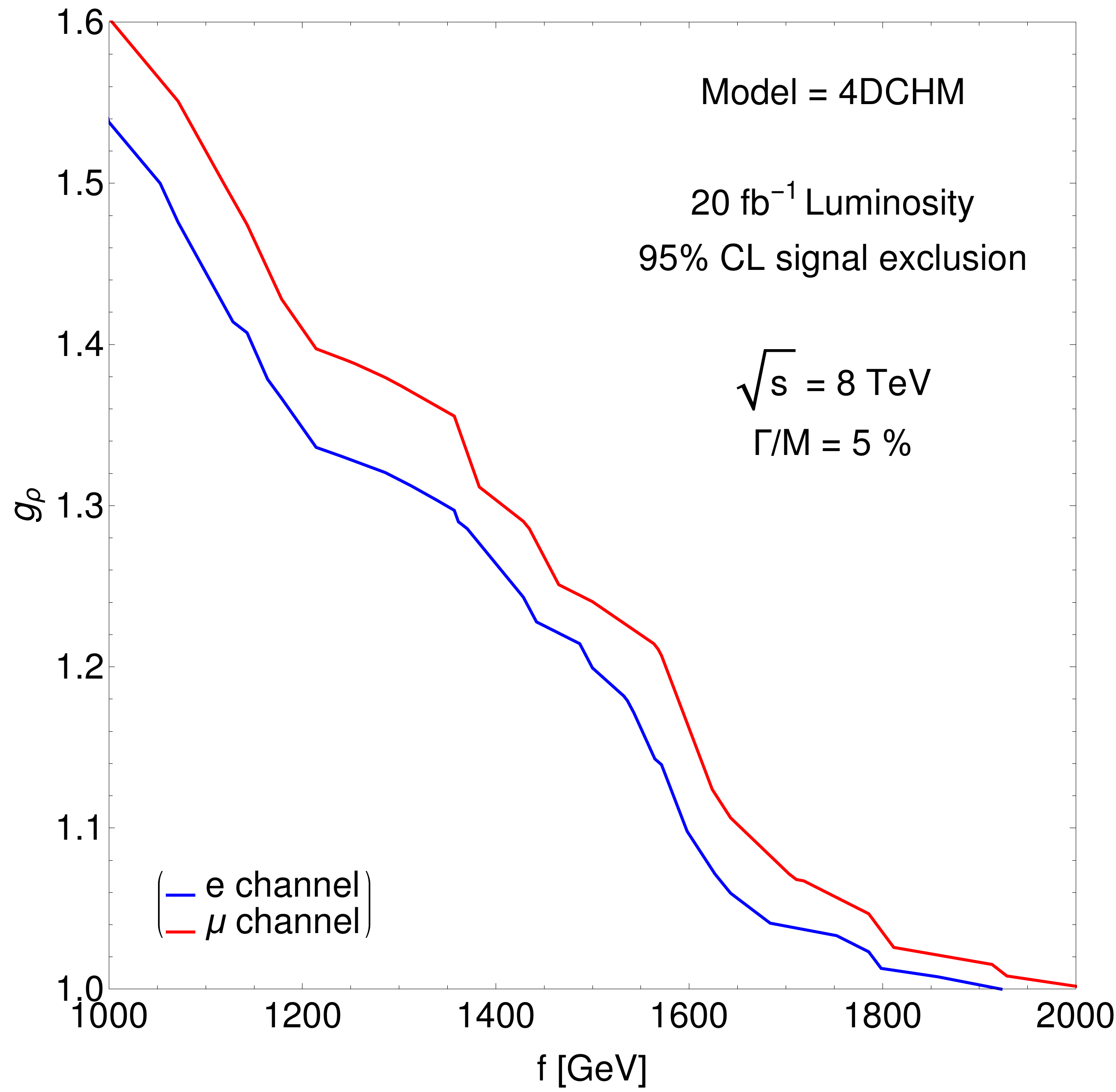}
\caption{95\% CL contour plot in the 4DCHM parameter space generated for the past LHC RunI with 8 TeV energy and luminosity ${\cal{L}}\simeq 20fb^{-1}$. The region above the solid lines is excluded by the dielectron channel (blue line) and the dimuon channel (red line). We have included NNLO QCD corrections and the appropriate acceptance-times-efficiency factor.}
\label{fig:Lum_signal_8TeV}
\end{figure}

Before illustrating the type of signatures that could appear at the LHC RunII, we need to extract the direct bounds on the 4DCHM parameter space coming from the data analysis performed at the past LHC RunI with 7, 8 TeV energy and ${\cal{L}}\simeq 20~fb^{-1}$. We apply the acceptance-times-efficiency factor for electrons and muons as defined in the last CMS analysis of dilepton final states \cite{Khachatryan:2014fba}. We compute the theoretical cross section by integrating over the invariant mass region whose extremes are the crossing point between signal and SM background on the left and three natural widths beyond the heavier resonance on the right. We include a mass scale dependent NNLO QCD correction. This prescription maximises the signal and is consistent with the experimental analyses, as we discussed previously. Using  Poisson statistics, we then compute the statistical significance of the 4DCHM signal and derive our limits on the parameter space specified by the plane ($g_\rho , f$) where $g_\rho$ is the gauge coupling of the $SO(5)$ group and $f$ is the scale of the spontaneous strong symmetry breaking. Their relation to the $Z'_{2,3}$-boson masses is given in Eq. \ref{eq:masses}.
The results are summarized in Fig. \ref{fig:Lum_signal_8TeV} where the blue(red) curve refers to the electron(muon) channel. (Muons have a worse mass resolution than electrons.) While 
the dielectron invariant mass resolution is $R_e\simeq 1.2\%$, rather constant over the entire mass spectrum, for muons the resolution depends sizeably on the mass scale and reaches the value $R_\mu\simeq 9\%$ for a dimuon invariant mass of the order of 3 TeV. This feature is however compensated by a better acceptance-times-efficiency factor with respect to electrons. The global result favours the muon channel which can then set the strongest limits, as shown in Fig.~\ref{fig:Lum_signal_8TeV}. In the next section, we'll see that the better mass resolution favours the electron channel in profiling the new resonances. The two channels are thus highly complementary within the 4DCHM.

We come now to an important issue which concerns the future search for spin-1 resonances at the ongoing LHC RunII. A key point to note is that the limits in Fig. \ref{fig:Lum_signal_8TeV} have been computed in-house, taking as external input only the CMS acceptance-times-efficiency factor for electrons and muons as a function of the dilepton invariant mass scale. This is because the limit setting procedure implemented by the experimental collaborations does not provide at the moment a multi-resonant signal hypothesis. Oppositely to what happens within the NUED models, where the first level KK-states of the extra dimensional tower are (almost) degenerate so that the multi-resonant structure collapses into a single "bump" standing far away from the dip induced by interference effects thus allowing a direct comparison with the experimental limits on the BSM cross section, within the 4DCHM we cannot extract mass/coupling bounds from present direct searches. The spectrum is in fact not degenerate, in general, and the peaking/dip structure can be quite compressed. We will continue this discussion with more detail in the next section, while projecting discovery and exclusion potential at the LHC RunII. For now, we assume the direct limits shown in Fig.~\ref{fig:Lum_signal_8TeV}. In the allowed region of the parameter space, we then select the three benchmark points listed in Tab.~\ref{tab:benchmarks} in order to illustrate the type of signatures one could expect at the ongoing LHC RunII. This will be done in the next sub-section.

\subsubsection{Multi-$Z'$ 4DCHM: signal shapes at the LHC RunII}\label{sec:4DCHM-signal}

\begin{figure}[t]
\centering
\subfigure[]{
\includegraphics[width=0.45\textwidth]{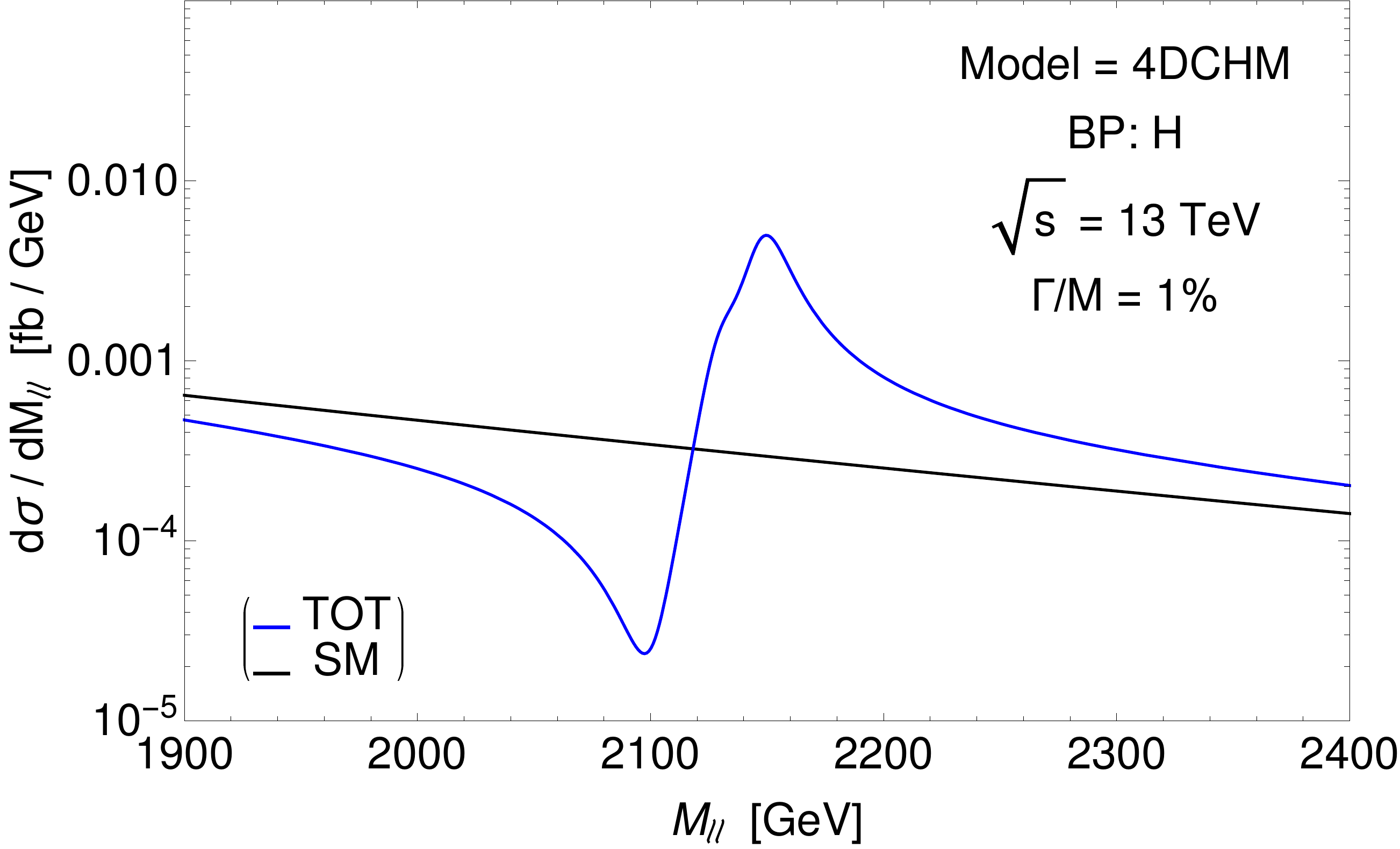}
\label{}
}
\subfigure[]{
\includegraphics[width=0.45\textwidth]{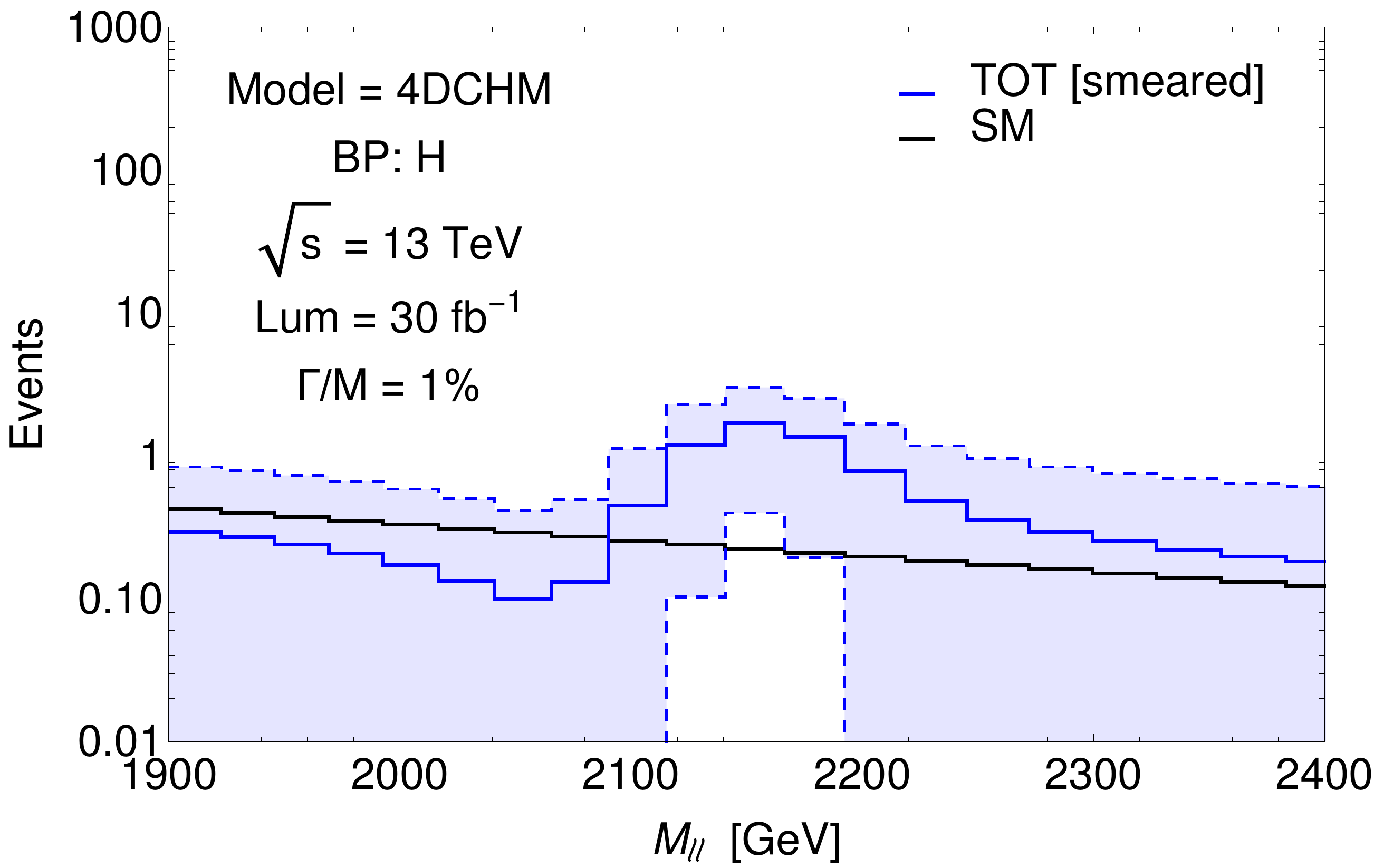}
\label{}
}
\subfigure[]{
\includegraphics[width=0.45\textwidth]{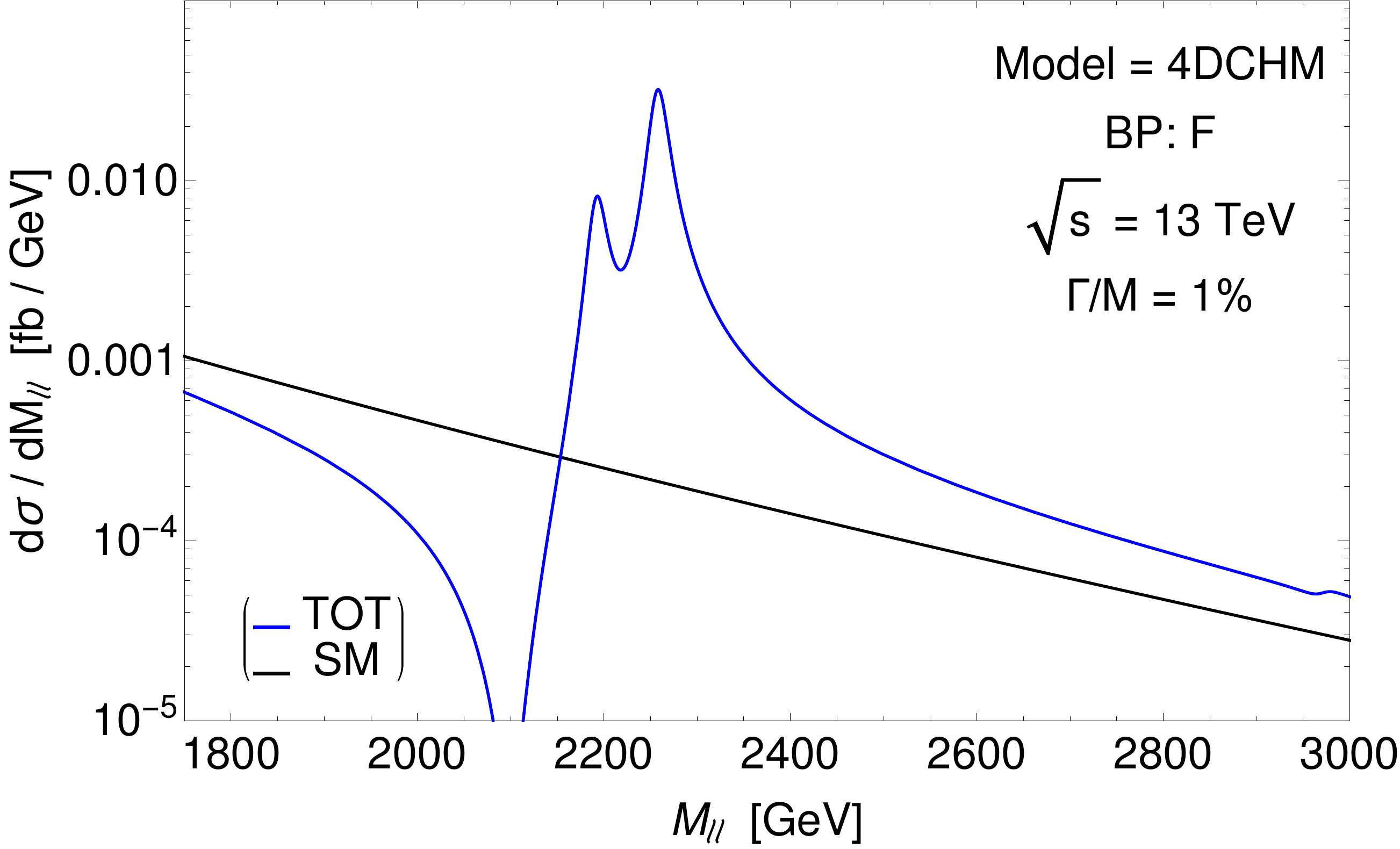}
\label{}
}
\subfigure[]{
\includegraphics[width=0.45\textwidth]{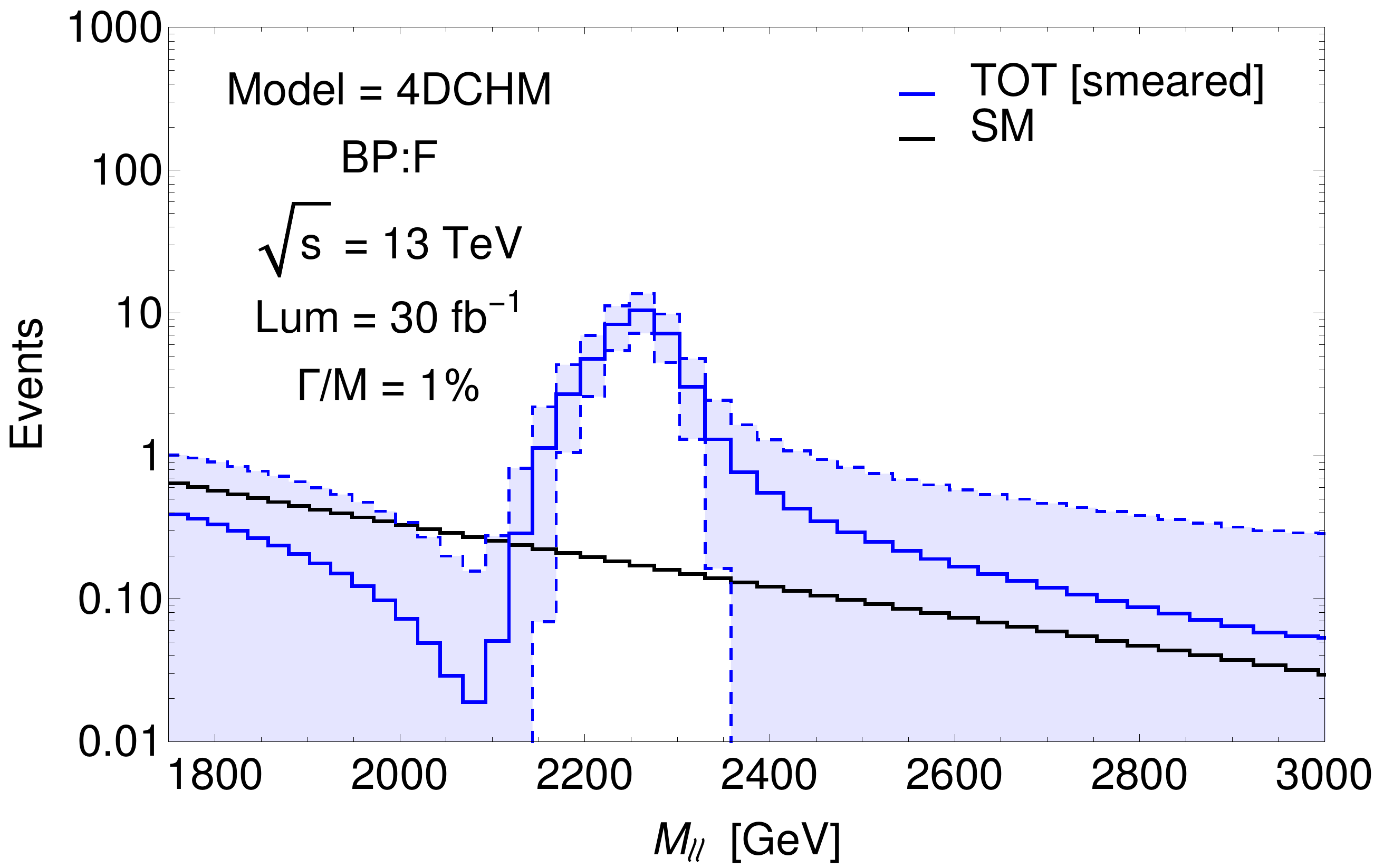}
\label{}
}
\subfigure[]{
\includegraphics[width=0.45\textwidth]{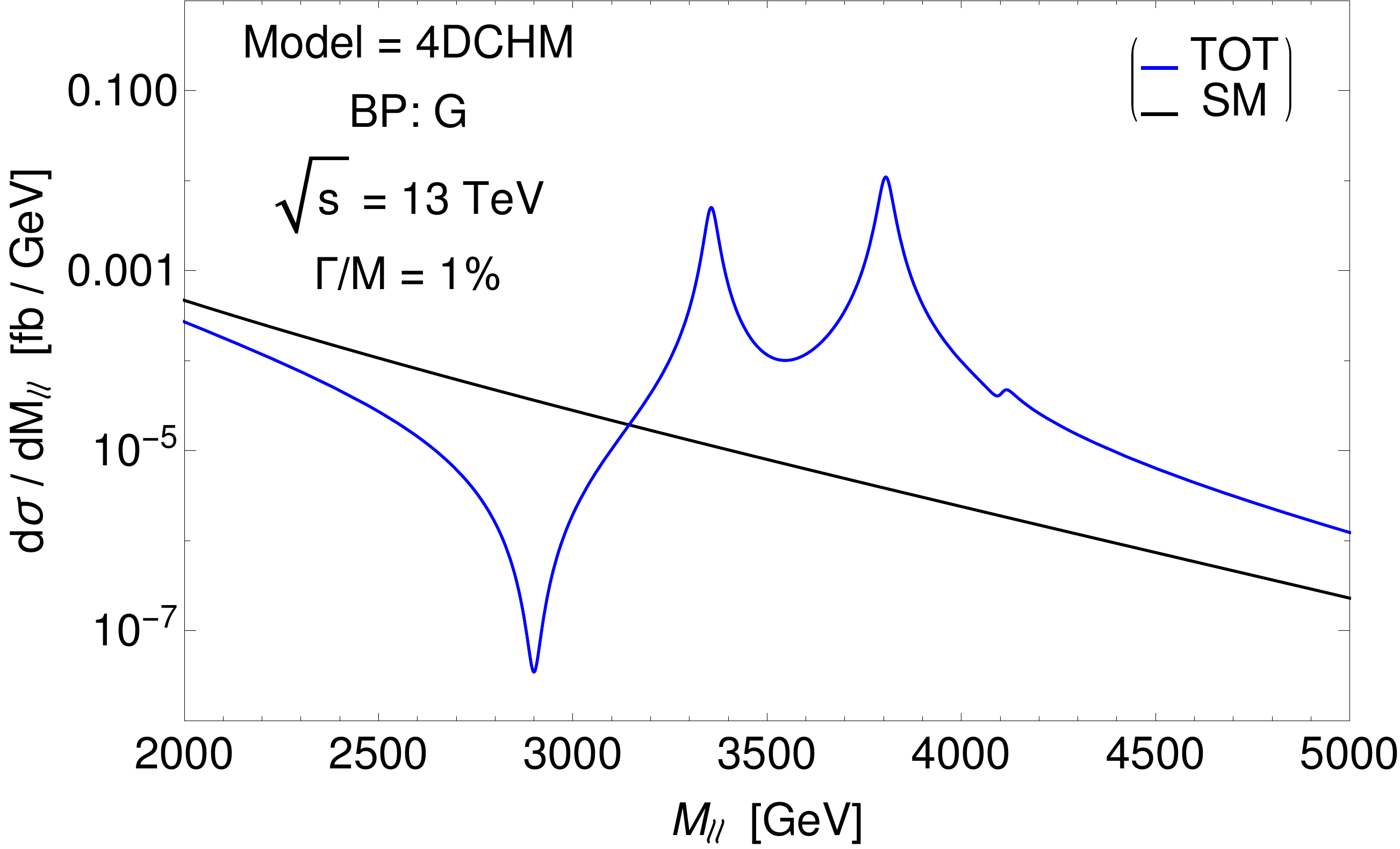}
\label{}
}
\subfigure[]{
\includegraphics[width=0.45\textwidth]{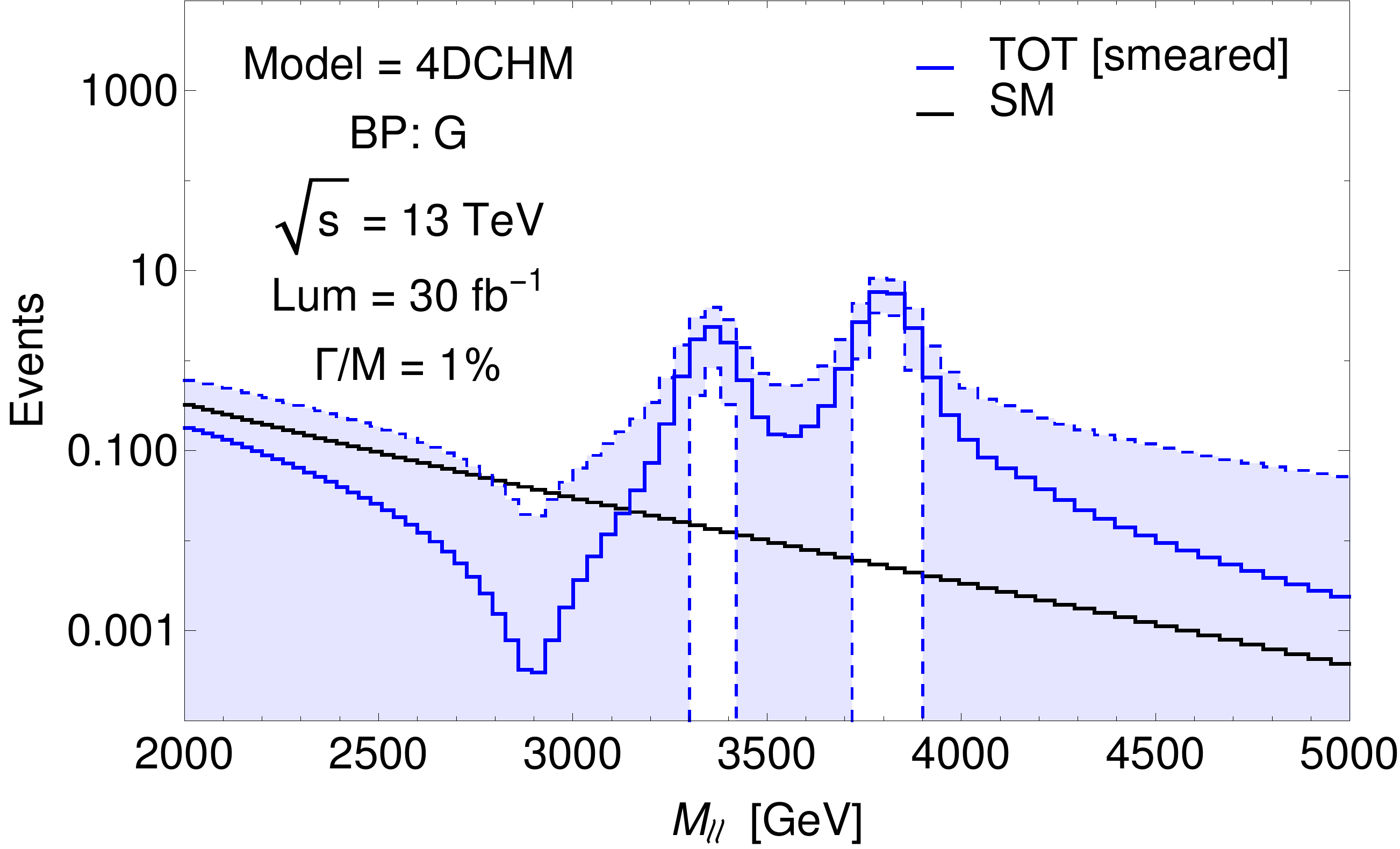}
\label{}
}
\caption{(a) Differential cross section in the dilepton invariant mass for the H benchmark point in Tab. \ref{tab:benchmarks}. We consider the LHC RunII at 13 TeV. (b) The same distribution after the smearing due to the finite detector resolution. The width of the Gaussian is fixed at $w$ = 25 GeV. We include the statistical error,  visualised by the blue bands, evaluated for an integrated luminosity of 30 fb$^{-1}$. (c) same as (a) for benchmark F. (d) Same as (b) for benchmark F with $w$ = 26 GeV. (e) same as (a) for benchmark G. (f) Same as (b) for benchmark G with $w$ = 38 GeV.}
\label{fig:smearFGH}
\end{figure}

In this section, we illustrate the 4DCHM multi-$Z'$ boson phenomenology at the ongoing LHC RunII with 13 TeV. In order to analyse the double resonant production of the new $Z^\prime_{2,3}$ bosons, a key variable is the distance between the two resonances. We start from the benchmark point H of Tab. \ref{tab:benchmarks},  representing the situation in which the two resonances are (almost) degenerate, quite  like in the NUED models. Fig.~\ref{fig:smearFGH}a displays the corresponding dilepton invariant mass spectrum for the ratio  $\Gamma_{Z^\prime_{2,3}}/M_{Z^\prime_{2,3}} = 1\%$. Here, the two $Z'$s are separated by a distance $d\simeq 0.4\% M_{Z^\prime_{2,3}}$ and, clearly, it is not possible to disentangle the two peaks in the differential cross section, even if quite narrow, because the difference between the two resonant masses is much smaller than the natural width. Furthermore, also the separation between the dip and the two degenerate peaks is of the same order. The peaking structure of the dilepton invariant mass spectrum can be therefore quite compressed. This is a distinctive feature of the 4DCHM as compared to the multi-resonant NUED model where the dip is far apart. This characteristics poses an even greater challenge insofar that the dilepton 
mass resulution which intervenes in sampling the mass spectrum may actually also include the negative dip, thereby blurring what sensible assumptions should be made in order to carry out an adequate 
statistical analysis. Even an integration around what would appear as a single peak might indeed paradoxically not produce any difference with respect to the SM background expectation, if accidentally one comes to capture also the dip, owing to experimental limitations in the mass resolution of the dilepton pairs. The resonance(s) will then appear totally invisible. (Obviously, this peculiar behavior is not contemplated at all in the NWA and FWA prescription.)

\begin{figure}[t]
\centering
\subfigure[]{
\includegraphics[width=0.45\textwidth]{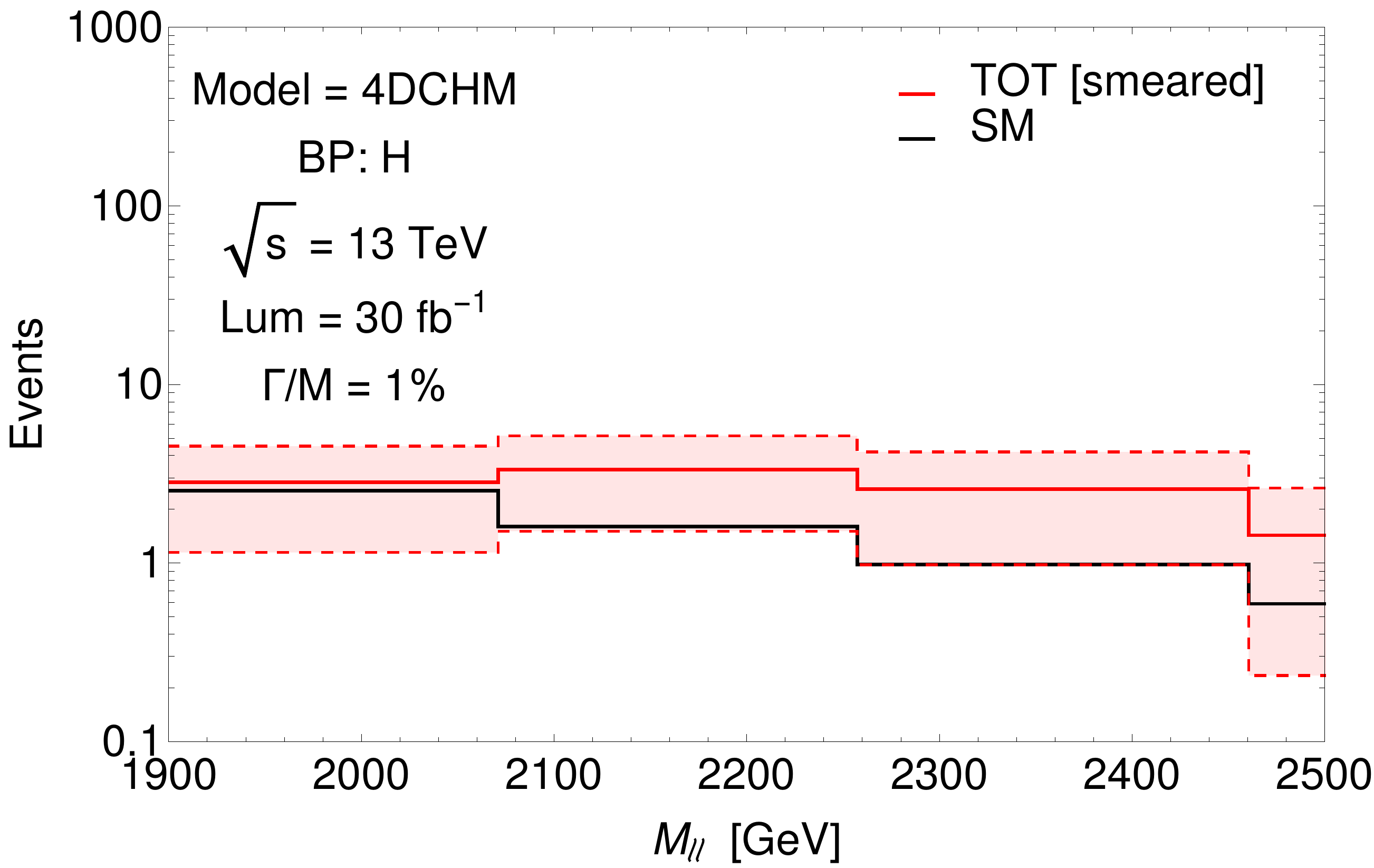}
\label{}
}
\subfigure[]{
\includegraphics[width=0.45\textwidth]{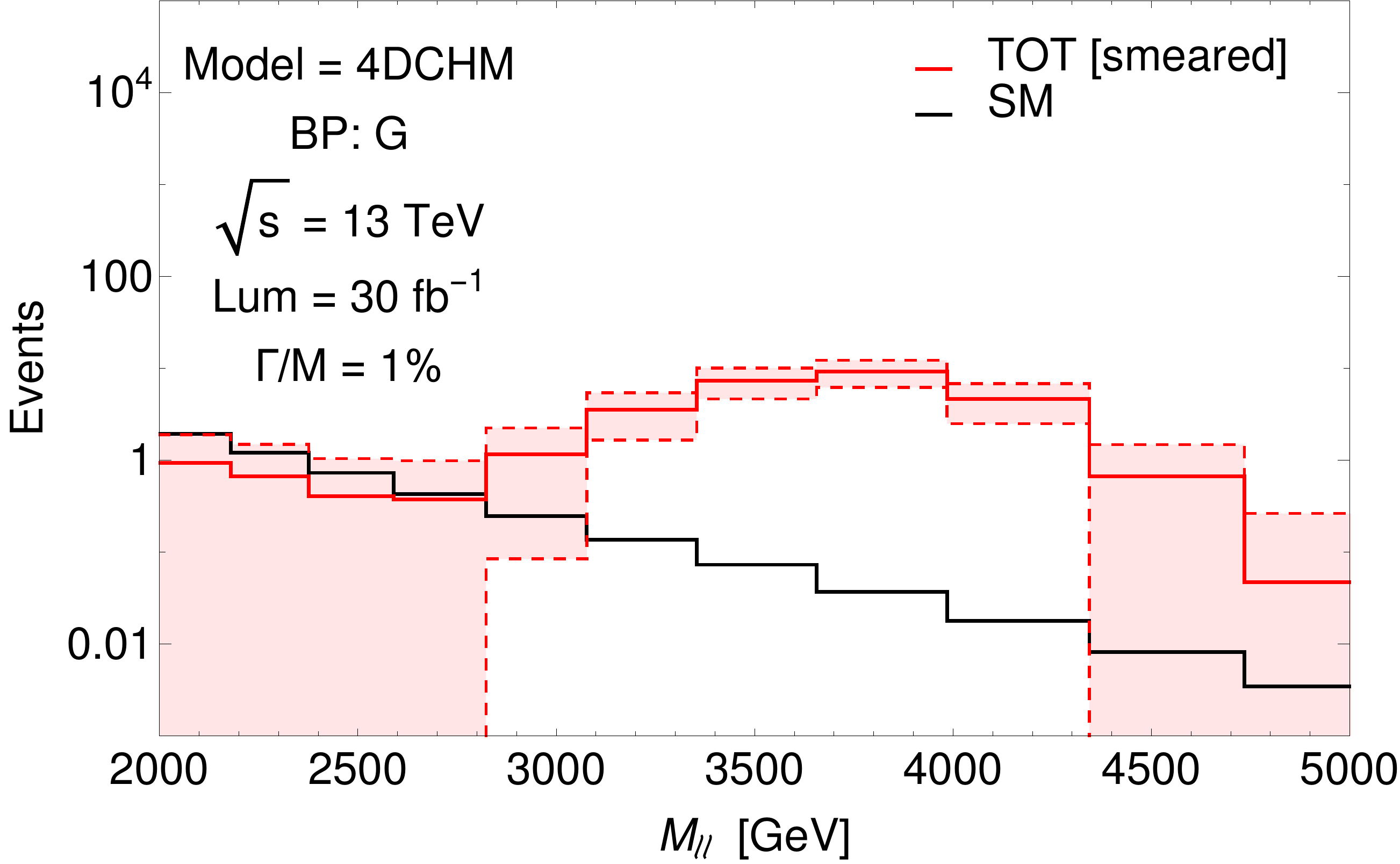}
\label{}
}
\caption{(a) Differential cross section in the dimuon invariant mass after the smearing due to detector resolution for the benchmark point H in Tab. \ref{tab:benchmarks} at the LHC RunII at 13 TeV energy. The width of the Gaussian is fixed at $w$ = 191 GeV.  We include the statistical error,  visualised by the blue bands, evaluated for an integrated luminosity of 30 fb$^{-1}$. (b) same as (a) for benchmark G. In this case, the width of the Gaussian is fixed at $w$ = 283 GeV.}
\label{fig:smearGH_muon}
\end{figure}

The quasi-natural degeneracy of the two $Z'$ bosons, discussed above, happens in a part of the parameter space characterised by large values of $g_\rho$ and small values of $f$ (top of Fig.~\ref{fig:Lum_signal_8TeV}). However, owing to the Gaussian smearing, also configurations in which the two resonances are separated by a distance bigger than the natural width but comparable to the dilepton mass resolution can actually appear as a single "bump". This is actually the most common scenario we can find in the 4DCHM. To illustrate this effect, it is instructive to re-create here a more realistic setup. To render the merging or otherwise of the two nearby $Z^\prime_2$ and $Z^\prime_3$ peaks quantitatively manifest, we have modeled the finite resolution of the detector by convoluting the signal with a Gaussian distribution chosen to reproduce the experimental environment. The width of the Gaussian shape thus has been fixed according to the CMS detector resolution for electron pairs, which is rougly 1.2\% of the dielectron invariant mass and is almost constant with the mass scale. In doing this exercise, we would like to see whether the smearing can change the multi-$Z'$ resonant structure qualitatively and, at the same time, whether the dip before such a peaking structure could be washed out or not. We thus take as effective mass to compute the Gaussian width the crossing point where the differential cross section in the dielectron invariant mass intersects the SM background expectation, i.e., after the dip and before the peak(s). The result is not very sensitive to the precise choice of the mass scale, though.  The effect of the smearing on the (quasi) degenerate scenario represented by the benchmark point H is displayed in Fig.~\ref{fig:smearFGH}b. There, we have included also the statistical error represented by the blue band. During the low luminosity run, the dip will not be statistically significant. However, the "bump" could be detected. We then select the benchmark point F in Tab. \ref{tab:benchmarks}, fixing the $Z'_{2,3}$ width to be $\Gamma_{Z^\prime_{2,3}}/M_{Z^\prime_{2,3}} = 1\%$. Oppositely to the case shown in Fig. \ref{fig3}(b), where we have the same benchmark F but with $\Gamma_{Z^\prime_{2,3}}/M_{Z^\prime_{2,3}} = 5\%$, now the two resonances are a priori clearly visible as displayed in Fig. \ref{fig:smearFGH}c. The distance between the two peaks, $d\simeq 75$ GeV, is in fact bigger than the natural width.

However, when we apply the smearing, the double resonant peaking structure of the signal is washed out, as shown in Fig. \ref{fig:smearFGH}(d). In a realistic setup, we are thus brought back to the single resonant case, effectively. This circumstance happens for all the points in the parameter space where the distance between the two peaks is smaller than about three times the Gaussian width. The parameter space of the model is large enough to find distribution profiles where the detector smearing is not sufficient to wash away the double resonant structure. This happens especially for points characterised by large $f$ and small $g_\rho$ values. An example is given in Figs. \ref{fig:smearFGH}e and f which correspond to the benchmark G in Tab. \ref{tab:benchmarks}. What is remarkable though, for both benchmarks F and G, is that the dip is substantially unaffected by the detector smearing, no matter whether the $Z^\prime_2$ and $Z^\prime_3$ peaks are resolved or otherwise. Up to now, we have applied the smearing to the dielectron channel, whose mass resolution is $R_e\simeq 0.012 M_{ee}$. The resolution is a key ingredient in detecting a 4DCHM signal, expecially because the peaking structure can be quite complicated and compressed. As already mentioned the muon channel is characterised by a resolution which is roughly 8 times the electron one at large mass scales: $R_\mu\simeq 0.09 M_{\mu\mu}$ for invariant masses above 2 TeV. Oppositely to the electron channel, where $R_e$ is almost constant with the mass range, the resolution $R_\mu$ increases with $M_{\mu\mu}$. For the considered spectrum, the situation then drastically changes compared to the electron case. 

In Fig.~\ref{fig:smearGH_muon}, we plot the result of the smearing for the muon channel. We consider the benchmark points H and G. As one can see, owing to the larger resolution, the signal for benchmark H is completely washed out. The wider resolution merges in fact dip and peak, avaraging over them.
The global number of events thus lays, evenly spread over the SM background, with no defined shape. As the depletion of events in the dip region, compared to the SM background, compensates for the excess of events under the resonant peak, the net result is not statistically distinguishable from a fluctuation of SM background. For the benchmark point G, in Fig.~\ref{fig:smearGH_muon}b, the muon channel cannot disantangle the double resonant structure. The signal would appear as an effective single broad "bump". Again, this is due to the worse resolution in the invariant mass of the muon pairs, as compared to the electron ones. The final message here is that for characterizing the 4DCHM signal shape, the muon channel is not efficient as it does not allow to resolve resonances and dips adequately. Also, in the interpretation of the data analysis results, a word of caution should be spent. If a signal is observed in the electron channel and has no counterpart in the muon channel during a run, this shouldn't be interpreted as a family non-universality but rather the hint that a complicated peaking structure of the observed signal is expected. This will be revealed in a successive run at higher luminosity, where the signal shape could be more likely fully reconstructed.   

\begin{figure}[t]
\centering
\subfigure[]{
\includegraphics[width=0.45\textwidth]{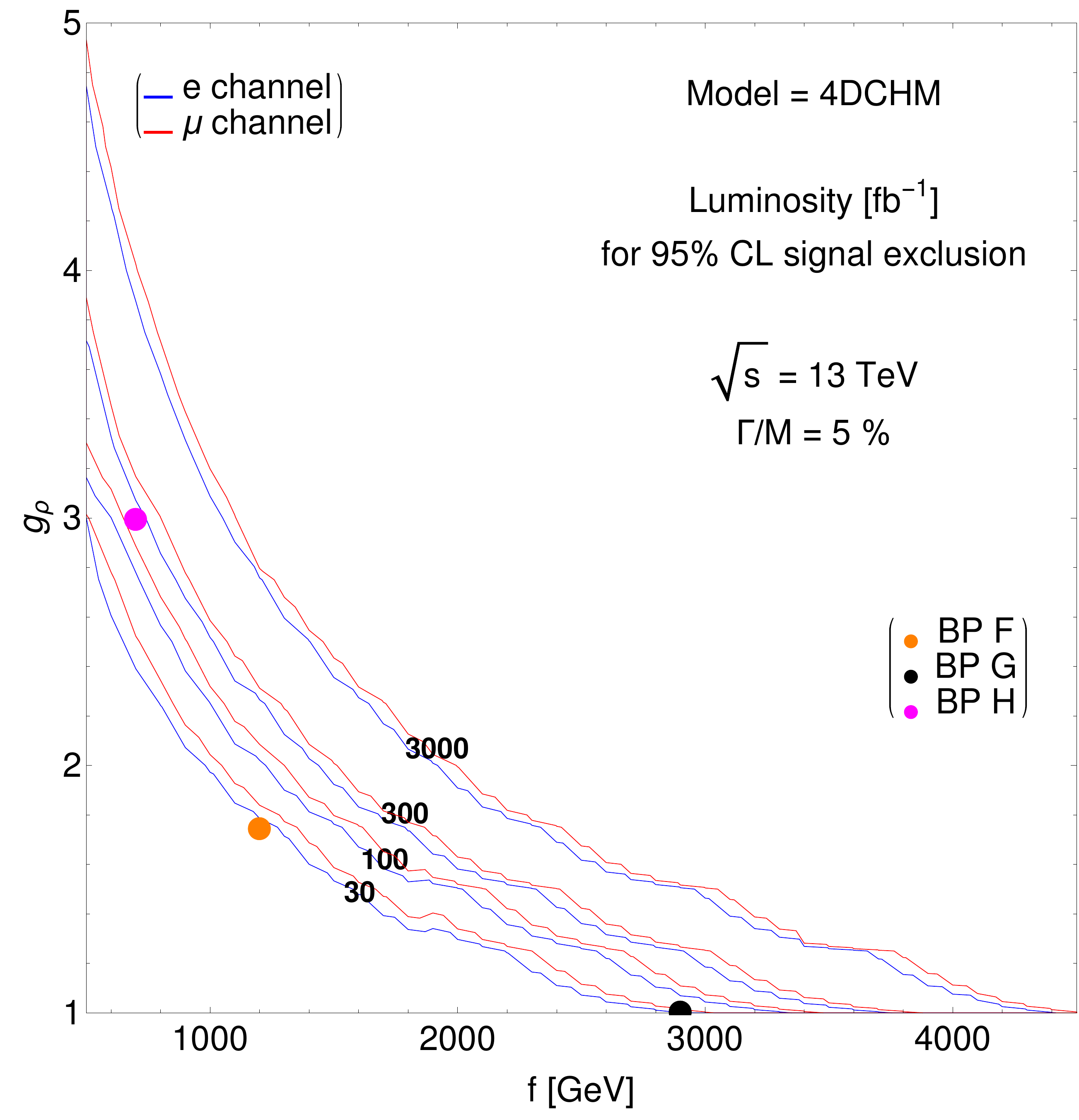}
\label{}
}
\caption{Projected 95\% CL exclusion limits at the LHC RunII with 13 TeV energy for different values of the integrated luminosity. The blue(red) contours refer to the electron(muon) channel. The dots represent the three benchmark points in Tab. \ref{tab:benchmarks}.}
\label{fig:limits_13TeV}
\end{figure}

In both Figs.~\ref{fig:smearFGH} and \ref{fig:smearGH_muon}, we have shown the statistical error expected at the ongoing LHC RunII with 13 TeV energy. The statistical analysis shows that, in the next couple of years when the collected luminosity will be ${\cal{L}}\simeq 30~fb^{-1}$, the LHC would acquire sensitivity to all these benchmarks. In order to have a complete projection of discovery or exclusion potential at the ongoing LHC RunII with 13 TeV energy, in Fig. \ref{fig:limits_13TeV} we  show the exclusion limits as contour plots in the parameter space defined by ($g_\rho , f$) for different values of the integrated luminosity, ranging from ${\cal{L}}\simeq 30~fb^{-1}$ (luminosity expected in the next two years) to ${\cal{L}}\simeq 3000~fb^{-1}$ (projected luminosity for a future run). We have assumed the same acceptance-times-efficiency factor for electrons and muons as for the past LHC RunI at 7, 8 TeV. We have moreover implemented the mass scale dependent NNLO QCD corrections. 
These are of course only theoretical projections. When data will become more copious, the experimental collaborations will perform the $Z'$ boson search in the leptonic DY channel. As already mentioned, within ED theories, mass limits on the KK states can directly  be extracted from the default 95\% CL upper bound on the BSM cross section derived from the experimental measurements. On the contrary, CHMs need an experimental analysis based on a modified approach. The present setup is indeed designed for single (or effectively single) $Z'$s. It would not be efficient in the limit setting procedure in presence of multi-$Z'$s.
The key variable is the distance between the two expected peaks. If the distance is bigger than the invariant mass range selected to normalize the SM background to the data, the default procedure could be applied twice and the results of the two likelihood fits could be combined. If the two resonances are rather separated but both lie within this mass interval, the standard likelihood function would interpret one of them as SM background, thus biasing the fitting procedure. A modified signal shape could then be inserted in the likelihood function in order to optimize the search for multi-$Z'$-bosons. A novel and dedicated analysis is advisable for general CHMs. Moreover, the dip before the peaks could become detectable as a (negative) deviation from the SM predictions for points in the parameter space similar to the F and G configurations shown in Fig. \ref{fig:smearFGH}. Again, an adequate statistical analysis would be necessary in order to classify this depletion of events happening before the "bump(s)" as evidence of a (rather complicated) signal, as opposed to a (downward) background fluctuation. This is the topic discussed in the next sub-section.

\subsubsection{Multi-$Z'$ 4DCHM: profiling the new resonances}\label{sec:4DCHM-profile} 

\begin{figure}[t]
\centering
\subfigure[]{
\includegraphics[width=0.47\textwidth]{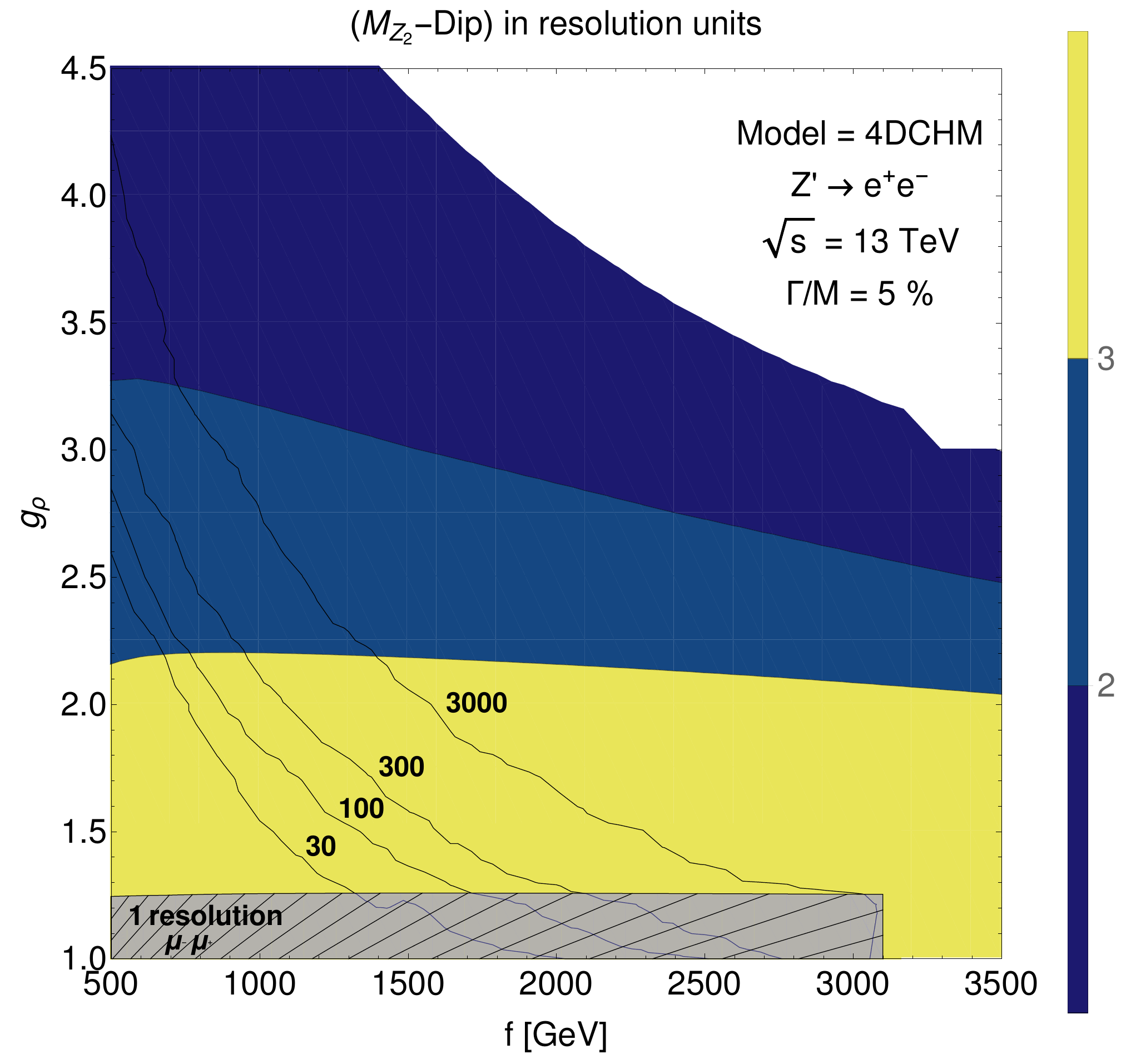}
\label{}
}
\caption{Separation between dip and first resonant peak of the 4DCHM in mass resolution units for the electron channel. From top to bottom, the colored areas represent regions in the parameter space with increasing separation. We assume a resonance width over mass ratio of 5\%. The labelled black contours indicate the luminosity needed to discard the SM background hypothesis at the 95\%~CL.
The light grey shaded area in the bottom represents the region where we have an unitary resolution separation between the dip and the $Z^\prime_2$ peak in the muon channel.}
\label{fig:separation}
\end{figure}

In the lucky event of a discovery in the usual "bump" hunt, the next question to be addressed would be the theoretical interpretation of the found resonance(s). In a successive run at higher integrated luminosities, one should then exploit all features of the observed events in the attempt to reach as a complete as possible reconstruction of the signal shape. A striking characteristic of multi-resonant models is the appearence of a sizeable depletion of events, compared to the SM background expectation, in the invariant mass region before the peaking structure. As done for NUED models in Sect. \ref{subsec:searches_ED}, also within the 4DCHM we should thus figure out in which region of the parameter space it is possible to successfully implement an optimised strategy, able to properly account for both the excess and depletion of events in order to reveal the presence of resonances likely induced by a CHM.

For instance, we could define the following variable:
\begin{equation}
\epsilon = \frac{M_{Z'_2}-M_{\rm dip}}{M_{Z'_2}},
\end{equation}
where $M_{\rm dip}$ is the value of the dilepton invariant mass corresponding to the minimum of the dip. The variable $\epsilon$ would quantify the relative distance between the (degenerate or otherwise) peaks, exemplified by the position of $M_{Z'_2}$, and the dip (or inverse peak). The role of $\epsilon$ is to discriminate a depletion from an excess of events. Should this variable be smaller than the detector resolution, we would never be able to disentangle the negative contribution of the interference from the excess on the peak(s). Under these circumstances, no experimental measurement would be able to underpin the CHM nature of the discovered resonance(s).
In Fig.~\ref{fig:separation}, we show the contour plot representing the condition $\epsilon = c\times R_e$, with $R_e$ the dielectron invariant mass resolution and $c$ a coefficient whose value can be read from the color legend on the righthand side of the plot. The colored parts of the parameter space represent the regions where $\epsilon\ge c\times R_e$. The entire colored region collapses to a narrow stripe, sitting at very low values of $g_\rho$, for the muon channel as the dimuon invariant mass resolution is much larger than the electron one.

The electron DY channel is thus particularly useful for profiling the discovered resonance(s), oppositely to the muon channel which is favoured for the actual search.
For values of the CHM free parameters where $\epsilon\ge R_e$, there is indeed the possibility of observing the peak(s) at some large dielectron invariant mass, simultaneously accompanied by a depletion of the SM background events expected at lower invariant mass values. This depletion should not be interpreted as a statistical fluctuation, nor as a negative correction (e.g., induced by large Sudakov logarithms ensuing from EW loop effects), rather it should be taken as an additional signal manifestation that a suitable statistical analysis would aim at extracting as such. 
\par\noindent
Finally, following again the suggestion given in Ref. \cite{Accomando:2015cfa}, we have explored the possiblity of using the forward-backward asymmetry to profile the resonances within this Composite Higgs scenario. Unfortunately, as for the NUED(s) model, the statistical significance of the AFB distribution in the dilepton invariant mass is subdominant with respect to the invariant mass peak evidence in all the explored parameter space of the model.

\section{Conclusions}\label{sec:summa}

In this paper, we have analysed the phenomenology of the Non-Universal ED and 4D Composite Higgs Models which are representative of two generic classes of multi-$Z^\prime$ scenarios, weakly and strongly interacting respectively. We have examined the consequences of both FW and interference effects on the signal shape and rate and, eventually, on the interpretation of the data analysis results, as such effects are not generally included in the time-honoured NWA approach adopted by the experimental collaborations in searching for narrow spin-1 resonances (thought some progress on this side has recently occurred).

Both contributions manifest themselves through a peculiar interplay which generically produces a large dip (almost an inverted peak, signicantly deeper than those seen in the case of narrow single $Z'$-boson scenarios) in the dilepton invariant mass spectrum that precedes the appearance of either a single degenerate peak (always for the NUED scenario and over certain combinations of the 4DCHM parameters) or a double resonant peaking structure (in the complementary parts of the 4DCHM parameter space). In the NUED case, such a dip appears in a mass region which is always resolvable from the peak one, for standard detector resolutions in the dilepton invariant mass. In contrast, for the 4DCHM, also the opposite situation can occur, when the dip and the peak interplay over the mass interval where a possible signal is sampled. 

Current statistical approaches implemented by the LHC collaborations do not allow one to model the signal as a composition of a dipping and peaking structure, so we concentrated  on describing the phenomenology emerging from treating the dip and the peak(s) separately. As the latter normally emerge(s) before the former as luminosity accrues, we have used the kinematic features related to the dip region as a characterising element of a possible discovery following
the extraction of the peak(s), with a twofold purpose. On the one hand, when the multiple $Z'$ peaks (two generally, in fact) merge into one (which can happen in both the NUED and 4DCHM scenarios), to make evident that the underlying BSM structure is not the standard single-$Z'$ one. On the other hand, when the two peaks are separable (as it can happen in the 4DCHM), to help one profiling the multiple $Z'$-boson signal in terms of masses, widths and quantum numbers of the new discovered resonances.

Regarding the multi-resonant peaking structure expected in both NUED and 4DCHM scenarios, our findings are the following. The two first level KK-modes predicted by the NUED model, $\gamma_{KK}^{(1)}$ and $Z_{KK}^{(1)}$, can be effectively regarded as any other single-$Z'$-boson, being (almost) degenerate and well separated from the dip. As for narrow $Z'$ models, the cross section is not affected by model-dependent FW and interference effects up to $O(10\% )$ accuracy if the optimal integration interval in the dilepton invariant mass proposed in Ref. \cite{Accomando:2013sfa} is applied. Within this setup, mass limits can thus be derived from the experimental 95\% C. L. upper bound on the BSM cross section, directly and unambiguously. In contrast, the 4DCHM might behave quite differently from the models considered in current analyses. If the predicted new resonances that are active, $Z'_2$ and 
$Z'_3$, are narrow ($\Gamma_{Z'_2,Z'_3}/M_{Z'_2,Z'_3}\le 3\%$) and well separated, the results of the  experimental data analysis, as currently performed, can be interpreted within the 4DCHM in the same consistent way as for NUED models. However, this situation happens in a limited part of the 4DCHM parameter space. The variety of possible shapes for the predicted signal that one can get by scanning over the full parameter space suggests a change in the standard search (or limit setting) procedure. The dilepton invariant mass spectrum can be rather complex, owing to different factors. First of all, the type of new resonances can range from very narrow to wide. Secondly, the multiple resonant peaks can be either (almost) degenerate or well separated depending on different factors: the specific point in the parameter space, the magnitude of the width(s) compared to the distance between the different peaks and the value of the dilepton mass resolution at that given mass scale. Finally, the dip that is caused by the interference between the new resonance(s) and the SM photon and $Z$-boson could appear in the close proximity of the first resonance, thus affecting the expected Breit-Wigner shape of the signal. So, often, the peaking structure of the spectrum can be very compressed, rendering the analysis challenging because of the detector resolution and other factors.

From the experimental point of view, these features can have in impact on the fitting procedure and on the way the results of the data analysis are interpreted within general CHMs. The presence of a dip in the close proximity of the first resonance could affect the normalization of the SM background, which is presently done in a selected invariant mass window around the hypothetical pole mass of the new vector boson(s). This would suggest to shift the SM background normalization region away from the probed $Z'$-boson mass(es). Coming now specifically to the $Z'$-boson(s) signal, in the case of a rather wide $Z'$-boson(s) characterized by a width larger than (or comparable to) the dilepton invariant mass resolution, the possible distorsion of its line-shape due to interference effects could affect the experimental fitting procedure in presence of data points. This would suggest the implementation of a modified experimental approach even when the spectrum is (almost) degenerate but wide. A multi-resonant spectrum might have even more sophisticated consequences on the experimental procedures and the bridging between data analysis and theoretical interpretation. When modelling a functional form for a doubly-resonant spectrum the distance between two resonances becomes a key variable. If the separation between two peaks is bigger than the optimised window selected for the SM background normalization, then the standard likelihood fit could still work. But, if the distance is smaller, one of the two resonances would be interpreted as SM background thus affecting the fitting procedure. This circumstance would suggest the implementation of a new signal line-shape, characterized by a double resonant peaking structure, in the likelihood function to optimize the multi-$Z'$ boson search (or limit setting) procedure. This would require the introduction of an additional variable in the likelihood fit, the distance between the peaks, that would in principle make its convergence to slow down. The extra variable could be however bounded by imposing a relation between such a variable and the number of events collected between the two peaks. The two points discussed up to now are of relevance for the direct $Z'$-boson(s) search (or limit setting) procedure. 

In the fortunate circumstance that an excess of events has been measured in the dilepton spectrum at some high invariant mass, the next point to be addressed would be profiling the new resonance(s) in the attempt to track down the underlying BSM theory. At this stage, an important role would be played by the dilepton invariant mass resolution. Concerning this issue, our findings are the following. In the dielectron channel, it would be likely that the multi-resonant structure could be detected, as the resolution is about $R_e\simeq 1.2\%$ and is almost constant over the mass scale. We have in fact shown that the distance between different peaks and between peaks and dip is larger than $R_e$ over a substantial part of the parameter space. In the dimuon channel, such a complicated structure would be completely obscured by the much worse dimuon invariant mass resolution ($R_\mu\simeq 9\%$ for $M_{\mu\mu}\ge$ 3 TeV). Summarizing, in the particular instance of this type of signal, both electrons and muons are useful for direct $Z'$-boson searches but, in profiling the resonance(s), the dielectron channel would be better being characterized by a much smaller mass resolution.

The size of the invariant mass resolution poses a further challenge. In case of a compressed spectrum, where dipping and peaking structures are at very short distances, the mass resolution might wash out the signal completely. Instead of a narrow resonance standing over a smooth background, the signal would appear as an excess of events evenly spread over the SM background being the result of an average of dip and peak(s). Such an excess could be reabsorbed within the SM background statistical fluctuation, thus excaping any observations. Therefore, for the parameter sets for which a signal cannot be established, FW and interference effects amongst multiple $Z'$-bosons can heavily affect the limits that can be extracted within the 4DCHM. These generally impinge on such an extraction in a way that leads one to believe that no positive (peak) signal is present, when in reality the signal is there but vanishes when combined with the negative effects (dip) due to interferences enabled by the FWs of the intervening $Z'$s. As a consequence, for some sets of masses and couplings, one could even observe a signal in the dielectron channel and not in the dimuon one. This should not be interpreted as a manifest family non-universality in the leptonic sector, but rather the hint that a complicated peaking/dipping structure is encoded in the signal observed in the dieletron invariant mass spectrum.

Finally, while carrying out the above analyses, we have tested the yield of our approach against that of alternative procedures existing in literature, normally dealing with single-$Z'$ scenarios, confirming that the latter are unapplicable to multi-$Z'$ ones. In fact, also the magic cut fails to enable one to carry out model-independent searches when multiple $Z'$ states are present, especially if their couplings to leptons are suppressed with respect to the SM interaction strength, so that model-dedicated analyses may be in order.

The models we have proposed here also present a rich phenomenology in the charged current sector. The NUED predicts indeed a tower of excitations of the SM $W$-boson, while the 4DCHM predicts three different $W^\prime$s. In this context, it would be recommended to include the charged sector in the analysis and possibly to combine this information with what we get from the neutral interaction, following the path proposed in Ref. ~\cite{deBlas:2012qp}.
\appendix
\section{Explicit expressions of the $\rho$ couplings to light fermions}
\label{app}

The couplings of the neutral gauge bosons to the light SM fermions can be expressed by the following Lagrangian
\begin{equation}
{\mathcal L}\supset\sum_f\big[ e\bar\psi^f \gamma_\mu Q^f \psi^f A^\mu+ \sum_{i=0}^5   (\bar\psi^f_L  g_{Z^\prime_i}^L(f) \gamma_\mu  \psi^f_L+\bar\psi^f_R  g_{Z^\prime_i}^R(f) \gamma_\mu  \psi^f_R ) Z^{\prime\mu}_i \big]
\label{eq:neutral-gauge-coupl-light}
\end{equation}
where $\psi_{L,R}=[(1\pm\gamma_5)/2]\psi$ and where $Z^\prime_0$ and $A$ corresponds to the neutral SM gauge bosons $Z$ and $\gamma$.
The photon field is coupled to the electromagnetic current in the standard way with the electric charge which is defined as
\begin{equation}
e=\frac{g_Lg_Y}{\sqrt{g_L^2+g_Y^2}}, \qquad g_L=g_0 c_\theta, \qquad g_Y = g_{0Y} c_\psi, \qquad tg_\theta=\frac{g_0}{g_\rho}, \qquad tg_\psi=\frac{\sqrt{2}g_{0Y}}{g_\rho}.
\label{eq:electric-charge}
\end{equation}
The $g^{L,R}_{Z^\prime_i}$ couplings have the following expression
\begin{equation}
g_{Z^\prime_i}^L(f)= A_{Z^\prime_i}T^3_L(f)+ B_{Z^\prime_i} Q^f, \quad\quad
g_{Z^\prime_i}^R(f)=  B_{Z^\prime_i}Q^f,
\end{equation}

where, at the leading order in the expansion parameter $\xi=v^2/f^2$, $A_{Z^\prime_i}$ and $B_{Z^\prime_i}$ read
\begin{equation}
\begin{alignedat}{2}
& A_{Z}\simeq \frac{e}{s_\omega c_\omega}\big[1+(c^2_\omega a_Z+ s^2_\omega b_Z)\xi\big], \qquad
&& B_{Z}\simeq -e\frac{s_\omega}{c_\omega}(1+b_Z\xi), \\
%%%
& A_{Z^\prime_1}=0, \qquad && B_{Z^\prime_1}=0,\\
%%%
& A_{Z^\prime_2}\simeq  -\frac{e}{c_\omega} \frac{s_\psi}{c_\psi} \Big[1+(\frac{c_\omega}{s_\omega} a_{Z^\prime_2}-b_{Z^\prime_2})\xi\Big], 
\qquad &&  B^\prime_{Z^\prime_2}\simeq \frac{e}{c_\omega} \frac{s_\psi}{c_\psi} \Big[1-b_{Z^\prime_2}\xi\Big],\\
%%%
& A_{Z^\prime_3}\simeq  -\frac{e}{s_\omega}\frac{s_\theta}{c_\theta}\big[1+(a_{Z^\prime_3}+\frac{s_\omega}{c_\omega} b_{Z^\prime_3})\xi\big]   
,\qquad &&  B_{Z^\prime_3}\simeq   \frac{e}{c_\omega} \frac{s_\theta}{c_\theta} b_{Z^\prime_3}\xi,\\
%%%
& A_{Z^\prime_4}=0, \qquad && B_{Z^\prime_4}=0,\\
%%%
& A_{Z^\prime_5}\simeq  e(\frac{1}{s_\omega} a_{Z^\prime_5}-\frac{1}{c_\omega} b_{Z^\prime_5})\sqrt{\xi},  
\qquad && B_{Z^\prime_5}\simeq  \frac{e}{c_\omega} b_{Z^\prime_5}\sqrt{\xi},
\label{eq:coup-neu-me}
\end{alignedat}
\end{equation}
with
\begin{equation}
\tan{\omega}=\frac{g_Y}{g_L}, \qquad e=g_L s_\omega = g_Y c_\omega, \qquad \frac{e}{s_\omega c_\omega}=\sqrt{g_L^2+g_Y^2},
\end{equation}
and
\begin{equation}
\begin{alignedat}{2}
& a_{Z}=  (2 s_\theta^2+s_\psi^2)(4 cˆ_\theta^2-1)/32, \qquad 
&& b_{Z}=  (2 s_\theta^2+s_\psi^2)(4 cˆ_\psi^2-1)/32, \\
& a_{Z^\prime_2}= \frac{\sqrt{2} s_\theta s_\psi c_\psi^6}{4(c_\psi^2-c_\theta^2)(2 c_\psi^2-1)}, 
\qquad && b_{Z^\prime_2}= \frac{c_\psi^4(2-7c_\psi^2+9 c_\psi^4-4 c_\psi^6)}{8s_\psi^2 (1-2 c_\psi^2)^2},\\
& a_{Z^\prime_3}=   \frac{-2 c_\theta^4+5 c_\theta^6-4 c_\theta^8}{4(1-2 c_\theta^2)^2},
\qquad && b_{Z^\prime_3}=  \frac{\sqrt{2} s_\theta s_\psi c_\theta^6}{4 (2 c_\theta^2-1)(c_\theta^2-c_\psi^2) },\\
& a_{Z^\prime_5}=   \frac{s_\theta}{2 \sqrt{2}(1-2 c_\theta^2)},
\qquad && b_{Z^\prime_5}= - \frac{s_\psi}{4(1-2 c_\psi^2)}.
\end{alignedat}
\end{equation}
%\clearpage
\section*{Acknowledgements}
\noindent
JF thanks the Galileo Galilei Institute for Theoretical Physics for the hospitality and support during the GGI lectures on the theory of fundamental interactions where the idea of this work started.  EA, JF and SM are financed in part through the NExT Institute.

\bibliographystyle{apsrev4-1}
\bibliography{bib}

\begin{thebibliography}{44}%
\makeatletter
\providecommand \@ifxundefined [1]{%
 \@ifx{#1\undefined}
}%
\providecommand \@ifnum [1]{%
 \ifnum #1\expandafter \@firstoftwo
 \else \expandafter \@secondoftwo
 \fi
}%
\providecommand \@ifx [1]{%
 \ifx #1\expandafter \@firstoftwo
 \else \expandafter \@secondoftwo
 \fi
}%
\providecommand \natexlab [1]{#1}%
\providecommand \enquote  [1]{``#1''}%
\providecommand \bibnamefont  [1]{#1}%
\providecommand \bibfnamefont [1]{#1}%
\providecommand \citenamefont [1]{#1}%
\providecommand \href@noop [0]{\@secondoftwo}%
\providecommand \href [0]{\begingroup \@sanitize@url \@href}%
\providecommand \@href[1]{\@@startlink{#1}\@@href}%
\providecommand \@@href[1]{\endgroup#1\@@endlink}%
\providecommand \@sanitize@url [0]{\catcode `\\12\catcode `\$12\catcode
  `\&12\catcode `\#12\catcode `\^12\catcode `\_12\catcode `\%12\relax}%
\providecommand \@@startlink[1]{}%
\providecommand \@@endlink[0]{}%
\providecommand \url  [0]{\begingroup\@sanitize@url \@url }%
\providecommand \@url [1]{\endgroup\@href {#1}{\urlprefix }}%
\providecommand \urlprefix  [0]{URL }%
\providecommand \Eprint [0]{\href }%
\providecommand \doibase [0]{http://dx.doi.org/}%
\providecommand \selectlanguage [0]{\@gobble}%
\providecommand \bibinfo  [0]{\@secondoftwo}%
\providecommand \bibfield  [0]{\@secondoftwo}%
\providecommand \translation [1]{[#1]}%
\providecommand \BibitemOpen [0]{}%
\providecommand \bibitemStop [0]{}%
\providecommand \bibitemNoStop [0]{.\EOS\space}%
\providecommand \EOS [0]{\spacefactor3000\relax}%
\providecommand \BibitemShut  [1]{\csname bibitem#1\endcsname}%
\let\auto@bib@innerbib\@empty
%</preamble>
\bibitem [{\citenamefont {Accomando}\ \emph {et~al.}(2011)\citenamefont
  {Accomando}, \citenamefont {Belyaev}, \citenamefont {Fedeli}, \citenamefont
  {King},\ and\ \citenamefont {Shepherd-Themistocleous}}]{Accomando:2010fz}%
  \BibitemOpen
  \bibfield  {author} {\bibinfo {author} {\bibfnamefont {E.}~\bibnamefont
  {Accomando}}, \bibinfo {author} {\bibfnamefont {A.}~\bibnamefont {Belyaev}},
  \bibinfo {author} {\bibfnamefont {L.}~\bibnamefont {Fedeli}}, \bibinfo
  {author} {\bibfnamefont {S.~F.}\ \bibnamefont {King}}, \ and\ \bibinfo
  {author} {\bibfnamefont {C.}~\bibnamefont {Shepherd-Themistocleous}},\ }\href
  {\doibase 10.1103/PhysRevD.83.075012} {\bibfield  {journal} {\bibinfo
  {journal} {Phys.Rev.}\ }\textbf {\bibinfo {volume} {D83}},\ \bibinfo {pages}
  {075012} (\bibinfo {year} {2011})},\ \Eprint {http://arxiv.org/abs/1010.6058}
  {arXiv:1010.6058 [hep-ph]} \BibitemShut {NoStop}%
\bibitem [{\citenamefont {Langacker}(2008)}]{Langacker:2008yv}%
  \BibitemOpen
  \bibfield  {author} {\bibinfo {author} {\bibfnamefont {P.}~\bibnamefont
  {Langacker}},\ }\href {\doibase 10.1103/RevModPhys.81.1199} {\bibfield
  {journal} {\bibinfo  {journal} {Rev. Mod. Phys.}\ }\textbf {\bibinfo {volume}
  {81}},\ \bibinfo {pages} {1199} (\bibinfo {year} {2008})},\ \Eprint
  {http://arxiv.org/abs/0801.1345} {arXiv:0801.1345 [hep-ph]} \BibitemShut
  {NoStop}%
%%CITATION = 0801.1345;%%
\bibitem [{\citenamefont {Antoniadis}(1990)}]{Antoniadis:1990ew}%
  \BibitemOpen
  \bibfield  {author} {\bibinfo {author} {\bibfnamefont {I.}~\bibnamefont
  {Antoniadis}},\ }\href {\doibase 10.1016/0370-2693(90)90617-F} {\bibfield
  {journal} {\bibinfo  {journal} {Phys.Lett.}\ }\textbf {\bibinfo {volume}
  {B246}},\ \bibinfo {pages} {377} (\bibinfo {year} {1990})}\BibitemShut
  {NoStop}%
%%CITATION = PHLTA,B246,377;%%
\bibitem [{\citenamefont {Antoniadis}\ and\ \citenamefont
  {Benakli}(1994)}]{Antoniadis:1993jp}%
  \BibitemOpen
  \bibfield  {author} {\bibinfo {author} {\bibfnamefont {I.}~\bibnamefont
  {Antoniadis}}\ and\ \bibinfo {author} {\bibfnamefont {K.}~\bibnamefont
  {Benakli}},\ }\href {\doibase 10.1016/0370-2693(94)91194-0} {\bibfield
  {journal} {\bibinfo  {journal} {Phys.Lett.}\ }\textbf {\bibinfo {volume}
  {B326}},\ \bibinfo {pages} {69} (\bibinfo {year} {1994})},\ \Eprint
  {http://arxiv.org/abs/hep-th/9310151} {arXiv:hep-th/9310151 [hep-th]}
  \BibitemShut {NoStop}%
%%CITATION = HEP-TH/9310151;%%
\bibitem [{\citenamefont {Antoniadis}\ \emph {et~al.}(1993)\citenamefont
  {Antoniadis}, \citenamefont {Munoz},\ and\ \citenamefont
  {Quiros}}]{Antoniadis:1992fh}%
  \BibitemOpen
  \bibfield  {author} {\bibinfo {author} {\bibfnamefont {I.}~\bibnamefont
  {Antoniadis}}, \bibinfo {author} {\bibfnamefont {C.}~\bibnamefont {Munoz}}, \
  and\ \bibinfo {author} {\bibfnamefont {M.}~\bibnamefont {Quiros}},\ }\href
  {\doibase 10.1016/0550-3213(93)90184-Q} {\bibfield  {journal} {\bibinfo
  {journal} {Nucl.Phys.}\ }\textbf {\bibinfo {volume} {B397}},\ \bibinfo
  {pages} {515} (\bibinfo {year} {1993})},\ \Eprint
  {http://arxiv.org/abs/hep-ph/9211309} {arXiv:hep-ph/9211309 [hep-ph]}
  \BibitemShut {NoStop}%
%%CITATION = HEP-PH/9211309;%%
\bibitem [{\citenamefont {Antoniadis}\ \emph {et~al.}(1994)\citenamefont
  {Antoniadis}, \citenamefont {Benakli},\ and\ \citenamefont
  {Quiros}}]{Antoniadis:1994yi}%
  \BibitemOpen
  \bibfield  {author} {\bibinfo {author} {\bibfnamefont {I.}~\bibnamefont
  {Antoniadis}}, \bibinfo {author} {\bibfnamefont {K.}~\bibnamefont {Benakli}},
  \ and\ \bibinfo {author} {\bibfnamefont {M.}~\bibnamefont {Quiros}},\ }\href
  {\doibase 10.1016/0370-2693(94)91058-8} {\bibfield  {journal} {\bibinfo
  {journal} {Phys.Lett.}\ }\textbf {\bibinfo {volume} {B331}},\ \bibinfo
  {pages} {313} (\bibinfo {year} {1994})},\ \Eprint
  {http://arxiv.org/abs/hep-ph/9403290} {arXiv:hep-ph/9403290 [hep-ph]}
  \BibitemShut {NoStop}%
%%CITATION = HEP-PH/9403290;%%
\bibitem [{\citenamefont {Benakli}(1996)}]{Benakli:1995ut}%
  \BibitemOpen
  \bibfield  {author} {\bibinfo {author} {\bibfnamefont {K.}~\bibnamefont
  {Benakli}},\ }\href {\doibase 10.1016/0370-2693(96)00902-1} {\bibfield
  {journal} {\bibinfo  {journal} {Phys.Lett.}\ }\textbf {\bibinfo {volume}
  {B386}},\ \bibinfo {pages} {106} (\bibinfo {year} {1996})},\ \Eprint
  {http://arxiv.org/abs/hep-th/9509115} {arXiv:hep-th/9509115 [hep-th]}
  \BibitemShut {NoStop}%
%%CITATION = HEP-TH/9509115;%%
\bibitem [{\citenamefont {Contino}(2011)}]{Contino:2010rs}%
  \BibitemOpen
  \bibfield  {author} {\bibinfo {author} {\bibfnamefont {R.}~\bibnamefont
  {Contino}},\ }in\ \href {\doibase 10.1142/9789814327183_0005} {\emph
  {\bibinfo {booktitle} {{Physics of the large and the small, TASI 09,
  proceedings of the Theoretical Advanced Study Institute in Elementary
  Particle Physics, Boulder, Colorado, USA, 1-26 June 2009}}}}\ (\bibinfo
  {year} {2011})\ pp.\ \bibinfo {pages} {235--306},\ \Eprint
  {http://arxiv.org/abs/1005.4269} {arXiv:1005.4269 [hep-ph]} \BibitemShut
  {NoStop}%
%%CITATION = ARXIV:1005.4269;%%
\bibitem [{\citenamefont {Belyaev}\ \emph {et~al.}(2009)\citenamefont {Belyaev}
  \emph {et~al.}}]{Belyaev:2008yj}%
  \BibitemOpen
  \bibfield  {author} {\bibinfo {author} {\bibfnamefont {A.}~\bibnamefont
  {Belyaev}} \emph {et~al.},\ }\href {\doibase 10.1103/PhysRevD.79.035006}
  {\bibfield  {journal} {\bibinfo  {journal} {Phys. Rev.}\ }\textbf {\bibinfo
  {volume} {D79}},\ \bibinfo {pages} {035006} (\bibinfo {year} {2009})},\
  \Eprint {http://arxiv.org/abs/0809.0793} {arXiv:0809.0793 [hep-ph]}
  \BibitemShut {NoStop}%
%%CITATION = 0809.0793;%%
\bibitem [{\citenamefont {Khachatryan}\ \emph {et~al.}(2014)\citenamefont
  {Khachatryan} \emph {et~al.}}]{Khachatryan:2014fba}%
  \BibitemOpen
  \bibfield  {author} {\bibinfo {author} {\bibfnamefont {V.}~\bibnamefont
  {Khachatryan}} \emph {et~al.} (\bibinfo {collaboration} {CMS
  Collaboration}),\ }\href@noop {} {\  (\bibinfo {year} {2014})},\ \Eprint
  {http://arxiv.org/abs/1412.6302} {arXiv:1412.6302 [hep-ex]} \BibitemShut
  {NoStop}%
%%CITATION = ARXIV:1412.6302;%%
\bibitem [{\citenamefont {Accomando}\ \emph
  {et~al.}(2013{\natexlab{a}})\citenamefont {Accomando}, \citenamefont
  {Becciolini}, \citenamefont {Belyaev}, \citenamefont {Moretti},\ and\
  \citenamefont {Shepherd-Themistocleous}}]{Accomando:2013sfa}%
  \BibitemOpen
  \bibfield  {author} {\bibinfo {author} {\bibfnamefont {E.}~\bibnamefont
  {Accomando}}, \bibinfo {author} {\bibfnamefont {D.}~\bibnamefont
  {Becciolini}}, \bibinfo {author} {\bibfnamefont {A.}~\bibnamefont {Belyaev}},
  \bibinfo {author} {\bibfnamefont {S.}~\bibnamefont {Moretti}}, \ and\
  \bibinfo {author} {\bibfnamefont {C.}~\bibnamefont
  {Shepherd-Themistocleous}},\ }\href {\doibase 10.1007/JHEP10(2013)153}
  {\bibfield  {journal} {\bibinfo  {journal} {JHEP}\ }\textbf {\bibinfo
  {volume} {1310}},\ \bibinfo {pages} {153} (\bibinfo {year}
  {2013}{\natexlab{a}})},\ \Eprint {http://arxiv.org/abs/1304.6700}
  {arXiv:1304.6700 [hep-ph]} \BibitemShut {NoStop}%
%%CITATION = ARXIV:1304.6700;%%
\bibitem [{\citenamefont {Dittmar}(1997)}]{Dittmar:1996my}%
  \BibitemOpen
  \bibfield  {author} {\bibinfo {author} {\bibfnamefont {M.}~\bibnamefont
  {Dittmar}},\ }\href {\doibase 10.1103/PhysRevD.55.161} {\bibfield  {journal}
  {\bibinfo  {journal} {Phys. Rev.}\ }\textbf {\bibinfo {volume} {D55}},\
  \bibinfo {pages} {161} (\bibinfo {year} {1997})},\ \Eprint
  {http://arxiv.org/abs/hep-ex/9606002} {arXiv:hep-ex/9606002} \BibitemShut
  {NoStop}%
%%CITATION = HEP-EX/9606002;%%
\bibitem [{\citenamefont {Rizzo}(2007)}]{Rizzo:2007xs}%
  \BibitemOpen
  \bibfield  {author} {\bibinfo {author} {\bibfnamefont {T.~G.}\ \bibnamefont
  {Rizzo}},\ }\href {\doibase 10.1088/1126-6708/2007/05/037} {\bibfield
  {journal} {\bibinfo  {journal} {JHEP}\ }\textbf {\bibinfo {volume} {05}},\
  \bibinfo {pages} {037} (\bibinfo {year} {2007})},\ \Eprint
  {http://arxiv.org/abs/0704.0235} {arXiv:0704.0235 [hep-ph]} \BibitemShut
  {NoStop}%
%%CITATION = ARXIV:0704.0235;%%
\bibitem [{\citenamefont {Petriello}\ and\ \citenamefont
  {Quackenbush}(2008)}]{Petriello:2008zr}%
  \BibitemOpen
  \bibfield  {author} {\bibinfo {author} {\bibfnamefont {F.}~\bibnamefont
  {Petriello}}\ and\ \bibinfo {author} {\bibfnamefont {S.}~\bibnamefont
  {Quackenbush}},\ }\href {\doibase 10.1103/PhysRevD.77.115004} {\bibfield
  {journal} {\bibinfo  {journal} {Phys. Rev.}\ }\textbf {\bibinfo {volume}
  {D77}},\ \bibinfo {pages} {115004} (\bibinfo {year} {2008})},\ \Eprint
  {http://arxiv.org/abs/0801.4389} {arXiv:0801.4389 [hep-ph]} \BibitemShut
  {NoStop}%
%%CITATION = 0801.4389;%%
\bibitem [{\citenamefont {Papaefstathiou}\ and\ \citenamefont
  {Latunde-Dada}(2009)}]{Papaefstathiou:2009sr}%
  \BibitemOpen
  \bibfield  {author} {\bibinfo {author} {\bibfnamefont {A.}~\bibnamefont
  {Papaefstathiou}}\ and\ \bibinfo {author} {\bibfnamefont {O.}~\bibnamefont
  {Latunde-Dada}},\ }\href {\doibase 10.1088/1126-6708/2009/07/044} {\bibfield
  {journal} {\bibinfo  {journal} {JHEP}\ }\textbf {\bibinfo {volume} {07}},\
  \bibinfo {pages} {044} (\bibinfo {year} {2009})},\ \Eprint
  {http://arxiv.org/abs/0901.3685} {arXiv:0901.3685 [hep-ph]} \BibitemShut
  {NoStop}%
%%CITATION = ARXIV:0901.3685;%%
\bibitem [{\citenamefont {Rizzo}(2009)}]{Rizzo:2009pu}%
  \BibitemOpen
  \bibfield  {author} {\bibinfo {author} {\bibfnamefont {T.~G.}\ \bibnamefont
  {Rizzo}},\ }\href {\doibase 10.1088/1126-6708/2009/08/082} {\bibfield
  {journal} {\bibinfo  {journal} {JHEP}\ }\textbf {\bibinfo {volume} {08}},\
  \bibinfo {pages} {082} (\bibinfo {year} {2009})},\ \Eprint
  {http://arxiv.org/abs/0904.2534} {arXiv:0904.2534 [hep-ph]} \BibitemShut
  {NoStop}%
%%CITATION = ARXIV:0904.2534;%%
\bibitem [{\citenamefont {Chiang}\ \emph {et~al.}(2012)\citenamefont {Chiang},
  \citenamefont {Christensen}, \citenamefont {Ding},\ and\ \citenamefont
  {Han}}]{Chiang:2011kq}%
  \BibitemOpen
  \bibfield  {author} {\bibinfo {author} {\bibfnamefont {C.-W.}\ \bibnamefont
  {Chiang}}, \bibinfo {author} {\bibfnamefont {N.~D.}\ \bibnamefont
  {Christensen}}, \bibinfo {author} {\bibfnamefont {G.-J.}\ \bibnamefont
  {Ding}}, \ and\ \bibinfo {author} {\bibfnamefont {T.}~\bibnamefont {Han}},\
  }\href {\doibase 10.1103/PhysRevD.85.015023} {\bibfield  {journal} {\bibinfo
  {journal} {Phys. Rev.}\ }\textbf {\bibinfo {volume} {D85}},\ \bibinfo {pages}
  {015023} (\bibinfo {year} {2012})},\ \Eprint {http://arxiv.org/abs/1107.5830}
  {arXiv:1107.5830 [hep-ph]} \BibitemShut {NoStop}%
%%CITATION = ARXIV:1107.5830;%%
\bibitem [{\citenamefont {Accomando}\ \emph {et~al.}(2012)\citenamefont
  {Accomando}, \citenamefont {Becciolini}, \citenamefont {De~Curtis},
  \citenamefont {Dominici}, \citenamefont {Fedeli},\ and\ \citenamefont
  {Shepherd-Themistocleous}}]{Accomando:2011eu}%
  \BibitemOpen
  \bibfield  {author} {\bibinfo {author} {\bibfnamefont {E.}~\bibnamefont
  {Accomando}}, \bibinfo {author} {\bibfnamefont {D.}~\bibnamefont
  {Becciolini}}, \bibinfo {author} {\bibfnamefont {S.}~\bibnamefont
  {De~Curtis}}, \bibinfo {author} {\bibfnamefont {D.}~\bibnamefont {Dominici}},
  \bibinfo {author} {\bibfnamefont {L.}~\bibnamefont {Fedeli}}, \ and\ \bibinfo
  {author} {\bibfnamefont {C.}~\bibnamefont {Shepherd-Themistocleous}},\ }\href
  {\doibase 10.1103/PhysRevD.85.115017} {\bibfield  {journal} {\bibinfo
  {journal} {Phys. Rev.}\ }\textbf {\bibinfo {volume} {D85}},\ \bibinfo {pages}
  {115017} (\bibinfo {year} {2012})},\ \Eprint {http://arxiv.org/abs/1110.0713}
  {arXiv:1110.0713 [hep-ph]} \BibitemShut {NoStop}%
%%CITATION = ARXIV:1110.0713;%%
\bibitem [{\citenamefont {Choudhury}\ \emph {et~al.}(2012)\citenamefont
  {Choudhury}, \citenamefont {Godbole},\ and\ \citenamefont
  {Saha}}]{Choudhury:2011cg}%
  \BibitemOpen
  \bibfield  {author} {\bibinfo {author} {\bibfnamefont {D.}~\bibnamefont
  {Choudhury}}, \bibinfo {author} {\bibfnamefont {R.~M.}\ \bibnamefont
  {Godbole}}, \ and\ \bibinfo {author} {\bibfnamefont {P.}~\bibnamefont
  {Saha}},\ }\href {\doibase 10.1007/JHEP01(2012)155} {\bibfield  {journal}
  {\bibinfo  {journal} {JHEP}\ }\textbf {\bibinfo {volume} {01}},\ \bibinfo
  {pages} {155} (\bibinfo {year} {2012})},\ \Eprint
  {http://arxiv.org/abs/1111.1054} {arXiv:1111.1054 [hep-ph]} \BibitemShut
  {NoStop}%
%%CITATION = ARXIV:1111.1054;%%
\bibitem [{\citenamefont {Boos}\ \emph {et~al.}(2007)\citenamefont {Boos},
  \citenamefont {Bunichev}, \citenamefont {Dudko},\ and\ \citenamefont
  {Perfilov}}]{Boos:2006xe}%
  \BibitemOpen
  \bibfield  {author} {\bibinfo {author} {\bibfnamefont {E.}~\bibnamefont
  {Boos}}, \bibinfo {author} {\bibfnamefont {V.}~\bibnamefont {Bunichev}},
  \bibinfo {author} {\bibfnamefont {L.}~\bibnamefont {Dudko}}, \ and\ \bibinfo
  {author} {\bibfnamefont {M.}~\bibnamefont {Perfilov}},\ }\href {\doibase
  10.1016/j.physletb.2007.03.064} {\bibfield  {journal} {\bibinfo  {journal}
  {Phys. Lett.}\ }\textbf {\bibinfo {volume} {B655}},\ \bibinfo {pages} {245}
  (\bibinfo {year} {2007})},\ \Eprint {http://arxiv.org/abs/hep-ph/0610080}
  {arXiv:hep-ph/0610080 [hep-ph]} \BibitemShut {NoStop}%
%%CITATION = HEP-PH/0610080;%%
\bibitem [{\citenamefont {de~Blas}\ \emph {et~al.}(2013)\citenamefont
  {de~Blas}, \citenamefont {Lizana},\ and\ \citenamefont
  {Perez-Victoria}}]{deBlas:2012qp}%
  \BibitemOpen
  \bibfield  {author} {\bibinfo {author} {\bibfnamefont {J.}~\bibnamefont
  {de~Blas}}, \bibinfo {author} {\bibfnamefont {J.~M.}\ \bibnamefont {Lizana}},
  \ and\ \bibinfo {author} {\bibfnamefont {M.}~\bibnamefont {Perez-Victoria}},\
  }\href {\doibase 10.1007/JHEP01(2013)166} {\bibfield  {journal} {\bibinfo
  {journal} {JHEP}\ }\textbf {\bibinfo {volume} {01}},\ \bibinfo {pages} {166}
  (\bibinfo {year} {2013})},\ \Eprint {http://arxiv.org/abs/1211.2229}
  {arXiv:1211.2229 [hep-ph]} \BibitemShut {NoStop}%
%%CITATION = ARXIV:1211.2229;%%
\bibitem [{\citenamefont {Aad}\ \emph {et~al.}(2014{\natexlab{a}})\citenamefont
  {Aad} \emph {et~al.}}]{Aad:2014cka}%
  \BibitemOpen
  \bibfield  {author} {\bibinfo {author} {\bibfnamefont {G.}~\bibnamefont
  {Aad}} \emph {et~al.} (\bibinfo {collaboration} {ATLAS}),\ }\href {\doibase
  10.1103/PhysRevD.90.052005} {\bibfield  {journal} {\bibinfo  {journal}
  {Phys.Rev.}\ }\textbf {\bibinfo {volume} {D90}},\ \bibinfo {pages} {052005}
  (\bibinfo {year} {2014}{\natexlab{a}})},\ \Eprint
  {http://arxiv.org/abs/1405.4123} {arXiv:1405.4123 [hep-ex]} \BibitemShut
  {NoStop}%
%%CITATION = ARXIV:1405.4123;%%
\bibitem [{\citenamefont {Accomando}\ \emph
  {et~al.}(2015{\natexlab{a}})\citenamefont {Accomando}, \citenamefont
  {Barducci}, \citenamefont {De~Curtis}, \citenamefont {Fiaschi}, \citenamefont
  {Moretti},\ and\ \citenamefont
  {Shepherd-Themistocleous}}]{Accomando:2015cva}%
  \BibitemOpen
  \bibfield  {author} {\bibinfo {author} {\bibfnamefont {E.}~\bibnamefont
  {Accomando}}, \bibinfo {author} {\bibfnamefont {D.}~\bibnamefont {Barducci}},
  \bibinfo {author} {\bibfnamefont {S.}~\bibnamefont {De~Curtis}}, \bibinfo
  {author} {\bibfnamefont {J.}~\bibnamefont {Fiaschi}}, \bibinfo {author}
  {\bibfnamefont {S.}~\bibnamefont {Moretti}}, \ and\ \bibinfo {author}
  {\bibfnamefont {C.}~\bibnamefont {Shepherd-Themistocleous}},\ }\href@noop {}
  {\  (\bibinfo {year} {2015}{\natexlab{a}})},\ \Eprint
  {http://arxiv.org/abs/1507.04245} {arXiv:1507.04245 [hep-ph]} \BibitemShut
  {NoStop}%
%%CITATION = ARXIV:1507.04245;%%
\bibitem [{\citenamefont {Arkani-Hamed}\ \emph {et~al.}(1998)\citenamefont
  {Arkani-Hamed}, \citenamefont {Dimopoulos},\ and\ \citenamefont
  {Dvali}}]{ArkaniHamed:1998rs}%
  \BibitemOpen
  \bibfield  {author} {\bibinfo {author} {\bibfnamefont {N.}~\bibnamefont
  {Arkani-Hamed}}, \bibinfo {author} {\bibfnamefont {S.}~\bibnamefont
  {Dimopoulos}}, \ and\ \bibinfo {author} {\bibfnamefont {G.~R.}\ \bibnamefont
  {Dvali}},\ }\href {\doibase 10.1016/S0370-2693(98)00466-3} {\bibfield
  {journal} {\bibinfo  {journal} {Phys. Lett.}\ }\textbf {\bibinfo {volume}
  {B429}},\ \bibinfo {pages} {263} (\bibinfo {year} {1998})},\ \Eprint
  {http://arxiv.org/abs/hep-ph/9803315} {arXiv:hep-ph/9803315} \BibitemShut
  {NoStop}%
%%CITATION = HEP-PH/9803315;%%
\bibitem [{\citenamefont {Antoniadis}\ \emph {et~al.}(1998)\citenamefont
  {Antoniadis}, \citenamefont {Arkani-Hamed}, \citenamefont {Dimopoulos},\ and\
  \citenamefont {Dvali}}]{Antoniadis:1998ig}%
  \BibitemOpen
  \bibfield  {author} {\bibinfo {author} {\bibfnamefont {I.}~\bibnamefont
  {Antoniadis}}, \bibinfo {author} {\bibfnamefont {N.}~\bibnamefont
  {Arkani-Hamed}}, \bibinfo {author} {\bibfnamefont {S.}~\bibnamefont
  {Dimopoulos}}, \ and\ \bibinfo {author} {\bibfnamefont {G.}~\bibnamefont
  {Dvali}},\ }\href {\doibase 10.1016/S0370-2693(98)00860-0} {\bibfield
  {journal} {\bibinfo  {journal} {Phys.Lett.}\ }\textbf {\bibinfo {volume}
  {B436}},\ \bibinfo {pages} {257} (\bibinfo {year} {1998})},\ \Eprint
  {http://arxiv.org/abs/hep-ph/9804398} {arXiv:hep-ph/9804398 [hep-ph]}
  \BibitemShut {NoStop}%
%%CITATION = HEP-PH/9804398;%%
\bibitem [{\citenamefont {Bella}\ \emph {et~al.}(2010)\citenamefont {Bella},
  \citenamefont {Etzion}, \citenamefont {Hod}, \citenamefont {Oz},
  \citenamefont {Silver} \emph {et~al.}}]{Bella:2010sc}%
  \BibitemOpen
  \bibfield  {author} {\bibinfo {author} {\bibfnamefont {G.}~\bibnamefont
  {Bella}}, \bibinfo {author} {\bibfnamefont {E.}~\bibnamefont {Etzion}},
  \bibinfo {author} {\bibfnamefont {N.}~\bibnamefont {Hod}}, \bibinfo {author}
  {\bibfnamefont {Y.}~\bibnamefont {Oz}}, \bibinfo {author} {\bibfnamefont
  {Y.}~\bibnamefont {Silver}},  \emph {et~al.},\ }\href {\doibase
  10.1007/JHEP09(2010)025} {\bibfield  {journal} {\bibinfo  {journal} {JHEP}\
  }\textbf {\bibinfo {volume} {1009}},\ \bibinfo {pages} {025} (\bibinfo {year}
  {2010})},\ \Eprint {http://arxiv.org/abs/1004.2432} {arXiv:1004.2432
  [hep-ex]} \BibitemShut {NoStop}%
%%CITATION = ARXIV:1004.2432;%%
\bibitem [{\citenamefont {Accomando}(2015)}]{Accomando:2015rsa}%
  \BibitemOpen
  \bibfield  {author} {\bibinfo {author} {\bibfnamefont {E.}~\bibnamefont
  {Accomando}},\ }\href {\doibase 10.1142/S0217732315400106} {\bibfield
  {journal} {\bibinfo  {journal} {Mod. Phys. Lett.}\ }\textbf {\bibinfo
  {volume} {A30}},\ \bibinfo {pages} {1540010} (\bibinfo {year}
  {2015})}\BibitemShut {NoStop}%
%%CITATION = MPLAE,A30,1540010;%%
\bibitem [{\citenamefont {Accomando}\ \emph {et~al.}(2000)\citenamefont
  {Accomando}, \citenamefont {Antoniadis},\ and\ \citenamefont
  {Benakli}}]{Accomando:1999sj}%
  \BibitemOpen
  \bibfield  {author} {\bibinfo {author} {\bibfnamefont {E.}~\bibnamefont
  {Accomando}}, \bibinfo {author} {\bibfnamefont {I.}~\bibnamefont
  {Antoniadis}}, \ and\ \bibinfo {author} {\bibfnamefont {K.}~\bibnamefont
  {Benakli}},\ }\href {\doibase 10.1016/S0550-3213(00)00123-1} {\bibfield
  {journal} {\bibinfo  {journal} {Nucl. Phys.}\ }\textbf {\bibinfo {volume}
  {B579}},\ \bibinfo {pages} {3} (\bibinfo {year} {2000})},\ \Eprint
  {http://arxiv.org/abs/hep-ph/9912287} {arXiv:hep-ph/9912287 [hep-ph]}
  \BibitemShut {NoStop}%
%%CITATION = HEP-PH/9912287;%%
\bibitem [{\citenamefont {Boos}\ \emph {et~al.}(2012)\citenamefont {Boos},
  \citenamefont {Volobuev}, \citenamefont {Perfilov},\ and\ \citenamefont
  {Smolyakov}}]{Boos:2011ib}%
  \BibitemOpen
  \bibfield  {author} {\bibinfo {author} {\bibfnamefont {E.~E.}\ \bibnamefont
  {Boos}}, \bibinfo {author} {\bibfnamefont {I.~P.}\ \bibnamefont {Volobuev}},
  \bibinfo {author} {\bibfnamefont {M.~A.}\ \bibnamefont {Perfilov}}, \ and\
  \bibinfo {author} {\bibfnamefont {M.~N.}\ \bibnamefont {Smolyakov}},\ }\href
  {\doibase 10.1007/s11232-012-0009-6} {\bibfield  {journal} {\bibinfo
  {journal} {Theor. Math. Phys.}\ }\textbf {\bibinfo {volume} {170}},\ \bibinfo
  {pages} {90} (\bibinfo {year} {2012})},\ \Eprint
  {http://arxiv.org/abs/1106.2400} {arXiv:1106.2400 [hep-ph]} \BibitemShut
  {NoStop}%
%%CITATION = ARXIV:1106.2400;%%
\bibitem [{\citenamefont {Accomando}\ \emph
  {et~al.}(2015{\natexlab{b}})\citenamefont {Accomando}, \citenamefont
  {Belyaev}, \citenamefont {Fiaschi}, \citenamefont {Mimasu}, \citenamefont
  {Moretti} \emph {et~al.}}]{Accomando:2015cfa}%
  \BibitemOpen
  \bibfield  {author} {\bibinfo {author} {\bibfnamefont {E.}~\bibnamefont
  {Accomando}}, \bibinfo {author} {\bibfnamefont {A.}~\bibnamefont {Belyaev}},
  \bibinfo {author} {\bibfnamefont {J.}~\bibnamefont {Fiaschi}}, \bibinfo
  {author} {\bibfnamefont {K.}~\bibnamefont {Mimasu}}, \bibinfo {author}
  {\bibfnamefont {S.}~\bibnamefont {Moretti}},  \emph {et~al.},\ }\href@noop {}
  {\  (\bibinfo {year} {2015}{\natexlab{b}})},\ \Eprint
  {http://arxiv.org/abs/1503.02672} {arXiv:1503.02672 [hep-ph]} \BibitemShut
  {NoStop}%
%%CITATION = ARXIV:1503.02672;%%
\bibitem [{\citenamefont {Accomando}\ \emph
  {et~al.}(2013{\natexlab{b}})\citenamefont {Accomando}, \citenamefont
  {Mimasu},\ and\ \citenamefont {Moretti}}]{Accomando:2013dia}%
  \BibitemOpen
  \bibfield  {author} {\bibinfo {author} {\bibfnamefont {E.}~\bibnamefont
  {Accomando}}, \bibinfo {author} {\bibfnamefont {K.}~\bibnamefont {Mimasu}}, \
  and\ \bibinfo {author} {\bibfnamefont {S.}~\bibnamefont {Moretti}},\ }\href
  {\doibase 10.1007/JHEP07(2013)154} {\bibfield  {journal} {\bibinfo  {journal}
  {JHEP}\ }\textbf {\bibinfo {volume} {07}},\ \bibinfo {pages} {154} (\bibinfo
  {year} {2013}{\natexlab{b}})},\ \Eprint {http://arxiv.org/abs/1304.4494}
  {arXiv:1304.4494 [hep-ph]} \BibitemShut {NoStop}%
%%CITATION = ARXIV:1304.4494;%%
\bibitem [{\citenamefont {Kaplan}\ and\ \citenamefont
  {Georgi}(1984)}]{Kaplan:1983fs}%
  \BibitemOpen
  \bibfield  {author} {\bibinfo {author} {\bibfnamefont {D.~B.}\ \bibnamefont
  {Kaplan}}\ and\ \bibinfo {author} {\bibfnamefont {H.}~\bibnamefont
  {Georgi}},\ }\href {\doibase 10.1016/0370-2693(84)91177-8} {\bibfield
  {journal} {\bibinfo  {journal} {Phys.Lett.}\ }\textbf {\bibinfo {volume}
  {B136}},\ \bibinfo {pages} {183} (\bibinfo {year} {1984})}\BibitemShut
  {NoStop}%
%%CITATION = PHLTA,B136,183;%%
\bibitem [{\citenamefont {Agashe}\ \emph {et~al.}(2005)\citenamefont {Agashe},
  \citenamefont {Contino},\ and\ \citenamefont {Pomarol}}]{Agashe:2004rs}%
  \BibitemOpen
  \bibfield  {author} {\bibinfo {author} {\bibfnamefont {K.}~\bibnamefont
  {Agashe}}, \bibinfo {author} {\bibfnamefont {R.}~\bibnamefont {Contino}}, \
  and\ \bibinfo {author} {\bibfnamefont {A.}~\bibnamefont {Pomarol}},\ }\href
  {\doibase 10.1016/j.nuclphysb.2005.04.035} {\bibfield  {journal} {\bibinfo
  {journal} {Nucl.Phys.}\ }\textbf {\bibinfo {volume} {B719}},\ \bibinfo
  {pages} {165} (\bibinfo {year} {2005})},\ \Eprint
  {http://arxiv.org/abs/hep-ph/0412089} {arXiv:hep-ph/0412089 [hep-ph]}
  \BibitemShut {NoStop}%
%%CITATION = HEP-PH/0412089;%%
\bibitem [{\citenamefont {De~Curtis}\ \emph {et~al.}(2012)\citenamefont
  {De~Curtis}, \citenamefont {Redi},\ and\ \citenamefont
  {Tesi}}]{DeCurtis:2011yx}%
  \BibitemOpen
  \bibfield  {author} {\bibinfo {author} {\bibfnamefont {S.}~\bibnamefont
  {De~Curtis}}, \bibinfo {author} {\bibfnamefont {M.}~\bibnamefont {Redi}}, \
  and\ \bibinfo {author} {\bibfnamefont {A.}~\bibnamefont {Tesi}},\ }\href
  {\doibase 10.1007/JHEP04(2012)042} {\bibfield  {journal} {\bibinfo  {journal}
  {JHEP}\ }\textbf {\bibinfo {volume} {1204}},\ \bibinfo {pages} {042}
  (\bibinfo {year} {2012})},\ \Eprint {http://arxiv.org/abs/1110.1613}
  {arXiv:1110.1613 [hep-ph]} \BibitemShut {NoStop}%
%%CITATION = ARXIV:1110.1613;%%
\bibitem [{\citenamefont {Barducci}\ \emph {et~al.}(2013)\citenamefont
  {Barducci}, \citenamefont {Belyaev}, \citenamefont {De~Curtis}, \citenamefont
  {Moretti},\ and\ \citenamefont {Pruna}}]{Barducci:2012kk}%
  \BibitemOpen
  \bibfield  {author} {\bibinfo {author} {\bibfnamefont {D.}~\bibnamefont
  {Barducci}}, \bibinfo {author} {\bibfnamefont {A.}~\bibnamefont {Belyaev}},
  \bibinfo {author} {\bibfnamefont {S.}~\bibnamefont {De~Curtis}}, \bibinfo
  {author} {\bibfnamefont {S.}~\bibnamefont {Moretti}}, \ and\ \bibinfo
  {author} {\bibfnamefont {G.~M.}\ \bibnamefont {Pruna}},\ }\href {\doibase
  10.1007/JHEP04(2013)152} {\bibfield  {journal} {\bibinfo  {journal} {JHEP}\
  }\textbf {\bibinfo {volume} {1304}},\ \bibinfo {pages} {152} (\bibinfo {year}
  {2013})},\ \Eprint {http://arxiv.org/abs/1210.2927} {arXiv:1210.2927
  [hep-ph]} \BibitemShut {NoStop}%
%%CITATION = ARXIV:1210.2927;%%
\bibitem [{\citenamefont {Grojean}\ \emph {et~al.}(2013)\citenamefont
  {Grojean}, \citenamefont {Matsedonskyi},\ and\ \citenamefont
  {Panico}}]{Grojean:2013qca}%
  \BibitemOpen
  \bibfield  {author} {\bibinfo {author} {\bibfnamefont {C.}~\bibnamefont
  {Grojean}}, \bibinfo {author} {\bibfnamefont {O.}~\bibnamefont
  {Matsedonskyi}}, \ and\ \bibinfo {author} {\bibfnamefont {G.}~\bibnamefont
  {Panico}},\ }\href {\doibase 10.1007/JHEP10(2013)160} {\bibfield  {journal}
  {\bibinfo  {journal} {JHEP}\ }\textbf {\bibinfo {volume} {1310}},\ \bibinfo
  {pages} {160} (\bibinfo {year} {2013})},\ \Eprint
  {http://arxiv.org/abs/1306.4655} {arXiv:1306.4655 [hep-ph]} \BibitemShut
  {NoStop}%
%%CITATION = ARXIV:1306.4655;%%
\bibitem [{\citenamefont {Aad}\ \emph {et~al.}(2014{\natexlab{b}})\citenamefont
  {Aad} \emph {et~al.}}]{Aad:2014pha}%
  \BibitemOpen
  \bibfield  {author} {\bibinfo {author} {\bibfnamefont {G.}~\bibnamefont
  {Aad}} \emph {et~al.} (\bibinfo {collaboration} {ATLAS}),\ }\href {\doibase
  10.1016/j.physletb.2014.08.039} {\bibfield  {journal} {\bibinfo  {journal}
  {Phys.Lett.}\ }\textbf {\bibinfo {volume} {B737}},\ \bibinfo {pages} {223}
  (\bibinfo {year} {2014}{\natexlab{b}})},\ \Eprint
  {http://arxiv.org/abs/1406.4456} {arXiv:1406.4456 [hep-ex]} \BibitemShut
  {NoStop}%
%%CITATION = ARXIV:1406.4456;%%
\bibitem [{\citenamefont {Chatrchyan}\ \emph
  {et~al.}(2014{\natexlab{a}})\citenamefont {Chatrchyan} \emph
  {et~al.}}]{Chatrchyan:2013wfa}%
  \BibitemOpen
  \bibfield  {author} {\bibinfo {author} {\bibfnamefont {S.}~\bibnamefont
  {Chatrchyan}} \emph {et~al.} (\bibinfo {collaboration} {CMS}),\ }\href
  {\doibase 10.1103/PhysRevLett.112.171801} {\bibfield  {journal} {\bibinfo
  {journal} {Phys.Rev.Lett.}\ }\textbf {\bibinfo {volume} {112}},\ \bibinfo
  {pages} {171801} (\bibinfo {year} {2014}{\natexlab{a}})},\ \Eprint
  {http://arxiv.org/abs/1312.2391} {arXiv:1312.2391 [hep-ex]} \BibitemShut
  {NoStop}%
%%CITATION = ARXIV:1312.2391;%%
\bibitem [{\citenamefont {Chatrchyan}\ \emph
  {et~al.}(2014{\natexlab{b}})\citenamefont {Chatrchyan} \emph
  {et~al.}}]{Chatrchyan:2013uxa}%
  \BibitemOpen
  \bibfield  {author} {\bibinfo {author} {\bibfnamefont {S.}~\bibnamefont
  {Chatrchyan}} \emph {et~al.} (\bibinfo {collaboration} {CMS}),\ }\href
  {\doibase 10.1016/j.physletb.2014.01.006} {\bibfield  {journal} {\bibinfo
  {journal} {Phys.Lett.}\ }\textbf {\bibinfo {volume} {B729}},\ \bibinfo
  {pages} {149} (\bibinfo {year} {2014}{\natexlab{b}})},\ \Eprint
  {http://arxiv.org/abs/1311.7667} {arXiv:1311.7667 [hep-ex]} \BibitemShut
  {NoStop}%
%%CITATION = ARXIV:1311.7667;%%
\bibitem [{CMS(2013)}]{CMS-PAS-B2G-13-003}%
  \BibitemOpen
  \href {http://cds.cern.ch/record/1629574} {\emph {\bibinfo {title} {{Search
  for Vector-Like b' Pair Production with Multilepton Final States in pp
  collisions at sqrt(s) = 8 TeV}}}},\ \bibinfo {type} {Tech. Rep.}\ \bibinfo
  {number} {CMS-PAS-B2G-13-003}\ (\bibinfo  {institution} {CERN},\ \bibinfo
  {address} {Geneva},\ \bibinfo {year} {2013})\BibitemShut {NoStop}%
\bibitem [{\citenamefont {Collaboration}(2015)}]{CMS:2015alb}%
  \BibitemOpen
  \bibfield  {author} {\bibinfo {author} {\bibfnamefont {C.}~\bibnamefont
  {Collaboration}} (\bibinfo {collaboration} {CMS}),\ }\href@noop {} {\
  (\bibinfo {year} {2015})}\BibitemShut {NoStop}%
%%CITATION = CMS-PAS-B2G-15-006;%%
\bibitem [{\citenamefont {Pappadopulo}\ \emph {et~al.}(2014)\citenamefont
  {Pappadopulo}, \citenamefont {Thamm}, \citenamefont {Torre},\ and\
  \citenamefont {Wulzer}}]{Pappadopulo:2014qza}%
  \BibitemOpen
  \bibfield  {author} {\bibinfo {author} {\bibfnamefont {D.}~\bibnamefont
  {Pappadopulo}}, \bibinfo {author} {\bibfnamefont {A.}~\bibnamefont {Thamm}},
  \bibinfo {author} {\bibfnamefont {R.}~\bibnamefont {Torre}}, \ and\ \bibinfo
  {author} {\bibfnamefont {A.}~\bibnamefont {Wulzer}},\ }\href {\doibase
  10.1007/JHEP09(2014)060} {\bibfield  {journal} {\bibinfo  {journal} {JHEP}\
  }\textbf {\bibinfo {volume} {1409}},\ \bibinfo {pages} {060} (\bibinfo {year}
  {2014})},\ \Eprint {http://arxiv.org/abs/1402.4431} {arXiv:1402.4431
  [hep-ph]} \BibitemShut {NoStop}%
%%CITATION = ARXIV:1402.4431;%%
\bibitem [{\citenamefont {del Aguila}\ \emph {et~al.}(2010)\citenamefont {del
  Aguila}, \citenamefont {de~Blas},\ and\ \citenamefont
  {Perez-Victoria}}]{delAguila:2010mx}%
  \BibitemOpen
  \bibfield  {author} {\bibinfo {author} {\bibfnamefont {F.}~\bibnamefont {del
  Aguila}}, \bibinfo {author} {\bibfnamefont {J.}~\bibnamefont {de~Blas}}, \
  and\ \bibinfo {author} {\bibfnamefont {M.}~\bibnamefont {Perez-Victoria}},\
  }\href {\doibase 10.1007/JHEP09(2010)033} {\bibfield  {journal} {\bibinfo
  {journal} {JHEP}\ }\textbf {\bibinfo {volume} {09}},\ \bibinfo {pages} {033}
  (\bibinfo {year} {2010})},\ \Eprint {http://arxiv.org/abs/1005.3998}
  {arXiv:1005.3998 [hep-ph]} \BibitemShut {NoStop}%
%%CITATION = ARXIV:1005.3998;%%
\bibitem [{\citenamefont {Langacker}\ \emph {et~al.}(1984)\citenamefont
  {Langacker}, \citenamefont {Robinett},\ and\ \citenamefont
  {Rosner}}]{Langacker:1984dc}%
  \BibitemOpen
  \bibfield  {author} {\bibinfo {author} {\bibfnamefont {P.}~\bibnamefont
  {Langacker}}, \bibinfo {author} {\bibfnamefont {R.~W.}\ \bibnamefont
  {Robinett}}, \ and\ \bibinfo {author} {\bibfnamefont {J.~L.}\ \bibnamefont
  {Rosner}},\ }\href {\doibase 10.1103/PhysRevD.30.1470} {\bibfield  {journal}
  {\bibinfo  {journal} {Phys. Rev.}\ }\textbf {\bibinfo {volume} {D30}},\
  \bibinfo {pages} {1470} (\bibinfo {year} {1984})}\BibitemShut {NoStop}%
%%CITATION = PHRVA,D30,1470;%%
\end{thebibliography}%

\end{document}